 \documentclass[preprint,3p,times]{elsarticle}

\usepackage{amsmath}
\usepackage{amssymb}
\usepackage{tensor}
\usepackage{stmaryrd}
\usepackage{tikz-cd}
\usepackage{amsthm}
\usepackage{abraces}
\usepackage{hyperref}

\usepackage{aliascnt}
\newtheorem{thm}{Theorem}

\newaliascnt{lem}{thm}
\newtheorem{lem}[lem]{Lemma}
\aliascntresetthe{lem}

\newaliascnt{prop}{thm}
\newtheorem{prop}[prop]{Proposition}
\aliascntresetthe{prop}

\newaliascnt{concl}{thm}
\newtheorem{concl}[concl]{Conclusion}
\aliascntresetthe{concl}

\newcommand{\wrt}{w.\,r.\,t.}
\newcommand{\eg}{e.\,g.}
\newcommand{\Eg}{E.\,g.}
\newcommand{\st}{s.\,t.}
\newcommand{\ie}{i.\,e.}

\newcommand{\cf}{cf.}

\newcommand{\R}{\mathbb{R}}
\newcommand{\stsurf}{\mathcal{M}}
\newcommand{\surf}{\mathcal{S}}
\newcommand{\timespace}{\mathcal{T}}

\newcommand{\tangent}[2][]{\tensor{\operatorname{T}\!}{#1}#2}
\newcommand{\tangentStar}[2][]{\tensor*{\operatorname{T}\!}{#1}#2}
\newcommand{\tangentS}[1][]{\tangent[#1]{\surf}}
\newcommand{\tangentR}[1][]{\tangent[#1]{\R^3|_{\surf}}}
\newcommand{\tangentstR}[1][]{\tangent[#1]{\R^4|_{\stsurf}}}
\newcommand{\tangentstS}[1][]{\tangent[#1]{\stsurf}}
\newcommand{\tangentstSStar}[1][]{\tangentStar[#1]{\stsurf}}

\newcommand{\symS}[1][]{\tensor{\operatorname{Sym}}{#1}\!\surf}

\newcommand{\QS}{\operatorname{\mathcal{Q}}\!\surf}
\newcommand{\TFS}{\operatorname{TF}\!\surf}
\newcommand{\SOS}{\operatorname{SO}\!\surf}

\newcommand{\symmapstS}[1][1]{\tensor*{\operatorname{T}\!}{^{#1}_{#1}}\stsurf}

\newcommand{\tanproj}[3][]{\operatorname{P}_{\!\!\!#2^{#1}}\!{#3}}
\newcommand{\taninststS}[1][]{\tanproj[#1]{\surf}\stsurf}
\newcommand{\tantransstS}[1][]{\tanproj[#1]{\transdir}\stsurf}
\newcommand{\tanshufstS}[2][]{\tanproj[#1]{#2}\stsurf}

\newcommand{\sttensorb}[2][]{\tensor{\operatorname{ST}\!}{#1}#2}
\newcommand{\sttensorbS}[1][]{\sttensorb[#1]{\surf}}

\newcommand{\shuffles}[2]{\operatorname{Sh}^{#1}_{#2}}

\newcommand{\stpara}{\boldsymbol{X}}
\newcommand{\paraC}{Z}
\newcommand{\para}{\boldsymbol{\paraC}}

\newcommand{\inner}[3][]{\left\langle #2 , #3  \right\rangle_{\!#1}}
\newcommand{\innerR}[3][]{\inner[{\tangentR[#1]}]{#2}{#3}}
\newcommand{\innerS}[3][]{\inner[{\tangentS[#1]}]{#2}{#3}}
\newcommand{\innerstR}[2]{\inner[\R^4]{#1}{#2}}
\newcommand{\innerstS}[3][]{\inner[{\tangentstS[#1]}]{#2}{#3}}

\newcommand{\stproj}[2][]{\operatorname{\boldsymbol{\mathfrak{P}}}^{#1}_{\!#2}}
\newcommand{\transproj}[1][]{\stproj[#1]{\transdir}}
\newcommand{\instproj}[1][]{\stproj[#1]{\surf}}

\newcommand{\proj}[1]{\pi_{#1}}
\newcommand{\projsurf}{\proj{\surf}}

\newcommand{\Id}[1][]{\operatorname{Id}_{#1}}
\newcommand{\IdS}{\Id[\surf]}
\newcommand{\IdstS}{\Id[\stsurf]}
\newcommand{\shId}[2]{\operatorname{Id}^{#1}_{#2}}

\newcommand{\matdst}[1][]{\operatorname{D}^{\m}_{#1}\!}
\newcommand{\matdhatst}[1]{\operatorname{\widehat{D}}^{\m}_{#1}\!}
\newcommand{\matd}[1][]{\operatorname{d}^{\m}_{#1}\!}

\newcommand{\totald}[1][]{\frac{\operatorname{D}\! #1}{\operatorname{D}\! t}}
\newcommand{\overdot}[1]{\dot{\aoverbrace[L1R]{#1}}}

\newcommand{\Lie}{\operatorname{\mathcal{L}}\!}
\newcommand{\instLie}{\operatorname{\mathfrak{L}}}
\newcommand{\condst}[2][]{\operatorname{L}^{\!\!#2}_{#1}\!}
\newcommand{\cond}[2][]{\operatorname{l}^{#2}_{#1}\!}

\newcommand{\instcond}[1]{\instLie^{#1}\!}

\newcommand{\jaudst}{\operatorname{J}\!}
\newcommand{\jaud}[1][]{\operatorname{j}_{#1}\!}
\newcommand{\instjaud}{\operatorname{\mathfrak{J}}\!}

\newcommand{\nablatan}{\nabla_{\!\operatorname{tan}}}

\newcommand{\shflat}[1][]{\flat_{#1}}
\newcommand{\shsharp}[1][]{\sharp_{#1}}

\newcommand{\rot}{\operatorname{rot}}
\newcommand{\trace}{\operatorname{tr}}

\newcommand{\underdot}[1][]{\operatornamewithlimits{\cdot}_{#1}}

\newcommand{\qmap}{\boldsymbol{\rho}}

\newcommand{\yb}{\boldsymbol{y}}
\newcommand{\wb}{\boldsymbol{w}}
\newcommand{\Wb}{\boldsymbol{W}}

\newcommand{\Eb}{\boldsymbol{E}}

\newcommand{\m}{\mathfrak{m}}
\newcommand{\e}{\mathfrak{e}}
\newcommand{\C}{\!\text{C}}

\newcommand{\tsigma}{\tilde{\sigma}}
\newcommand{\talpha}{\tilde{\alpha}}
\newcommand{\bsigma}{\check{\sigma}}
\newcommand{\balpha}{\check{\alpha}}
\newcommand{\tbeta}{\tilde{\beta}}
\newcommand{\bbeta}{\bar{\beta}}
\DeclareRobustCommand{\rbeta}{\text{\rotatebox[]{180}{$\beta$}}}

\newcommand{\lct}{\boldsymbol{\epsilon}}
\newcommand{\lctC}{\epsilon}

\newcommand{\shuffbase}[2]{\tensor*{\boldsymbol{E}}{^{#1}_{#2}}}
\newcommand{\sttenbase}[1]{\boldsymbol{e}^{#1}}

\newcommand{\normalC}{N}
\newcommand{\normal}{\boldsymbol{\normalC}}

\newcommand{\velonor}{\nu}

\newcommand{\obveloC}{V}
\newcommand{\obvelo}{\boldsymbol{\obveloC}}
\newcommand{\obveloSC}{v}
\newcommand{\obveloS}{\boldsymbol{\obveloSC}}

\newcommand{\obaccelC}{A}
\newcommand{\obaccel}{\boldsymbol{\obaccelC}}
\newcommand{\obaccelSC}{a}
\newcommand{\obaccelS}{\boldsymbol{\obaccelSC}}
\newcommand{\obaccelnor}{\lambda}

\newcommand{\matveloC}{V_{\!\m}}
\newcommand{\matvelo}{\boldsymbol{V}_{\!\m}}
\newcommand{\matveloSC}{v_\m}
\newcommand{\matveloS}{\boldsymbol{v}_\m}

\newcommand{\mataccelC}{A_\m}
\newcommand{\mataccel}{\boldsymbol{A}_\m}

\newcommand{\mataccelS}{\boldsymbol{a}_\m}
\newcommand{\mataccelnor}{\lambda_{\m}}

\newcommand{\relveloC}{u}
\newcommand{\relvelo}{\boldsymbol{u}}

\newcommand{\shopC}{I\!I}
\newcommand{\shop}{\boldsymbol{\shopC}}

\newcommand{\BopC}{\mathcal{B}}
\newcommand{\Bop}{\boldsymbol{\BopC}}

\newcommand{\bopC}{b}
\newcommand{\bop}{\boldsymbol{\bopC}}

\newcommand{\transdirC}{\tau}
\newcommand{\transdir}{\boldsymbol{\transdirC}}

\newcommand{\matdir}{\boldsymbol{\tau}_{\!\!\m}}

\newcommand{\stgC}{\eta}
\newcommand{\stg}{\boldsymbol{\stgC}}
\newcommand{\gC}{g}
\newcommand{\g}{\boldsymbol{\gC}}

\newcommand{\starbtC}{R}
\newcommand{\starbt}{\boldsymbol{\starbtC}}
\newcommand{\indexstarbt}[2][]{\tensor*{\starbtC}{_{#1}^{#2}}}
\newcommand{\hstarbt}{\widehat{\starbt}}
\newcommand{\indexhstarbt}[2][]{\tensor*{\widehat{\starbtC}}{_{#1}^{#2}}}

\newcommand{\arbtC}{r}
\newcommand{\arbt}{\boldsymbol{\arbtC}}

\newcommand{\varstarbtC}{Q}
\newcommand{\varstarbt}{\boldsymbol{\varstarbtC}}
\newcommand{\vararbtC}{q}
\newcommand{\vararbt}{\boldsymbol{\vararbtC}}

\newcommand{\formComma}{\,\text{,}}
\newcommand{\formPeriod}{\,\text{.}}

\newcommand{\indexten}[2]{\tensor{\left[#1\right]}{#2}}
\newcommand{\indexsymten}[2]{\tensor*{\left[#1\right]}{#2}}

\newcommand{\bbrackets}[2][]{\left\llbracket #2 \right\rrbracket_{#1}}
\newcommand{\btensor}[1]{\bbrackets{\begin{matrix}#1\end{matrix}}}
\newcommand{\bsmalltensor}[1]{\bbrackets{\begin{smallmatrix}#1\end{smallmatrix}}}
\newcommand{\tensorstS}[2][]{\begin{bmatrix} #2 \end{bmatrix}_{\tangentstS[#1]}}
\newcommand{\tensorstR}[2][]{\begin{bmatrix} #2 \end{bmatrix}_{\tangentstR[#1]}}
\newcommand{\tensorS}[2][]{\begin{bmatrix} #2 \end{bmatrix}_{\tangentS[#1]}}

\journal{Journal of Geometry and Physics}

\begin{document}

\begin{frontmatter}

\title{Observer-invariant time derivatives on moving surfaces}

\author[1]{Ingo Nitschke\corref{cor1}}
\cortext[cor1]{Corresponding author: ingo.nitschke@tu-dresden.de}
\author[1,2,3]{Axel Voigt}

\address[1]{Institut f{\"u}r Wissenschaftliches Rechnen, Technische Universit{\"a}t Dresden, 01062 Dresden, Germany}
\address[2]{Dresden Center for Computational Materials Science (DCMS), Technische Universit{\"a}t Dresden, 01062 Dresden, Germany}
\address[3]{Center for Systems Biology Dresden (CSBD), Pfotenhauerstr. 108, 01307 Dresden, Germany}

\begin{abstract}
 Observer-invariance is regarded as a minimum requirement for an appropriate definition of time derivatives. We derive various time derivatives systematically from a spacetime setting, where observer-invariance is a special case of a covariance principle and covered by Ricci-calculus. The analysis is considered for tangential $n$-tensor fields on moving surfaces and provides formulations which are applicable for numerical computations. For various special cases, e.g., vector fields ($n = 1$) and symmetric and trace-less tensor fields ($n = 2$) we compare material and convected derivatives and demonstrate the different underlying physics.
\end{abstract}


\end{frontmatter}


\section{Introduction}
    Observer-invariant time derivatives are inevitable for dealing with general equations of motion in an unsteady material domain.
    They comprise not only specific rates of change independently of their observation, but also transport mechanism reflecting a certain inertia in
    the considered quantity induced by material motions.
    We are mainly interested in observer-invariance \wrt\ a \emph{moving surface} $\surf$, 
    \ie\ a 2-dimensional smooth orientable Riemannian manifold embedded in the Euclidean space $ \R^3 $. Equations of motion on moving surfaces are of interest in various disciplines. Prominent examples are thin elastic films with stress tensors as quantities of interest, see, e.g., \cite{Chapelle2003,Bischoff2017}. Other examples are fluidic interfaces, with the tangential fluid velocity and pressure/surface tension as unknowns \cite{Scriven1960,Arroyo2009,Torres-Sanchez2020,Reuther2020}, 
    or surface polar and nematic liquid crystals, with tangential director and Q-tensor fields as unknowns \cite{Nitschke2019a,Nitschke2019}, which is e.g. used to model the cellular cortex or epithelia tissue \cite{Doostmohammadi2018}. In addition there are problems with surface scalar quantities, such as concentrations, e.g. of surfactants, proteins or lipids, see e.g. \cite{Teigen2011,Marth2014,Elliott2010}. But also higher order surface tensor fields are found in applications, e.g. in graphics applications, such as surface parameterization and remeshing, painterly rendering and pen-and-ink sketching, and texture synthesis \cite{Palacios2007}. With the exception of the last examples, which are not determined by physics, and the surface scalar quantities for which transport is described by the Leibniz formula/transport theorem, see e.g. \cite{Dziuk2007}, the different underlying physics for surface vector- and tensor-fields in these examples imply different transport mechanisms and thus different time derivatives. The goal of this paper is to systematically develop, analyse and compare observer-invariant time derivatives for the mentioned examples and more general n-tensor fields on moving surfaces. We thereby regard observer-invariance as a minimum requirement on the calculus.

    There are plenty of principles and conceptualities concerning transformation properties for physical materials and/or experimenting observers, \eg\ (material) frame-independence, objectivity, form-invariance, Galilean-invariance, etc., with terminology itself not being unambiguous in parts depending on restrictions, spacetime settings and even applying authors, which leads to
    misunderstandings, confusion and lasting controversies, see \cite{Frewer2009} for an extensive and enlightening discussion about this topic. We thus first need to clarify our understanding on observer-invariance. We think of an \emph{observer} as an unbodied being
    capable of sensing the whole considered physical situation or experiment without any influencing of physical states.
    Therefore, the physical space is independent of their observers, albeit the opposite does not have to be true. 
    Thus, we call a physical statement, \eg\ an identity or numerical term, \emph{observer-invariant}, 
    if this statement is considered equal by every two arbitrary admissible observer \wrt\ their communication. 

    The moving surface $\surf$ is shaped by a continuum of moving material particles in $ \R^3 $. We assume that material particles do not overlay each other and the motion of them is smooth in time and space. The \emph{material observer}, short for all material particles in motion, is sufficient to shape the moving surface, but it is not necessary.
    Therefore, the material observer is one representative of an equivalence class of \emph{observer}, which are sufficient for $ \surf $. 
    Hence the material observer occupy the Lagrangian perspective, whereas for surfaces with stationary shape a stationary observer reveal
    the Eulerian perspective.
    We generalize the latter example to general unsteady surfaces by the \emph{transversal observer}, whereby observer particles are only moving in normal direction
    \wrt\ to the surface, \ie\ for a 2-dimensional inhabitant of $ \surf $ this might appears as an Eulerian perspective.
    Unfortunately, the equivalence class of observers is not consistent \wrt\ differential calculus concerning time $ t $,
    \eg\ the partial time derivative $ \partial_t $ of a quantity differs between different observers of the same class, since 
    every observer has their own relative insight how things change in time.
    We call an operator \emph{observer-invariant} if it is invariant within the observer class depicturing a moving surface $\surf$.
    
    The main issue to develop observer-invariant time derivatives is that the time $ t $ is not a coordinate of $ \surf $, but rather
    a parameter to describe time-dependencies \wrt\ an observer and the relation between time and space.
    Nevertheless, considering an observer within a spacetime locally, \st\ time $ t $ is a genuine local coordinate of this spacetime, can be a game changer
    if the spacetime is pseudo-Riemannian at least.
    In this situation observer-invariance is a special case of a covariance principle \wrt\ the choice of spacetime coordinates. For tensorial quantities this
    is already covered by Ricci calculus.
    All of these spacetime observer, originated from the observer class, form again a equivalence class of spacetime observer, which are sufficient
    to shape a (2+1)-dimensional pseudo-Riemannian spacetime $ \stsurf $.
    For instance in context of Einsteins general theory of relativity, $ \stsurf $ would be a Lorentzian manifold. 
    But this would go a bit over the top in many situations, where motions are much slower than the speed of light and changes in gravity are negligible, as in all mentioned examples above.
    We thus consider $ \stsurf $ embeddable in a (3+1)-dimensional Euclidean spacetime, 
    \ie\ $ \stsurf\subset\R^4 $ is a \emph{spacetime surface} analogously to the definition of surfaces above.
    Hence time is a global measurement, \ie\ every \emph{event}, which is defined as a point in spacetime, is equipped with exactly the same clock and all clocks stay instantaneously synchronized.
    As we will see in \autoref{sec:spacetime_surface_and_tensor_bundle} $ \stsurf $ can be seen as a (2+1)-dimensional curvilinear version
    of a classical Newtonian spacetime, see \cite{Frewer2009,Friedrichs1928,Rodrigues1995}
    .
    One side effect considering the spacetime surface $ \stsurf $ instead of the moving surface $ \surf $ is 
    that $ n $-tensor fields have $ (3^n - 2^n) $  more degrees of freedom.
    There is not a general approach to augment given \emph{surface tensor fields} in $ \tangentS[^n] $ to 
    \emph{spacetime tensor fields} in $ \tangentstS[^n] $. However, to establish a connection between both will be useful 
    and is derived in \autoref{sec:spacetime_tensor_bundles} by an orthogonal decomposition of $\tangentstS[^n]$
    
    We call a tensor field \emph{instantaneous}, if it is instantaneous observable at a fixed time $ t $.
    For instance, the vector field of directors of polar liquid crystals is instantaneous, also known as space-like vector field.
    Instantaneous spacetime tensor fields are characterized numerically by vanishing contravariant coordinate functions, which are related to
    the time base vector.
    Apparently a spacetime vector field describing the tangential fluid velocity cannot be instantaneous, since moving particles have to traverse time and perhaps space,
    \ie\ at fixed time $ t $, where no information about past and future is available, we cannot determine a velocity vector.
    A spacetime vector field is called \emph{transversal}, also known as time-like, if it is orthogonal to a instantaneous spacetime  vector field,
    \eg\ locally, the spacetime velocity direction of a transversal or Eulerian observer particle is transversal.
    As a consequence, every spacetime vector field in $ \tangentstS $ has its unambiguous instantaneous and transversal part.
    We show in \autoref{sec:spacetime_surface_and_tensor_bundle} that this is also generalizable to spacetime $ n $-tensor fields.
    
    One could note that developing observer-invariant operators on $ \tangentstS[^n] $ is a trivial task, since observer-invariance
    is only a special case of coordinates-invariance covered by Ricci-calculus on $ \stsurf $.
    However, this is only of limited practical value for solving the equations of motion in the above examples, as numerical approaches in $ \tangentstS[^n] $ are uncommon. More common are time-discrete identities on $ \tangentS[^n] $. We thus need to bring the observer-invariant operators from $ \tangentstS[^n] $ to $ \tangentS[^n] $. In the case of spacetime vector fields we can exploit the orthogonal decomposition above, where we
    associate the instantaneous part with a tangent vector field in $ \tangentS $ and the transversal part with a scalar field in $ \tangentS[^0] $
    independent of each other.
    Afterwards we rejoin both surface fields in the spacetime bundle $ \sttensorbS := \tangentS\times\tangentS[^0]\cong\tangentstS $.
    Note that the properties ``instantaneous'' and ``transversal'' are not depending on the choice of an observer. Therefore also this decomposition can be called observer-invariant. 
    Such a decomposition becomes much more complex in its combinatorics for general spacetime $ n $-tensor fields 
    in $ \sttensorbS[^n] \cong \tangentstS[^n] $, as will be outlined in \autoref{sec:spacetime_tensor_bundles}. However, with the isomorphism $ \bbrackets{\cdot}:\tangentstS[^n] \rightarrow \sttensorbS[^n]$, to be defined in \autoref{sec:spacetime_tensor_bundles}, an observer-invariant operator $ \operatorname{op} $ on $ \sttensorbS[^n] $ commutes the diagram
    \begin{equation}\label{eq:tikzcd_invariant_operator}
            \begin{tikzcd}
                    \tangentstS[^n] \arrow[rrrr,"\operatorname{Op}","\textit{(observer-invariant)}"'] \arrow[d,"\bbrackets{\cdot}"'] 
                            &&&& \tangentstS[^m]\arrow[d,"\bbrackets{\cdot}"]\\
                    \sttensorbS[^n] \arrow[rrrr,"\operatorname{op}","\textit{(observer-invariant)}"']           				     
                            &&&& \sttensorbS[^m]
            \end{tikzcd}
    \end{equation}
    for an convenient observer-invariant operator $ \operatorname{Op} $ on $\tangentstS[^n]$. This allows the systematic derivation of observer-invariant operators on $ \sttensorbS[^n] $, which are suitable for numerical purposes.
    
    Time derivatives along a material motion are intrinsic transport mechanism. \Eg\ for a force free material rigid body motion of $ \surf $,
    a spacetime tensor field could be parallel transported or ``frozen'' along the motion in dependency of the modeling aspects for the tensor field.
    The first one coincides with the directional covariant derivative $ \nabla_{\matdir}:\tangentstS[^n]\rightarrow\tangentstS[^n]$, 
    where $ \matdir\in\tangentstS $ is the \emph{(spacetime) material (velocity) direction}
    and we call $ \nabla_{\matdir} $ as well as 
    $ \bbrackets{\cdot}\circ\nabla_{\matdir}\circ\bbrackets{\cdot}^{-1}: \sttensorbS[^n] \rightarrow \sttensorbS[^n]$ the
    \emph{material derivative}. It is commonly used to model surface fluids on moving surfaces, see e.g. \cite{Nitschke2019a,Torres-SanchezMillanArroyo_JoFM_2019,Reuther2020}, and will be derived in \autoref{sec:material_derivative}.
    The second case coincide with Lie-derivatives $ \Lie_{\matdir} $ and we call them \emph{convected derivatives}
    for considerations \wrt\ $ \tangentstS[^n] $ as well as on $ \sttensorbS[^n] $.
    Since Lie-derivatives are not metric compatible, several varieties of convected derivatives \wrt\ musical isomorphism
    $ \flat $ (flat) and $ \sharp $ (sharp) known for lowering and rising tensor indices, arise. 
    We develop all possible convected derivatives \wrt\ these isomorphism and a special averaging, resulting in a Jaumann derivative, in \autoref{sec:convected_derivatives}. These derivatives are typically used in constitutive relations in continuum mechanics, see e.g. \cite{Morozov_Springer_2015,Oldroyd_PRSA_1950,Thiffeault_JPA_2001,Snoeijer_PRSA_2020}. The consideration of an arbitrary tensor rank $ n $ in \autoref{sec:spacetime_tensor_bundles}, \autoref{sec:material_derivative}
    and \autoref{sec:convected_derivatives} yields a very technical and combinatorial proceeding.
    We encourage readers, who are more interested in better readable results for lower rank tensors, to skip these sections.
    In \autoref{sec:scalar_fields} we consider the special case of scalar fields, where all versions boil down to the well-known total, or substantial, derivative. In \autoref{sec:vector_fileds} and \autoref{sec:two_tensor_fields}, respectively, we summarize the results for vector and 2-tensor fields and give some simple illustrating examples for different time derivatives 
    for instantaneous vector and symmetric and trace-less 2-tensor fields. 
    We further provide a computational tool to explore more general instantaneous 2-tensor fields.

\section{Notation}
    In this section we only clarify the basics of notation used in this paper.
    Detailed definitions, if they are present, can be found in corresponding sections.
    The basic framework of calculus is Ricci calculus, see \eg\ \cite{Schouten1954}.
    Within index notations we mainly use Latin indices to name tensor (proxy) components.
    To be more precise, capitals $ I,J,K,L $ indicate components \wrt\ (2+1)-dimensional curved spacetime $ \stsurf $ and 
    lower case letters $ i,j,k,l $ \wrt\ 2-dimensional curved spatial space $ \surf $.
    Components concerning associated Euclidean embedding spaces bear Latin indices $ A,B,C $ for $ \R^{(3+1)}=\R^4\supset\stsurf $
    and $ a,b,c $ for $ \R^3\supset\surf $.
    Note that capitals expand lower case letters by a temporal index $ t $, \eg\ either $ I $ equals $ t $ or spatial index $ i $.
    The covariant derivative in the spacetime index notation is highlighted by a semicolon ``$ ; $'', whereas a bar ``$ | $'' is used \wrt\ 
    spatial space notation.
    Sometimes switching between index and index-free notation is of advantage. 
    For this purpose we make use of square brackets ``$ [\,] $'' and braces ``$ \{\,\} $'', 
    \eg\ $ \tensor{[\nabla\starbt]}{^{IJ}_{K}} = \tensor{\starbtC}{^{IJ}_{;K}} $ for $ \nabla\starbt\in\tangentstS[^2_1] $
    and $ \{ \tensor{\arbtC}{^{ij}_{|k}} \} = \nabla\arbt\in\tangentS[^2_1] $, where we established that the order of evaluation 
    within braces has to be alphanumerically.
    Only indices of symmetric tensors are allowed to lay on top of each other, 
    \eg\  $ \tensor*{\arbt}{^i_j}$, since symmetry $\tensor{\arbt}{^i_j} = \tensor{\arbt}{_j^i}$ does not effect ambiguities.
    For compacting the notation of spacetime vector and 2-tensor fields, we employ a semi vector or matrix proxy notation,
    where temporal and spatial (or blends of both) components are considered separately.
    This notation comprises square brackets and a frame giving space for evaluation,
    \eg\ $ \starbtC^A\Eb_A=[\starbtC^t, \{\starbtC^a\}]'_{\tangent{\R^4}}= \starbtC^t\Eb_t + \starbtC^a\Eb_a $ in the Euclidean spacetime,
    where $ \Eb_A $ are the usual Euclidean unit basis vectors for $ \tangent{\R^4} $ with $ a=x,y,z $.
    Similarly double strokes square brackets $ \bbrackets{} $ in semi proxy notations separate transversal from instantaneous components. This applies exclusively for the spacetime $ \sttensorbS $ and $ \sttensorbS[^2] $.
    Note that the index ``1'' is omitted in naming the tangent bundle of vector fields contravariantly, \eg\ $ \tangent{\R^4} = \tangent[^1]{\R^4} $.
    The stroke ``$ ' $'' is the syntactical transpose operator for column vector proxies to use them as row vector proxies.
    However, the upper index $ T $ is the transpose operator, especially used for 2-tensor fields.
    This operator is generalized to $ T_\sigma $ for $ n $-tensor fields \wrt\ permutations $ \sigma\in S_n $, \eg\ 
    $ [\starbt^{T_{(2\, 1)}}]^{IJ} = [\starbt^{T}]^{IJ} = \starbtC^{JI}$ for $ \starbt\in\tangentstS[^2] $ or
    $ [\arbt^{T_\sigma}]^{i_1 \ldots i_n} = \arbtC^{i_{\sigma(1)}\ldots i_{\sigma(n)}} $ for $ \arbt\in\tangentS[^n] $.
    The inner product is characterized by angle brackets and the underlying space,
    \eg\ $ \innerS{\cdot}{\cdot}: \tangentS\times\tangentS\rightarrow\R $,
    and norms are deduced from this.
    The dot operator $ \underdot[l] $ denotes the inner multiplication of an $ n $-tensor field by a 2-tensor field \wrt\ 
    $ l $-th tensor dimension, \cf\ \ref{sec:linear_maps}.
    We omit this dot operator for $ l=1 $.
    The widehat ``$ \widehat{\quad} $'' above an expression means to omit this term.
    Occasionally, quantities like tensor fields wear an extra index referring to shuffles, which are a special kind of permutations, 
    see \ref{sec:shuffles}, or ``$ \m $'' to clarify relation to the underlying material, \eg\ material velocity $ \matvelo $.
    
    For a better reading flow, we summarize frequently used quantities and abbreviations in \autoref{tab:freq_quants}.
    Their definitions take place in the course of this paper.
    
    \begin{table}[]
        \centering
        \footnotesize
        \begin{tabular}{|c|c|c|c|} \hline 
            $\stpara$ & event, spacetime parameterization in $\stsurf$
                & $\para$ & point, spatial parameterization in $\surf$ \\\hline
            $\stg\in\tangentstS[_2]$ & spacetime metric tensor
                & $\g\in\tangentS[_2]$ & spatial metric tensor \\\hline
            $\gamma^I_{JK}$ & spacetime Christoffel symbols
                & $\Gamma^i_{jk}$ & spatial Christoffel symbols \\\hline
            $\matdir,\transdir\in\tangentstS$ & material, transversal direction
                & $\matvelo,\obvelo\in\tangentR$ & material, observer velocity \\\hline
            $\zeta = \frac{1}{1 + \velonor^2}$ & frequently used factor
                & $\velonor\in\tangentS[^0]$ & normal velocity \\\hline
            $\normal\bot\tangentS$ & spatial normal field
                & $\matveloS,\obveloS,\relvelo=\matveloS-\obveloS\in\tangentS$ & tangential material, observer, relativ velocity \\\hline
            $\shop\in\tangentS[_2]$ & 2nd fundamental form (shape operator)
                & $\bop_{[\m]} = \nabla\velonor + \shop\obveloS_{\![\m]} \in\tangentS[_1]$ & for observer[material] velocity, see \ref{sec:Bop_N_bop} \\\hline
            $\Bop_{[\m]} + \Bop^T_{[\m]}$ & rate of observer[material] deformation
                & $\Bop_{[\m]} = \nabla\obveloS_{\![\m]} - \velonor\shop\in\tangentS[_2]$ & for observer[material] velocity, see \ref{sec:Bop_N_bop} \\\hline
            $*_\beta = -\lct\cdot_\beta$ & tensorial Hodge operator 
                & $\lct=\{\sqrt{\det\g}\varepsilon_{ij}\}\in\tangentS[_2]$ & Levi-Civita tensor \\\hline
            $\nabla$ & covariant derivative acc.\! to context
                & $\rot\matveloS = -<\nabla\matveloS,\lct>$ & curl of tangential material velocity \\\hline
        \end{tabular}
        \caption{Some frequently used quantities.}
        \label{tab:freq_quants}
    \end{table}

\section{Spacetime surface and tensor bundles}\label{sec:spacetime_surface_and_tensor_bundle}
	Let $ \surf_t\subset\R^3 $ be a surface at fixed time $ t\in\timespace $.
	We realize this surface by an arbitrary instantaneous parametrization $ \para_t=\para_t(y^1,y^2) $ with local coordinates $ [y^1,y^2]'_{\R^2} $
    patch-wisely in a common way, see \eg\ \cite{Abraham2012}.
	Hence, the associated \emph{moving surface} is time-depending and defined by $ \surf=\surf(t):=\surf_t $ with parametrization
	$ \para=\para(t,y^1,y^2) = \para_t(y^1,y^2) $. 
	The \emph{spacetime surface} $ \stsurf\subset\R^4 $ consists of a moving surface $ \surf $ and the time interval $ \timespace $, 
	\st\ a fixed time $ t\in\timespace $ yields $ \stsurf|_t = \{t\}\times\surf_t $.
	The related parametrization is $ \stpara=\stpara(t,y^1,y^2)=[t,\para(t,y^1,y^2)]'_{\R^4} $.
	The choice of parametrizations $ \para $ to generate $ \stsurf $ is not unique, hence we call $ \para $ an \emph{observer parametrization}.
	This is justified by the behavior, that for fixing local coordinates $ \yb:=[y^1,y^2]'_{\R^2} $ the curve 
	$ \para_{\yb}= \para_{\yb}(t)$ describe the path experienced by a single observer particle as time passed.
	The \emph{observer velocity} field $ \obvelo:=\partial_t\para\in\tangentR $ can be decomposed into tangential part $ \obveloS\in\tangentS $ and normal
	part $ \velonor\in\tangentS[^0] $, \st\ $ \obvelo= \obveloS + \velonor\normal $ holds with time-depending normal field $ \normal\bot\partial_i\para $.
	Since for a fixed single event $ \stpara\in\stsurf $ the normal $ \normal|_{\stpara} $ does not depend on the choice of an observer, the normal velocity $ \velonor|_{\stpara} $ neither does. 
	As part of this, we introduce the observer independent 
    \emph{transversal direction} $ \transdir = [1, -\obveloS]'_{\tangentstS} = [1, \velonor\normal]'_{\R^4} $
    and scalar field $ \zeta^{-1}:= \|\transdir\|^{2}= 1+\velonor^2  $.
	By the availed isometric embedding the \emph{spacetime metric tensor} components $ \stgC_{IJ}=\innerstR{\partial_I\stpara}{\partial_J\stpara}$ 
	and their inverses $ \stgC^{IJ} $ are
	\begin{align}\label{eq:spacetime_metric}
		\stg &= \begin{bmatrix} \left\| \obveloS \right\|_{\surf}^{2} + \zeta^{-1} & \obveloS^{\flat} \\
								\obveloS^{\flat} & \g \end{bmatrix}_{\tangentstS[_2]}
		&&\text{and}
			& \stg^{-1}
					&= \begin{bmatrix} \zeta & -\zeta \obveloS \\
						-\zeta\obveloS & \g^{-1} + \zeta \obveloS\otimes\obveloS \end{bmatrix}_{\tangentstS[^2]} \formPeriod
	\end{align}
	There are two symmetric endomorphism to emphasize, 
	the orthogonal \emph{transversal projection} $ \transproj\in\symmapstS $ and 
	the orthogonal \emph{instantaneous projection} $ \instproj\in\symmapstS $ given by
	\begin{align*}
		\transproj &= \begin{bmatrix}
							1 & 0 \\ -\obveloS & 0
						\end{bmatrix}_{\tangentstSStar[^1_1]}
                   =\zeta\begin{bmatrix}
                                          1 & \velonor\normal \\ \velonor\normal & \velonor^2 \normal\otimes\normal  
                                   \end{bmatrix}_{\tangentstR[^2]} \quad \text{and} \quad
		\instproj = \begin{bmatrix}
						0 & 0 \\ \obveloS & \IdS
					   \end{bmatrix}_{\tangentstSStar[^1_1]}
                  = \begin{bmatrix}
                        0 & 0 \\ 0 &\Id[\R^3] - \normal\otimes\normal
                  \end{bmatrix}_{\tangentstR[^2]}
          \formPeriod
	\end{align*}
	Their image spaces are the \emph{transversal bundle} $ \tantransstS:= \transproj(\tangentstS) < \tangentstS $
	and the \emph{instantaneous bundle} $ \taninststS:= \instproj(\tangentstS) < \tangentstS $,
	which decompose the tangential bundle $ \tangentstS=\tantransstS\oplus\taninststS $ into linear subbundles, orthogonally and independently 
    \wrt\ the choice of an observer.
	We recognize an orthogonal decomposition of the identity $ \IdstS= \transproj + \instproj$ as well, since 
	$ \innerstS{\transproj}{\instproj}=0 $, which allows us to measure transversal and instantaneous parts, separately.
	This means that the Riemannian spacetime manifold $ (\stsurf,\stg) $ is basically a curved classical Newtonian spacetime 
	$ (\stsurf,\transproj[\flat],\instproj[\sharp]) $, where $ \transproj[\flat]=\zeta^{-1}dX^t\otimes dX^t $ and
	$ \instproj[\sharp] = \gC^{ij}\partial_{i}\stpara\otimes\partial_j\stpara $.
	
	The orthogonal tangential bundle decomposition is extendable to $ \tangentstS[^n] $.
	For this purpose we introduce the orthogonal \emph{shuffled projection} $ \stproj{\sigma}\in\symmapstS[n] $ by means of
	\begin{align}
		\indexsymten{\stproj{\sigma}}{^{I_1\ldots I_n}_{J_1\ldots J_n}}
			&:= \indexsymten{\transproj}{^{I_{\sigma(1)}}_{J_{\sigma(1)}}}
			    \cdots\indexsymten{\transproj}{^{I_{\sigma(\alpha)}}_{J_{\sigma(\alpha)}}}
			    \indexsymten{\instproj}{^{I_{\sigma(\alpha+1)}}_{J_{\sigma(\alpha+1)}}}
			    \cdots\indexsymten{\instproj}{^{I_{\sigma(\alpha+n)}}_{J_{\sigma(\alpha+n)}}}
	\end{align}
	for all shuffles $ \sigma\in\shuffles{n}{\alpha} $, see \ref{sec:shuffles}. 
    For $ n=0 $ we claim $ \stproj{\sigma} \equiv 1 $.
    Note that all $ \stproj{\sigma} $ are pair-wise distinguishable.
    This would not be true if we consider all permutations in $ \operatorname{S}_n\supset\shuffles{n}{\alpha} $, due to arising symmetries.
	Additionally, we obtain an orthogonal system featured $ \innerstS{\stproj{\sigma}}{\stproj{\tilde{\sigma}}}= 0 $
	for all $ \sigma\neq\tilde{\sigma} $.
    This leads to the orthogonal and observer-invariant decomposition
    \begin{align}\label{eq:orthogonal_decomposition_of_spacetime_surface}
        \tangentstS[^n] &= \bigoplus_{\alpha=0}^n\bigoplus_{\sigma\in\shuffles{n}{\alpha}}\tanshufstS{\sigma}
    \end{align}
    with \emph{shuffled bundle} $ \tanshufstS{\sigma}:= \stproj{\sigma}(\tangentstS[^n]) < \tangentstS[^n] $.
    
\section{Spacetime tensor bundles at moving surface}\label{sec:spacetime_tensor_bundles}
    In this section we develop the observer-invariant \emph{spacetime tensor bundle} $ \sttensorbS[^n] $ as 
    $ 2^n $-ary Cartesian product of surface tensor bundles $ \tangentS[^m] $ for  miscellaneous $ m\le n $,
    which covered all information of $ \tangentstS[^n] $, \ie\ the dimensionality reveals
    $ \dim_{\R}\tangentstS[^n]|_{\stpara}=\dim_{\R}\sttensorbS[^n]|_{t,\para} = 3^n $ at event $ \stpara=[t,\para]'_{\{t\}\times\surf}\in\stsurf $,
    particularly to preserve the amount of degrees of freedom.
    We also present a corresponding isomorphism 
    $ \bbrackets{\cdot}:\tangentstS[^n] \rightarrow \sttensorbS[^n] $.
    
    The decomposition in eq. \eqref{eq:orthogonal_decomposition_of_spacetime_surface} provides the opportunity to treat the considerably simpler shuffled bundles. 
    Let us consider the commutative diagram
    \begin{equation}\label{eq:tikzcd_shuffle_spaces}
        \forall\sigma\in\shuffles{n}{\alpha}:\quad
        \begin{tikzcd}
            \tanshufstS{\sigma}\arrow[rd, "\phi_{\sigma}"']\arrow[r, "{\bbrackets[\sigma]{\cdot}}"] 
                        & \tangentS[^{n-\alpha}] \\
                        & \taninststS[^{n-\alpha}] \arrow[u, "\iota"']
        \end{tikzcd}
        \formComma\quad
        \begin{tikzcd}
               \starbt_{\sigma}\arrow[rd, mapsto, "\phi_{\sigma}"']\arrow[r, mapsto, "{\bbrackets[\sigma]{\cdot}}"] 
                         & \arbt_{\sigma} \\
                         & \phi_{\sigma}(\starbt_{\sigma}) \arrow[u, mapsto, "\iota"']
        \end{tikzcd} 
        \formComma
    \end{equation}
    \ie\ $ \arbt_{\sigma} = \bbrackets[\sigma]{\starbt_{\sigma}} = (\phi_{\sigma}\circ\iota)(\starbt_{\sigma}) $.
    Since $ \phi_{\sigma}(\starbt_{\sigma}) $ is an instantaneous $ (n-\alpha) $-tensor all time-like components are zero,
    \ie\ $ \indexten{\phi_{\sigma}(\starbt_{\sigma})}{^{I_1\ldots I_{n-\alpha}}} = 0 $ if there exists a $  k\le n-\alpha $ 
    \st\ $ I_k=t $. 
    Therefore, we define the $ \sigma $-independent isomorphism $ \iota $ merely by cutting off the zeros throughout all time-like components, 
    \ie\ $ \indexten{\arbt_{\sigma}}{^{i_1\ldots i_{n-\alpha}}} = \indexten{\phi_{\sigma}(\starbt_{\sigma})}{^{i_1\ldots i_{n-\alpha}}} $
    for $ \arbt_{\sigma}= \iota(\phi_{\sigma}(\starbt_{\sigma})) $. 
    A tighter examination of the shuffled bundle $ \tanshufstS{\sigma} $ reveals that it is spanned by 
    \begin{align}\label{eq:shuffel_base} 
       \shuffbase{\sigma}{i_1 \ldots i_{n-\alpha}}
        &:=\left(\left( \bigotimes_{k=1}^\alpha\transdir \right)\otimes\partial_{i_1}\stpara\otimes\cdots\otimes\partial_{i_{n-\alpha}}\stpara\right)^{T_{\sigma}}
    \end{align}
    event-wisely for all $ i_1,\ldots,i_{n-\alpha} $.
    The $ \alpha $-fold outer product of the transversal direction $ \transdir $ is redundant though.
    Singly, it is able to retain nothing but scalar-valued information, which is absorbable by the remaining spatial basis tensor parts.
    Therefore, we let $ \phi_{\sigma} $ test off all transversal parts of $ \starbt_{\sigma} $ by $ \frac{1}{\| \transdir \|^2}\transdir=\zeta\transdir $, \ie
    \begin{align}\label{eq:phi_sigma}
        \begin{aligned}
            \forall\sigma\in\shuffles{n}{\alpha}, \starbt_{\sigma}\in\tanshufstS{\sigma}:&&
                    \indexten{\phi_{\sigma}(\starbt_{\sigma})}{^{I_{\sigma(\alpha+1)}\ldots I_{\sigma(n)}}}
                        &= \zeta^{\alpha}\transdirC_{I_{\sigma(1)}}\!\!\!\cdots\transdirC_{I_{\sigma(\alpha)}}
                                    \indexsymten{\starbt_{\sigma}}{^{I_1\ldots I_n}} \formComma\\
                    \text{resp. }\forall\hat{\starbt}_{\sigma}\in\taninststS[^{n-\alpha}]:&&
                    \indexsymten{\phi_{\sigma}^{-1}(\hat{\starbt}_{\sigma})}{^{I_1\ldots I_n}}
                        &= \transdirC^{I_{\sigma(1)}}\!\cdots\transdirC^{I_{\sigma(\alpha)}}
                                    \indexten{\hat{\starbt}_{\sigma}}{^{I_{\sigma(\alpha+1)}\ldots I_{\sigma(n)}}} \formPeriod
        \end{aligned}
    \end{align}
    Eventually, the isomorphism 
    $ \bbrackets[\sigma]{\cdot}:\tanshufstS{\sigma} \leftrightarrow \tangentS[^{n-\alpha}]:\bbrackets[\sigma]{\cdot}^{-1} $ 
    and its inverse are defined entirely.
    
    Since the decomposition in eq. \eqref{eq:orthogonal_decomposition_of_spacetime_surface} is orthogonal, for $ \starbt\in\tangentstS[^n]  $ all parts
    $ \starbt_{\sigma}=\stproj{\sigma}(\starbt)\in\tanshufstS{\sigma} $ are determined uniquely and are orthogonal.
    Therefore all $ 2^n $ images $ \bbrackets[\sigma]{\starbt_{\sigma}}$ can be unified and remain disjointed.
    We realize the emerging Cartesian product with orthogonal basis vectors $ \sttenbase{\sigma} $ formally, \st\ 
    \begin{align}
        \left(\partial_{i_1}\para\otimes\cdots\otimes\partial_{i_{n-\alpha}}\para\right)\sttenbase{\sigma}
                &= \bbrackets{\shuffbase{\sigma}{i_1 \ldots i_{n-\alpha}}} \in \sttensorbS[^n]\formPeriod
    \end{align} 
    In this way $ \sttenbase{\sigma} $ and every linear combination for all $ \sigma\in\shuffles{n}{\alpha} $
    could be depicted as a simple vector of length $ 2^n $, with components in $ \tangentS[^{n-\alpha}] $, in predefined order or as a hypermatrix of rank $ n $ with 
    2 entries in all $ n $ dimensions, see \eg\ \autoref{sec:vector_fileds} or \autoref{sec:two_tensor_fields}.   
    With $ |\shuffles{n}{\alpha}|=\binom{n}{\alpha} $ we summarize in conclusion that
    \begin{align}
        \begin{aligned}
            \bbrackets{\cdot}
                        &= \sum_{\alpha=0}^n\sum_{\sigma\in\shuffles{n}{\alpha}} \bbrackets[\sigma]{\stproj{\sigma}(\cdot)}\sttenbase{\sigma}:
                            &\tangentstS[^n] 
                                &\longrightarrow \prod_{\alpha=0}^{n}\left( \tangentS[^{n-\alpha}] \right)^{\binom{n}{\alpha}} = \sttensorbS[^n] 
                  \formComma\\
                    \bbrackets{\cdot}^{-1}
                        &= \sum_{\alpha=0}^n\sum_{\sigma\in\shuffles{n}{\alpha}}\bbrackets[\sigma]{\cdot}^{-1}:
                            &\sttensorbS[^n]
                            &\longrightarrow \bigoplus_{\alpha=0}^n\bigoplus_{\sigma\in\shuffles{n}{\alpha}}\tanshufstS{\sigma} = \tangentstS[^n]
                  \formPeriod
        \end{aligned}
    \end{align}
    Note that a spacetime tensor field $ \arbt= \sum_{\alpha=0}^n\sum_{\sigma\in\shuffles{n}{\alpha}} \arbt_{\sigma}\sttenbase{\sigma}\in\sttensorbS[^n]$
    is observer-invariant in the sense that all surface tensor components $ \arbt_{\sigma}\in\tangentS[^{n-\alpha}] $ are.
    Their proxies $ \arbtC_{\sigma}^{i_1 \ldots i_{n-\alpha}} $ can depend on observer parametrization $ \para $ though.
    
\section{Material derivative}\label{sec:material_derivative}

    In recognition of the general principle of covariance in $ \stsurf $, we formulate the material derivative as covariant derivative along the spacetime \emph{material direction} $ \matdir\in\tangentstS $.  
    This direction is given by $ \matdir=\left[ 1, \relvelo \right]'_{\tangentstS} $,
    with \emph{relative velocity} $ \relvelo = \matveloS - \obveloS\in\tangentS $ depending on the given
    \emph{tangential material velocity} $ \matveloS\in\tangentS  $ and chosen \emph{tangential observer velocity} $\obveloS\in\tangentS $,
    see \autoref{sec:material_direction} for greater details.
    The tensor-valued material derivative 
    \begin{align}\label{eq:material_derivative_basic}
        \matdst:\ \tangentstS[^n] &\rightarrow \tangentstS[^n] \formComma
                                    &\starbt  &\mapsto \matdst\starbt:= \nabla_{\!\matdir}\starbt
                                          = \left\{ \tensor{\starbtC}{^{I_1\ldots I_n}_{;t}} + \relveloC^k\tensor{\starbtC}{^{I_1\ldots I_n}_{;k}} \right\}
    \end{align}
    thereby is well-defined.
    In virtue of eq. \eqref{eq:tikzcd_invariant_operator} the \emph{material derivative} on the observer independent spacetime bundle at
    the moving surface is
    \begin{align*}
        \matd:=\bbrackets{\cdot}\circ\matdst\circ\bbrackets{\cdot}^{-1}:\ \sttensorbS[^n] &\rightarrow \sttensorbS[^n]\formComma
            &\arbt&\mapsto\matd\arbt = \bbrackets{\matdst\bbrackets{\arbt}^{-1}}\formPeriod
    \end{align*}
    Taking eq. \eqref{eq:tikzcd_shuffle_spaces} into account, we constitute the commuting diagram
    \begin{equation}\label{eq:tikzcd_shuffle_material_derivative}
        \begin{tikzcd}
            \tangentstS[^n] \arrow[rrrr,"{\matdst[\sigma]:=\stproj{\sigma}\circ\matdst}"] \arrow[rrd, "{\matdhatst{\sigma}}"]\arrow[dd, "{\bbrackets{\cdot}}"']
                                    &&&&\tanshufstS{\sigma} \arrow[lld, "{\phi_{\sigma}}"'] \arrow[dd, "{\bbrackets[\sigma]{\cdot}}"]\\
                        &&\taninststS[n-\alpha] \arrow[rrd, "{\iota}"] &&\\
            \sttensorbS[^n] \arrow[rrrr, "{\matd[\sigma]}"]
                                    &&&&\tangentS[^{n-\alpha}]
        \end{tikzcd}\formComma
    \end{equation} 
    which defines the mappings $ \matdst[\sigma], \matdhatst{\sigma} $ and $\matd[\sigma]$ sufficiently.
    As a consequence, this yields the desirable decomposition behaviors 
    \begin{align*}
        \forall\starbt\in\tangentstS[^n]: 
                    &&\matdst\starbt
                            &= \sum_{\alpha=0}^n\sum_{\sigma\in\shuffles{n}{\alpha}}\matdst[\sigma]\starbt
                             = \sum_{\alpha=0}^n\sum_{\sigma\in\shuffles{n}{\alpha}}
                                    \indexten{\matdhatst{\sigma}\starbt}{^{i_1\ldots i_{n-\alpha}}}\shuffbase{\sigma}{i_1 \ldots i_{n-\alpha}} \\
        \arbt=\bbrackets{\starbt}\in\sttensorbS[^n]:
                    &&\matd\arbt
                            &= \bbrackets{\matdst\starbt}
                             = \sum_{\alpha=0}^n\sum_{\sigma\in\shuffles{n}{\alpha}} \left( \matd[\sigma]\arbt \right)\sttenbase{\sigma}\formPeriod
    \end{align*}
    The trivial situation $ n=0 $, where $ f\in\tangentstS[^0] $ and hence $ \bbrackets{f}=f $, reveals the well-known material derivative
    \begin{align}\label{eq:dotf}
        \dot{f} &:= \matd f = \matdst f = \partial_t f + \nabla_{\relvelo}f \formPeriod
    \end{align}
    We now consider the case $ n>0 $.
    Before bringing $ \matd[\sigma] $ in a convenient form to determine $ \matd $, we investigate two helpful special cases: the transversal direction and instantaneous tensors.
    
    We calculate the material derivative of transversal direction field $ \transdir=[1,-\obveloS]'_{\tangentstS}\in\tanshufstS{\tau}\subset\tangentstS $
    directly by eq.  \eqref{eq:material_derivative_basic} using Christoffel symbols, see \ref{sec:christoffel_symbols}, 
    substituting acceleration and handle velocity tangential gradient terms with the aid of \ref{sec:accel} and \ref{sec:Bop_N_bop}, 
    \ie\ the temporal and spatial part yield
    \begin{align*}
        \indexten{\matdst\transdir}{^t}
            &= \gamma_{tt}^t + \left( \relveloC^k - \obveloSC^k \right)\gamma_{kt}^t - \relveloC^k\obveloSC^j\gamma_{kj}^t
             = \zeta\velonor\left( \obaccelnor + \nabla_{\obveloS}\velonor - \innerS{\bop}{\obveloS} \right)
             = \zeta\velonor\dot{\velonor} \\
        \indexten{\matdst\transdir}{^i}
            &= -\partial_t\obveloSC^i +\gamma_{tt}^i + \left( \relveloC^k - \obveloSC^k \right)\gamma_{kt}^i
                -\relveloC^k\left( \partial_k\obveloSC^i + \gamma_{kj}^i\obveloSC^j \right)\\
            &= \obaccelSC^i + \indexsymten{\Bop(\relvelo - \obveloS)}{^i}
                -\left( \partial_t\obveloSC^i + \indexten{\nabla_{\relvelo}\obveloS}{^i} + \indexten{\matdst\transdir}{^t}\obveloSC^i \right)
             = -\velonor\left( \bopC_{\m}^i + \zeta\dot{\velonor}\obveloSC^i \right) \formPeriod
    \end{align*}
    Therefore, it holds
    $\matdst\transdir = \left(\matdhatst{\transdir}\transdir\right)\transdir + \matdhatst{\surf}\transdir
                        = \zeta\velonor\dot{\velonor}\transdir - [0 , \velonor\bop_{\m}]'_{\tangentstS}$.
    Together with eq. \eqref{eq:tikzcd_shuffle_material_derivative} the following lemma holds.
    \begin{lem}\label{lem:material_derivative_of_transversal_direction}
        The material derivative of the transversal direction $ \sttenbase{\transdir}=\bbrackets{\transdir}\in\sttensorbS $ in the spacetime vector bundle is
        \begin{align*}
            \matd\transdir 
                    &= \left( \matd[\transdir]\sttenbase{\transdir} \right)\sttenbase{\transdir}
                       +\left( \matd[\surf]\sttenbase{\transdir} \right)\sttenbase{\surf}
                     = \velonor\left( \zeta\dot{\velonor}\sttenbase{\transdir} - \bopC_{\m}\sttenbase{\surf} \right)\in\sttensorbS\formPeriod
        \end{align*}
    \end{lem}
    
    Similarly to above, we also calculate the material derivative of an instantaneous $ n$-tensor field 
    $ \starbt\in\taninststS[^n]\subset\tangentstS[^n] $ directly.
    It is clear that parts of $ \matdst\starbt $ with more than one temporal dimension vanish, since the Christoffel symbols are only able to catch one temporal index, where $ \starbtC^{I_1 \ldots I_n} $ would vanish at the same time,
    \ie\ it holds $\matdst\starbt\in\taninststS[^n]\oplus\bigoplus_{\beta=1}^n\tanshufstS[n]{\surf_\beta}  $ at least.
    For the non-vanishing single temporal afflicted and pure spatial parts we obtain
    \begin{align*}
         \indexten{\matdst\starbt}{^{i_1 \ldots i_{\beta-1} t  i_{\beta+1} \ldots i_n}}
            &=\left( \gamma_{tj}^t + \relveloC^k\gamma_{kj}^t \right)\starbtC^{i_1 \ldots i_{\beta-1} j  i_{\beta+1} \ldots i_n}
             =\zeta\velonor\bopC_{\m}^j \tensor{\starbtC}{^{i_1 \ldots i_{\beta-1}}_j^{i_{\beta+1} \ldots i_n}} \\
         \indexten{\matdst\starbt}{^{i_1 \ldots i_n}}
            &= \partial_t\starbtC^{i_1 \ldots i_n} + \relveloC^k\partial_k\starbtC^{i_1 \ldots i_n}
                +\sum_{\beta=1}^{n}\left( \gamma_{tj}^{i_{\beta}} + \relveloC^k\gamma_{kj}^{i_{\beta}} \right)
                                \starbtC^{i_1 \ldots i_{\beta-1} j  i_{\beta+1} \ldots i_n} \\
            &= \partial_t\starbtC^{i_1 \ldots i_n} + \indexten{\nabla_{\relvelo}\starbt}{^{i_1 \ldots i_n}}
                +\sum_{\beta=1}^{n}\indexten{\Bop - \zeta\velonor\obveloS\otimes\bop_{\m}}{^{i_{\beta}}_{j}}
                                \starbtC^{i_1 \ldots i_{\beta-1} j  i_{\beta+1} \ldots i_n}\formPeriod
    \end{align*}
    This means we can formulate the material derivation in terms of $ \matdhatst{\surf^n} $, $  \matdhatst{\surf^n_\beta} $ and the
    associated orthogonal basis tensors in eq. \eqref{eq:shuffel_base}, namely
    \begin{align*}
        \matdst\starbt &= \left( \partial_t\starbtC^{i_1 \ldots i_n} + \indexten{\nabla_{\relvelo}\starbt}{^{i_1 \ldots i_n}} 
                                + \sum_{\beta=1}^{n}\tensor{\BopC}{^{i_{\beta}}_{j}}
                                       \starbtC^{i_1 \ldots i_{\beta-1} j  i_{\beta+1} \ldots i_n}\right)\shuffbase{\surf^n}{i_1 \ldots i_n}
                                       +\zeta\velonor\sum_{\beta=1}^{n}
                                \bopC_{\m}^j \tensor{\starbtC}{^{i_1 \ldots i_{\beta-1}}_j^{i_{\beta+1} \ldots i_n}}
                                            \shuffbase{\surf^n_\beta}{i_1 \ldots \widehat{i_{\beta}} \ldots i_n} \formPeriod
    \end{align*}
    Since $ \starbt $ is instantaneous, it holds $ \bbrackets[\surf^n]{\starbt}^{i_1\ldots i_n}= \starbtC^{i_1\ldots i_n} $,
    which leads to the following lemma.
    \begin{lem}\label{lem:material_derivative_of_instantaneous_tensors}
        Assuming $ \arbt_{\!\surf^n}\in\tangentS[^n] $,
        the material derivative of an instantaneous tensor field $ \arbt=\arbt_{\!\surf^n}\sttenbase{\surf^n}\in\sttensorbS[^n] $
        in the spacetime bundle is
        \begin{align*}
            &&\matd\arbt &= \left(\matd[\surf^n]\arbt\right) \sttenbase{\surf^n}
                            +\sum_{\beta=1}^{n} \left(\matd[\surf^n_{\beta}]\arbt\right) \sttenbase{\surf^n_{\beta}}\formComma\\
           &\text{where}&
           \indexten{\matd[\surf^n]\arbt}{^{i_1\ldots i_n}}
                      &= \partial_t\arbtC_{\!\surf^n}^{i_1 \ldots i_n} + \indexten{\nabla_{\relvelo}\arbt_{\!\surf^n}}{^{i_1 \ldots i_n}} 
                                    + \sum_{\beta=1}^{n}\tensor{\BopC}{^{i_{\beta}}_{j}}
                                            \arbtC_{\!\surf^n}^{i_1 \ldots i_{\beta-1} j  i_{\beta+1} \ldots i_n} \\
           &\text{and}&
           \indexten{\matd[\surf^n_{\beta}]\arbt}{^{i_1 \ldots \widehat{i_{\beta}} \ldots i_n}}
                      &= \zeta\velonor \indexten{\bop_{\m}}{_j} \arbtC_{\!\surf^n}^{i_1 \ldots i_{\beta-1} j i_{\beta+1} \ldots i_n}
              \formPeriod
        \end{align*} 
    \end{lem}
    
    As we see from this, derivatives $ \matdst[\sigma]\circ\stproj{\surf^n} $ have a trivial image except for 
    $ \sigma=\surf^n $ or $ \sigma=\surf^n_{\beta} $, we thus ask about supports of 
    the occurring derivatives in the orthogonal decomposition
    $ \matdst=\sum_{\alpha=0}^n\sum_{\talpha=0}^n\sum_{\sigma\in\shuffles{n}{\alpha}}\sum_{\tsigma\in\shuffles{n}{\talpha}} 
               \matdst[\sigma]\circ\stproj{\tsigma} $ \wrt\ the image as well as the domain.
    For this we assume $ \starbt\in\tangentstS[^n] $, $ \starbt_{\tsigma}=\stproj{\tsigma}\starbt $ and 
    $ \hstarbt_{\tsigma}=\phi_{\tsigma}(\starbt_{\tsigma}) $, 
    \ie\ $ \indexstarbt[\tsigma]{I_1\ldots I_n} = \transdirC^{I_{\tsigma(1)}}\cdots\transdirC^{I_{\tsigma(\talpha)}} 
                                \indexhstarbt[\tsigma]{I_{\tsigma(\talpha+1)}\ldots I_{\tsigma(n)}}$, see eq. \eqref{eq:phi_sigma}.
    Hence product rule leads to
    \begin{align*}
        \indexten{\matdst[\sigma]\starbt_{\tsigma}}{^{I_1\ldots I_n}}
            = \indexsymten{\transproj}{^{I_{\sigma(1)}}_{J_{\sigma(1)}}}
            			    \cdots\indexsymten{\transproj}{^{I_{\sigma(\alpha)}}_{J_{\sigma(\alpha)}}}
            			    &\indexsymten{\instproj}{^{I_{\sigma(\alpha+1)}}_{J_{\sigma(\alpha+1)}}}
            			    \cdots\indexsymten{\instproj}{^{I_{\sigma(\alpha+n)}}_{J_{\sigma(\alpha+n)}}} 
                            \Big( \transdirC^{J_{\tsigma(1)}}\cdots\transdirC^{J_{\tsigma(\talpha)}} 
                                                     \indexten{\matdst\hstarbt_{\tsigma}}{^{J_{\tsigma(\talpha+1)}\ldots J_{\tsigma(n)}}}\\
           & + \indexhstarbt[\tsigma]{J_{\tsigma(\talpha+1)}\ldots J_{\tsigma(n)}}
                        \sum_{\tbeta=1}^{\talpha} \indexten{\matdst\transdir}{^{J_{\tsigma(\tbeta)}}}
                                 \transdirC^{J_{\tsigma(1)}}\cdots\widehat{\transdirC^{J_{\tsigma(\tbeta)}}}\cdots\transdirC^{J_{\tsigma(\talpha)}}  
                         \Big)\formPeriod        
    \end{align*}
    Since $ \transdir $ is transversal and $ \hstarbt_{\tsigma} $ instantaneous, all non-vanishing $ \matdst[\sigma]\starbt_{\tsigma} $ require a shuffle $ \tsigma $ broadly similar to $ \sigma $,
    more precisely, $ \tsigma $ has to be equal $ \sigma $, $ \sigma^{\beta} $ or $ \sigma_{\beta} $ for all applicable $ 0 \le \beta \le n $
    according to eq. \eqref{eq:shuffle_relations}.
    We work through these cases hereafter.
    For $ \tsigma=\sigma $ the projection of the first summand contains $\matdst[\surf^{n-\alpha}]\hstarbt_{\sigma}=  \matdhatst{\surf^{n-\alpha}}\hstarbt_{\sigma} $ and each of the remaining summands expose
    $ \matdst[\transdir]\transdir = (\matdhatst{\transdir}\transdir)\transdir $.
    Applying $ \phi_{\sigma} $ yields
    \begin{align}\label{eq:qwertzI}
        \matdhatst{\sigma}\starbt_{\sigma} 
            &= \matdhatst{\surf^{n-\alpha}}\hstarbt_{\sigma} + \alpha(\matdhatst{\transdir}\transdir)\hstarbt_{\sigma}
              \in \taninststS[n-\alpha] \formPeriod
    \end{align}
    For $ \tsigma=\sigma^{\beta} $ the entire rear sum vanish, since one transversal projection encounters 
    the instantaneous $ \hstarbt_{\sigma^\beta}\in\taninststS[n-\alpha+1] $.
    The material derivative in the front sum is projected \wrt\ shuffle
    \begin{align*}
        \left( \sigma(\beta)\mid \sigma(\alpha+1) \ldots \sigma(n) \right)
            &= \left( \sigma^\beta(\alpha+\rbeta) \mid \sigma^{\beta}(\alpha+1) \ldots \sigma^{\beta}(n) \right) \formComma
    \end{align*}
    which is $ \surf^{n-\alpha+1}_{\rbeta}\in\shuffles{n-\alpha+1}{1} $ effectively for $ \rbeta $ \st\ $ (\sigma^{\beta})_{\rbeta}= \sigma $,
    counting from $ \alpha+1 $ though.
    Hence, $ \phi_{\sigma} $ archives
    \begin{align}\label{eq:qwertzII}
        \forall\beta:\ 1\le\beta\le\alpha:&&
        \matdhatst{\sigma}\starbt_{\sigma^{\beta}} 
            &= \matdhatst{\surf^{n-\alpha}_{\rbeta}}\hstarbt_{\sigma^{\beta}}\in \taninststS[n-\alpha] \formPeriod
    \end{align}
    For $ \tsigma=\sigma_{\beta} $ almost all summands are fading, since one single instantaneous projections faces $ \transdir $.
    Only for $ \tbeta=\rbeta $ \st\  $ (\sigma_{\beta})^{\rbeta}= \sigma $, \ie\  $ \sigma_{\rbeta}(\alpha+\beta)=\sigma $, the term
    \begin{align*}
        \transdirC^{I_{\sigma(1)}}\cdots\transdirC^{I_{\sigma(\alpha)}}
                        \indexten{\matdst[\surf]\transdir}{^{I_{\sigma(\alpha+\beta)}}}
                                \indexhstarbt[\sigma_{\beta}]{I_{\sigma(\alpha+1)}\ldots \widehat{I_{\sigma(\alpha+\beta)}} \ldots I_{\sigma(n)}}
    \end{align*}
    survive, which results in
    \begin{align}\label{eq:qwertzIII}
        \forall\beta:\ 1\le\beta\le n-\alpha:&&
        \matdhatst{\sigma}\starbt_{\sigma_{\beta}} 
              &= \left( \left( \matdhatst{\surf}\transdir \right) \otimes \hstarbt_{\sigma_{\beta}} \right)^{T_{\surf^{n-\alpha}_{\beta}}} 
                      \in \taninststS[n-\alpha]
    \end{align}
    using $ \phi_{\sigma} $.
    The representations of pure instantaneous identities eqs. \eqref{eq:qwertzI}, \eqref{eq:qwertzII} and \eqref{eq:qwertzIII} 
    are fully determined in $ \sttensorbS[^{n-\alpha}] $ 
    by \autoref{lem:material_derivative_of_transversal_direction} and \autoref{lem:material_derivative_of_instantaneous_tensors}.
    Using the narrow support, \ie\  for all considered $ \tsigma $ cases above 
    $ \matdhatst{\sigma}\starbt=\sum_{\tsigma}\matdhatst{\sigma}\starbt_{\tsigma} $ 
    is sufficient,
    the isomorphism $ \iota $ on $ \taninststS[n-\alpha] $ yields
    \begin{align*}
        \matd[\sigma]\arbt
            &= \matd[\surf^{n-\alpha}]\left( \arbt_{\sigma}\sttenbase{\surf^{n-\alpha}} \right)
                    +\alpha\left( \matd[\transdir]\sttenbase{\transdir} \right)\arbt_{\sigma}
                    +\sum_{\beta=1}^{\alpha}\matd[\surf^{n-\alpha+1}_{\rbeta}]\left( \arbt_{\sigma^{\beta}}\sttenbase{\surf^{n-\alpha+1}} \right) +\sum_{\beta=1}^{n-\alpha}\left( \left( \matd[\surf]\sttenbase{\transdir}\right)
                                \otimes \arbt_{\sigma_{\beta}} \right)^{T_{\surf^{n-\alpha}_{\beta}}}
                 \in\tangentS[^{n-\alpha}]                    
    \end{align*}
    for $ \arbt=\bbrackets{\starbt}\in\sttensorbS[^n] $  and we can formulate the following theorem.
    \begin{thm}[Material derivative]\label{thm:material_derivative}
        Assuming $ \arbt=\sum_{\alpha=0}^n\sum_{\sigma\in\shuffles{n}{\alpha}}\arbt_{\sigma}\sttenbase{\sigma}\in\sttensorbS[^n] $,
        where $ \arbt_{\sigma}\in\tangentS[^{n-\alpha}] $, the material derivative of $ \arbt $ is
        \begin{align*}
            \matd\arbt &= \sum_{\alpha=0}^n\sum_{\sigma\in\shuffles{n}{\alpha}}\left(\matd[\sigma]\arbt\right)\sttenbase{\sigma}\in\sttensorbS[^n]\\
            \indexten{\matd[\sigma]\arbt}{^{i_1\ldots i_{n-\alpha}}}
                       &= \partial_t\arbtC_{\sigma}^{^{i_1\ldots i_{n-\alpha}}}
                                \! + \! \indexten{\nabla_{\relvelo}\arbt_{\sigma}}{^{i_1\ldots i_{n-\alpha}}}
                                \! + \! \alpha\zeta\velonor\dot{\velonor}\arbtC_{\sigma}^{^{i_1\ldots i_{n-\alpha}}}
                                \! + \! \zeta\velonor\indexten{\bop_{\m}}{_{k}} 
                                   \sum_{\beta=1}^{\alpha} \arbtC_{\sigma^{\beta}}^{i_{1}\ldots i_{\rbeta-1} k i_{\rbeta} \ldots i_{n-\alpha}} \! + \!\sum_{\beta=1}^{n-\alpha} \left(
                                \tensor{\BopC}{^{i_{\beta}}_{k}}\arbtC_{\sigma}^{i_{1}\ldots i_{\beta-1} k i_{\beta+1} \ldots i_{n-\alpha}}
                               -\velonor \bopC_{\m}^{i_{\beta}} \arbtC_{\sigma_{\beta}}^{^{i_1\ldots \widehat{i_{\beta}} \ldots i_{n-\alpha}}}
                                                                \right) \formComma
        \end{align*}
        where $ \matd[\sigma]\arbt \in\tangentS[^{n-\alpha}]$ for $ \sigma\in\shuffles{n}{\alpha} $ and
        $ \rbeta $ is given by $ (\sigma^{\beta})_{\rbeta}= \sigma $ implicitly. 
    \end{thm}

        For the remaining section we devote some investigations to instantaneous action of the material derivative.
        We restrict the treatment to instantaneous $ n $-tensor fields,
        \ie\ $ \matd[\surf^n]|_{\operatorname{Span}_{\tangentS[^n]}\{\sttenbase{\surf^{n}}\}} $ 
        is the object of interest.
        This restriction on domain and image gives an intrinsic derivative, which generalizes the common 
        total derivative $ \totald $ established in two-dimensional flat space.
        Proceeding from spatial embedding space $ \R^3 $ yields an Eulerian perspective on the moving surface $ \surf $,
        hence the total derivative of a surface scalar fields $ f\in\tangentS[^0] $ gives
        \begin{align}\label{eq:dotfII}
            \totald[f] &= \partial_t f + \matveloC^a\partial_a f
                        = \transdirC_{\m}^A \partial_A f 
                        = \matd f
                        = \dot{f} \formComma
        \end{align}
        \cf eq. \eqref{eq:dotf}.
        Therefore the total derivative in $ \R^3|_{\surf} $ equals the material derivative
        $ \dot{f} = \partial_t f + \nabla_{\relvelo}f $ on the moving surface.  
        Applying this insight to Euclidean components of an observer frame yields
        $ \totald \partial_i \paraC^a = \partial_i\obveloC^a + \relveloC^k\partial_k\partial_i\paraC^a $.
        Obviously, this cannot results in a tangential vector field generally, hence the total derivative alone is not convenient
        for an intrinsic surface calculus. Its tangential part is though.
        Taking up the reasoning, we introduce the \emph{tangential total derivative} identified by the dot-operator, 
        \ie\ $ \dot{\vararbt} := (\projsurf\circ\totald)\vararbt\in\tangentS[^n] $ for all $ \vararbt\in\tangentS[^n] $,
        where $ \projsurf: \tangentR[^n]\rightarrow\tangentS[^n] $ is the orthogonal projection into the tangential tensor bundle.
        With the tangential derivative given in \ref{sec:Bop_N_bop} \wrt\ to an observer frame it holds
        \begin{align}\label{eq:dotframe}
            \overdot{\partial_{i}\para}
                &=g^{jl}\delta_{ab}\left( \totald \partial_i \paraC^a \right)\partial_{l}\paraC^{b}\partial_j \para
                 = g^{jl}\left( \innerR{\partial_{i}\obvelo}{\partial_{l}\para} + \relveloC^k\Gamma_{kil} \right) \partial_j \para
                 = \left( \tensor{\BopC}{^j_i} + \relveloC^k\Gamma_{ki}^{j} \right)\partial_j \para \formPeriod
        \end{align}
        Performing the product rule for $ \vararbt=\vararbtC^{i_1\ldots i_n} \bigotimes_{\beta=1}^{n} \partial_{i_{\beta}}\para\in\tangentS[^n] $ 
        results in
        \begin{align*}
            \dot{\vararbt}
                    &= \overdot{\vararbtC^{i_1\ldots i_n}} \bigotimes_{\beta=1}^{n} \partial_{i_{\beta}}\para
                        +\vararbtC^{i_1\ldots i_n} \sum_{\beta=1}^{n} \partial_{i_{1}}\para\otimes \ldots\otimes\partial_{i_{\beta-1}}\para  
                                                \otimes\overdot{\partial_{i_{\beta}}\para}\otimes
                                                \partial_{i_{\beta+1}}\para\otimes  \ldots \otimes\partial_{i_{n}}\para
                   \formPeriod
        \end{align*}
        Using eq. \eqref{eq:dotfII} for the proxy function and eq. \eqref{eq:dotframe} for the frame, reveals
        \begin{align}\label{eq:tangential_total_derivative}
            \indexten{\dot{\vararbt}}{^{i_1\ldots i_n}}
                &= \partial_t\vararbtC^{i_1 \ldots i_n} + \indexten{\nabla_{\relvelo}\vararbt}{^{i_1 \ldots i_n}} 
                                                    + \sum_{\beta=1}^{n}\tensor{\BopC}{^{i_{\beta}}_{j}}
                                                            \vararbtC^{i_1 \ldots i_{\beta-1} j  i_{\beta+1} \ldots i_n}
        \end{align}
        Regarding \autoref{lem:material_derivative_of_instantaneous_tensors} let us formulate the following proposition.
        \begin{prop}\label{prop:instantaneous_material_derivative}
            For all $\vararbt\in\tangentS[^n]$ with $ \surf $ embedded in $ \R^3 $ and 
            total derivative $ \totald: \tangentS[^n] \rightarrow \tangentR[^n] $ holds
            \begin{align*}
                \dot{\vararbt} &= \projsurf\left( \totald[\vararbt] \right)
                                = \matd[\surf^n]\left( \vararbt\sttenbase{\surf^{n}} \right) \formPeriod
            \end{align*}
        \end{prop}
This means that the tangential total derivative equals the fully instantaneous part of the material derivative on instanteneous tensor fields. This provides a simple way to implement the material derivative, see e.g. \cite{Reuther2020}.
Moreover if the surface stays Cartesian, \ie\ flat, and the observer is transversal, \ie\ Eulerean, then the proposition results in
$\dot{\vararbt}= \matd[\surf^n]\left( \vararbt\sttenbase{\surf^{n}} \right)= \partial_t\vararbt + \nabla_{\matveloS}\vararbt$,
which is the wellknown material derivative in Cartesian spaces.

\section{Convected derivatives}\label{sec:convected_derivatives}
    In this section we consider convected derivatives, which are similar to the material derivative, but based on the Lie derivatives
    $ \Lie_{\matdir} $ in material spacetime direction $ \matdir\in\tangentstS $ instead of the covariant directional derivative $ \nabla_{\!\matdir} $.
    
    Unlike the latter derivative, Lie derivatives are not metric compatible. 
    With this in mind we introduce the \emph{shuffled flat operator}
    \begin{align*}
        \shflat[\tsigma]:\ \tangentstS[^n] 
                &\rightarrow \left( \tangentstS[^\talpha_{n-\talpha}] \right)^{T_{\tsigma}}\\
         \starbt
                &\mapsto \starbt^{\shflat[\tsigma]}
                      = \left\{ \tensor{\starbtC}{^{I_{1}\ldots I_{n}}_{I_{\talpha+1}\ldots I_{n}}}  \right\}^{T_{\tsigma}}
                      = \left\{ \delta^{I_{\tsigma(1)}}_{J_{\tsigma(1)}}\ldots \delta^{I_{\tsigma(\talpha)}}_{J_{\tsigma(\talpha)}}  
                                \stgC_{I_{\tsigma(\talpha+1)}J_{\tsigma(\talpha+1)}} \ldots \stgC_{I_{\tsigma(n)}J_{\tsigma(n)}}
                                \starbtC^{I_1 \ldots I_n}\right\} 
    \end{align*}
    for shuffles $ \tsigma\in\shuffles{n}{\talpha} $, see \ref{sec:shuffles}.
    This is an isomorphism and we denote its inverse by $ \shsharp[\tsigma]:=\shflat[\tsigma]^{-1} $.
    Note that the reason for using shuffles instead of permutations $ S_n $ is to omit merely transposing indices, which would not have any effects on upcoming convected derivatives below.
    The material derivative is invariant \wrt\ isomorphism $ \shflat[\tsigma] $, \ie\ 
    $ \matdst = \nabla_{\!\matdir} =\shsharp[\tsigma]\circ\nabla_{\!\matdir}\circ\shflat[\tsigma] $ for all $ \tsigma\in\shuffles{n}{\talpha} $.
    This feature can not arise from Lie derivatives 
    $ \Lie_{\matdir}:\ \left( \tangentstS[^\talpha_{n-\talpha}] \right)^{T_{\tsigma}} \rightarrow \left( \tangentstS[^\talpha_{n-\talpha}] \right)^{T_{\tsigma}} $.
    Therefor we define, for distinguishing in dependency of $ \tsigma\in\shuffles{n}{\talpha} $, the 
    \emph{shuffled convected derivatives} as 
    $ \condst{\shflat[\tsigma]} :=  \shsharp[\tsigma]\circ\Lie_{\matdir}\circ\shflat[\tsigma] := \tangentstS[^n] \rightarrow \tangentstS[^n] $.
    Let $ \starbt\in\tangentstS[^n] $, invariances \wrt\ transpositions yield
    $ (\Lie_{\matdir}\starbt^{\shflat[\tsigma]})^{T_{\tsigma^{-1}}} =  \Lie_{\matdir}\starbt^{\shflat[\shId{n}{\talpha}]} \in \tangentstS[^\talpha_{n-\talpha}]$
    and are given by \cite[Ch.~5.3]{Abraham2012}, \ie
    \begin{alignat*}{2}
        \indexten{\Lie_{\matdir}\starbt^{\shflat[\shId{n}{\talpha}]}}{^{I_{1}\ldots I_{\talpha}}_{I_{\talpha+1}\ldots I_{n}}}
            &= \transdirC^{K}_{\m}\partial_K \tensor{\starbtC}{^{I_{1}\ldots I_{\talpha}}_{I_{\talpha+1}\ldots I_{n}}}
                &&-\sum_{\beta=1}^{\talpha} \left( \partial_J \transdirC_{\m}^{I_{\beta}} \right)
                           \tensor{\starbtC}{^{I_{1}\ldots I_{\beta-1} J I_{\beta+1} \ldots I_{n}}_{I_{\talpha+1}\ldots I_{n}}} +\sum_{\beta=\talpha + 1}^{n} \left( \partial_{I_{\beta}} \transdir_{\m}^{J} \right)  
                         \tensor{\starbtC}{^{I_{1}\ldots I_{\talpha}}_{I_{\talpha+1}\ldots I_{\beta-1} J I_{\beta+1}\ldots I_{n}}} \formPeriod
    \end{alignat*}
    One feature of Lie-derivatives is that we are able to substitute partial derivatives by covariant derivatives, since the added Christoffel symbols are 
    extinguishing each other.
    Therefore it holds
    \begin{align}\label{eq:convected_derivative_raw}
        \condst{\shflat[\tsigma]}\starbt
            &= \matdst\starbt 
               - \sum_{\beta=1}^{\talpha} \left( \nabla\matdir \right) \underdot[\tsigma(\beta)] \starbt
               + \sum_{\beta=\talpha + 1}^{n} \left( \nabla\matdir \right)^{T} \underdot[\tsigma(\beta)] \starbt
    \end{align}  
    Apart from the issue that every weighted sum 
    $ \sum_{\talpha=0}^n\sum_{\tsigma\in\shuffles{n}{\talpha}}\omega_{\tsigma}\condst{\shflat[\tsigma]}:\ \tangentstS[^n] \rightarrow \tangentstS[^n] $ 
    could give 
    a noteworthy derivative as long as the weights $ \omega_{\tsigma}\in\R $  comply with 
    $\sum_{\talpha=0}^n\sum_{\tsigma\in\shuffles{n}{\talpha}}\omega_{\tsigma} = 1$,
    we emphasize here only the \emph{Jaumann derivative} $ \jaudst:=\frac{1}{2}(\condst{\sharp^{n}} + \condst{\flat^{n}})  $
    containing the (fully) \emph{upper convected derivative} $ \condst{\sharp^{n}} $ 
    and (fully) \emph{lower convected derivative} $ \condst{\flat^{n}} $.
    
    Similar to the material derivative we use orthogonal components 
    $ \cond[\sigma]{\shflat[\tsigma]} := \bbrackets[\sigma]{\cdot}\circ\condst{\shflat[\tsigma]}\circ\bbrackets{\cdot}^{-1} 
       :\ \sttensorbS[^n]\rightarrow\tangentS[^{n-\alpha}]$ to determine the \emph{shuffled convected derivatives}
    $ \cond{\shflat[\tsigma]}:= \bbrackets{\cdot}\circ\condst{\shflat[\tsigma]}\circ\bbrackets{\cdot}^{-1}
       :\ \sttensorbS[^n]\rightarrow  \sttensorbS[^n] $ on spacetime n-tensor bundles, \cf\ eq. \eqref{eq:tikzcd_invariant_operator}, 
    \ie\ for all $ \arbt\in\sttensorbS[^n] $ holds
    $ \cond{\shflat[\tsigma]}\arbt = \sum_{\alpha=0}^n\sum_{\sigma\in\shuffles{n}{\alpha}}(\cond[\sigma]{\shflat[\tsigma]}\arbt)\sttenbase{\sigma} $.
    The first summand in eq. \eqref{eq:convected_derivative_raw} yields the material derivative according to \autoref{thm:material_derivative} and the 
    remaining sum is determined by the rule of shuffled sum in \autoref{lem:sumofQdotsR} in conjunction with representation eq. \eqref{eq:gradmatdir} of
    the gradient of material direction.
    Adding this up results in
    \begin{align}\label{eq:convected_derivative_tmp}
        \begin{aligned}
            \indexten{\cond[\sigma]{\shflat[\tsigma]}\arbt}{^{i_1\ldots i_{n-\alpha}}}
                        &= \partial_t \arbtC_{\sigma}^{i_1\ldots i_{n-\alpha}} + \indexten{\nabla_{\relvelo}\arbt_{\sigma}}{^{i_1\ldots i_{n-\alpha}}} 
                            +\sum_{\beta=1}^{n-\alpha}\arbtC_{\sigma}^{i_1 \ldots i_{\beta-1} k i_{\beta+1} \ldots i_{n-\alpha}}
                                \begin{cases}
                                    -\tensor{\relveloC}{^{i_{\beta}}_{|k}} & \text{, if } (\tsigma^{-1}\circ\sigma)(\alpha+\beta) \le \talpha \\
                                    \indexsymten{\Bop+\Bop^{T}}{^{i_{\beta}}_{k}} + \tensor{\relveloC}{_{k}^{|i_{\beta}}} & \text{, otherwise}
                                \end{cases}\\
                        &\quad\quad+\zeta\hspace{-1.3em}\sum_{\substack{\beta=1\\ (\tsigma^{-1}\circ\sigma)(\beta) > \talpha}}^{\alpha}\hspace{-1.3em}
                           \left( \indexten{\instcond{\sharp}\matveloS}{_k} \arbtC_{\sigma^{\beta}}^{i_1\ldots i_{\rbeta-1}k i_{\rbeta} \ldots i_{n-\alpha}} 
                                  + 2\velonor\dot{\velonor}\arbtC_{\sigma}^{i_1\ldots i_{n-\alpha}}\right)
                           -\hspace{-1.3em}\sum_{\substack{\beta=1\\ (\tsigma^{-1}\circ\sigma)(\alpha+\beta) \le \talpha}}^{n-\alpha}\hspace{-1.3em}
                                    \indexten{\instcond{\sharp}\matveloS}{^{i_{\beta}}} \arbtC_{\sigma_{\beta}}^{i_1 \ldots \widehat{i_{\beta}} \ldots i_{n-\alpha}}
                        \formComma
        \end{aligned}
    \end{align}
    since $ \Bop-\Bop_{\m}=-\nabla\relvelo $ and $ \Bop+\Bop_{\m}^T = \Bop +\Bop^T + (\nabla\relvelo)^T $ is valid.
    We generalize the already predefined derivative $ \instcond{\sharp}\matveloS $  in \ref{sec:gradmatdir} 
    by \emph{shuffled instantaneous convected derivatives} 
    $ \instcond{\bsigma}:=\shsharp[\bsigma]\circ(\partial_t + \Lie_{\relvelo})\circ\shflat[\bsigma] 
        :\tangentS[^n] \rightarrow \tangentS[^n]$ for all $ \bsigma\in\shuffles{n}{\balpha} $,
    where $ (\partial_t + \Lie_{\relvelo}): (\tangentS[^{\balpha}_{n-\balpha}])^{T_{\bsigma}} \rightarrow (\tangentS[^{\balpha}_{n-\balpha}])^{T_{\bsigma}} $
    and $ \partial_t $ operates on the proxy functions in $ (\tangentS[^{\balpha}_{n-\balpha}])^{T_{\bsigma}} $.
    This derivative is \eg\ defined as Lie derivative on (possible time-dependent) tensor fields \wrt\ time-dependent relative velocity $ \relvelo $ in
    \cite[Ch.~1.6]{Marsden1994}. 
    Investigating the first line in the right-hand side of eq. \eqref{eq:convected_derivative_tmp} reveals that all terms containing $ \relvelo $
    are composing $ (\Lie_{\relvelo}\arbt_{\sigma}^{\shflat[\bsigma]})^{\shsharp[\bsigma]} $ for 
    $ \bsigma=\tsigma\setminus\sigma|_{\{1,\ldots,\alpha\}}\in\shuffles{n-\alpha}{\balpha} $, \cf\ \ref{sec:shuffles}.
    Moreover, the partial time derivative and terms of twice the observer frame deformation, 
    \ie\ $ \partial_t g_{ij} = \BopC_{ij} + \BopC_{ji}$, see \cite{Arroyo2009}, are sum up to
    $ \big(\partial_t\indexsymten{\arbtC_{\sigma}^{\shflat[\bsigma]}}{^{\cdots}_{\cdots}}\big)^{\shsharp[\bsigma]} $,
    where $ \indexsymten{\arbt_{\sigma}^{\shflat[\bsigma]}}{^{\cdots}_{\cdots}} $ means mixed co- and contravariant proxy functions
    matching the space $ (\tangentS[^{\balpha}_{n-\balpha}])^{T_{\bsigma}} $.
    Finally, we can formulate the following theorem, which determine 
    $ \cond{\shflat[\tsigma]} = \bbrackets{\cdot}\circ\condst{\shflat[\tsigma]}\circ\bbrackets{\cdot}^{-1}  $.
    \begin{thm}[Convected derivatives]\label{thm:convected_derivatives}
        Assuming $ \arbt=\sum_{\alpha=0}^n\sum_{\sigma\in\shuffles{n}{\alpha}}\arbt_{\sigma}\sttenbase{\sigma}\in\sttensorbS[^n] $,
        where $ \arbt_{\sigma}\in\tangentS[^{n-\alpha}] $, the shuffled convected derivative of $ \arbt $ \wrt\ 
        $ \tsigma\in\shuffles{n}{\talpha} $ is
        \begin{align*}
            \cond{\shflat[\tsigma]}\arbt 
              &= \sum_{\alpha=0}^n\sum_{\sigma\in\shuffles{n}{\alpha}}\left(\cond[\sigma]{\shflat[\tsigma]}\arbt\right)\sttenbase{\sigma}\in\sttensorbS[^n]
                    \formComma\\
            \indexten{\cond[\sigma]{\shflat[\tsigma]}\arbt}{^{i_1\ldots i_{n-\alpha}}}
                       &=  \indexten{\instcond{\shflat[\bsigma]}\arbt_{\sigma}}{^{i_1\ldots i_{n-\alpha}}}
                       +\zeta\hspace{-1.3em}\sum_{\substack{\beta=1\\ (\tsigma^{-1}\circ\sigma)(\beta) > \talpha}}^{\alpha}\hspace{-1.3em}
                                                  \left( \indexten{\instcond{\sharp}\matveloS}{_k} \arbtC_{\sigma^{\beta}}^{i_1\ldots i_{\rbeta-1}k i_{\rbeta} \ldots i_{n-\alpha}} 
                                                         + 2\velonor\dot{\velonor}\arbtC_{\sigma}^{i_1\ldots i_{n-\alpha}}\right)
                                                  -\hspace{-1.3em}\sum_{\substack{\beta=1\\ (\tsigma^{-1}\circ\sigma)(\alpha+\beta) \le \talpha}}^{n-\alpha}\hspace{-1.3em}
                                                           \indexten{\instcond{\sharp}\matveloS}{^{i_{\beta}}} \arbtC_{\sigma_{\beta}}^{i_1 \ldots \widehat{i_{\beta}} \ldots i_{n-\alpha}}\formComma \\
           \indexten{\instcond{\sharp}\matveloS}{^i}
                       &= \partial\matveloSC^{i} + \indexten{\nabla_{\relvelo}\matveloS - \nabla_{\matveloS}\relvelo}{^i} \formComma \\
           \instcond{\shflat[\bsigma]}\arbt_{\sigma}
                &= \partial_t^{\shflat[\tsigma]}\arbt_{\sigma} + \nabla_{\relvelo}\arbt_{\sigma}
                    - \sum_{\substack{\beta=1\\\mathcal{E}(\beta)}}^{n-\alpha} \left( \nabla\relvelo \right)\underdot[\beta]\arbt_{\sigma}
                    + \sum_{\substack{\beta=1\\\neg\mathcal{E}(\beta)}}^{n-\alpha} \left( \nabla\relvelo \right)^T\underdot[\beta]\arbt_{\sigma}
                 = \dot{\arbt}_{\sigma}
                     - \sum_{\substack{\beta=1\\\mathcal{E}(\beta)}}^{n-\alpha} \Bop_{\m}\underdot[\beta]\arbt_{\sigma}
                     + \sum_{\substack{\beta=1\\\neg\mathcal{E}(\beta)}}^{n-\alpha} \Bop_{\m}^T\underdot[\beta]\arbt_{\sigma}\formComma\\
           \partial_t^{\shflat[\bsigma]}\arbt_{\sigma}
                &:= \left\{ \partial_t \indexsymten{\arbt_{\sigma}^{\shflat[\bsigma]}}{^{\cdots}_{\cdots}} \right\}^{\shsharp[\bsigma]}
                 = \left\{\partial_t \arbtC_{\sigma}^{i_1\cdots i_{n-\alpha}}\right\}
                    +\sum_{\substack{\beta=1\\\neg\mathcal{E}(\beta)}}^{n-\alpha} 
                                \left( \Bop + \Bop^{T} \right)\underdot[\beta]\arbt_{\sigma}\formComma \\
           \mathcal{E}(\beta) &:= \left( (\tsigma^{-1}\circ\sigma)(\alpha+\beta) \le \talpha \right)
                                = \left( \bsigma^{-1}(\beta) \le \balpha \right) 
                    \text{, where } \bsigma=\tsigma\setminus\sigma|_{\{1,\ldots,\alpha\}}\in\shuffles{n-\alpha}{\balpha} 
        \end{align*}
        and $ \rbeta $ is given by $ (\sigma^{\beta})_{\rbeta}= \sigma $ implicitly. 
        Especially, the upper convected derivative $ \cond{\sharp^n} $ and lower convected derivative $ \cond{\flat^n} $ yield
        \begin{align*}
            \indexten{\cond[\sigma]{\sharp^n}\arbt}{^{i_1\ldots i_{n-\alpha}}}
                    &= \indexten{\instcond{\sharp^{n-\alpha}}\arbt_{\sigma}}{^{i_1\ldots i_{n-\alpha}}}
                        -\sum_{\beta=1}^{n-\alpha}  \indexten{\instcond{\sharp}\matveloS}{^{i_{\beta}}} 
                                                \arbtC_{\sigma_{\beta}}^{i_1 \ldots \widehat{i_{\beta}} \ldots i_{n-\alpha}} \formComma\\
            \indexten{\cond[\sigma]{\flat^n}\arbt}{^{i_1\ldots i_{n-\alpha}}}
                    &= \indexten{\instcond{\flat^{n-\alpha}}\arbt_{\sigma} 
                                    + 2\alpha\zeta\velonor\dot{\velonor}\arbt_{\sigma}}{^{i_1\ldots i_{n-\alpha}}}
                        +\zeta\indexten{\instcond{\sharp}\matveloS}{_k}\sum_{\beta=1}^{\alpha}
                                                    \arbtC_{\sigma^{\beta}}^{i_1\ldots i_{\rbeta-1}k i_{\rbeta} \ldots i_{n-\alpha}}\formComma\\
            \indexten{\instcond{\sharp^n}\arbt_{\sigma}}{^{i_1\ldots i_{n-\alpha}}}
                    &= \partial_t \arbtC_{\sigma}^{i_1\ldots i_{n-\alpha}}
                       + \indexten{\nabla_{\relvelo}\arbt_{\sigma} 
                                     - \sum_{\beta=1}^{n-\alpha} \left( \nabla\relvelo \right)\underdot[\beta]\arbt_{\sigma}}
                                  {^{i_1\ldots i_{n-\alpha}}}
                    &&=\indexten{\dot{\arbt}_{\sigma}
                                                     - \sum_{\beta=1}^{n-\alpha} \Bop_{\m}\underdot[\beta]\arbt_{\sigma}}{^{i_1\ldots i_{n-\alpha}}}\formComma\\
           \indexten{\instcond{\flat^n}\arbt_{\sigma}}{^{i_1\ldots i_{n-\alpha}}}
                    &= g^{i_1 j_1} \cdots g^{i_{n-\alpha}j_{n-\alpha}} \partial_t \indexten{\arbt_{\sigma}}{_{j_1\ldots j_{n-\alpha}}}
                        + \indexten{\nabla_{\relvelo}\arbt_{\sigma} 
                                      + \sum_{\beta=1}^{n-\alpha} \left( \nabla\relvelo \right)^T\underdot[\beta]\arbt_{\sigma}}
                                   {^{i_1\ldots i_{n-\alpha}}}
                    &&=  \indexten{\dot{\arbt}_{\sigma}
                                 + \sum_{\beta=1}^{n-\alpha} \Bop_{\m}^T\underdot[\beta]\arbt_{\sigma}}{^{i_1\ldots i_{n-\alpha}}}\formComma
        \end{align*}
        where $ \dot{\arbt}_{\sigma}\in\tangentS[^{n-\alpha}] $ is the tangential total derivative \eqref{eq:tangential_total_derivative} of 
        $ \arbt_{\sigma}\in\tangentS[^{n-\alpha}] $.
        The Jaumann derivative $ \jaud = \bbrackets{\cdot}\circ \jaudst \circ\bbrackets{\cdot}^{-1} $ is given by
        $ \jaud\arbt=\sum_{\alpha=0}^n\sum_{\sigma\in\shuffles{n}{\alpha}}\left(\jaud[\sigma]\arbt\right)\sttenbase{\sigma}\in\sttensorbS[^n] $ and
        \begin{align*}
            \indexten{\jaud[\sigma]\arbt}{^{i_1\ldots i_{n-\alpha}}}
                &= \frac{1}{2}\indexten{\cond[\sigma]{\sharp^n}\arbt + \cond[\sigma]{\flat^n}\arbt}{^{i_1\ldots i_{n-\alpha}}}\\
                &= \indexten{\dot{\arbt}_{\sigma} + \alpha\zeta\velonor\dot{\velonor}\arbt_{\sigma}
                           - \frac{\rot\matveloS}{2}\sum_{\beta=1}^{n-\alpha} *_{\beta}\arbt_{\sigma}}{^{i_1\ldots i_{n-\alpha}}}
                    +\frac{\zeta}{2}\indexten{\instcond{\sharp}\matveloS}{_k}\sum_{\beta=1}^{\alpha}
                                             \arbtC_{\sigma^{\beta}}^{i_1\ldots i_{\rbeta-1}k i_{\rbeta} \ldots i_{n-\alpha}}
                    -\frac{1}{2}\sum_{\beta=1}^{n-\alpha}  \indexten{\instcond{\sharp}\matveloS}{^{i_{\beta}}} 
                                             \arbtC_{\sigma_{\beta}}^{i_1 \ldots \widehat{i_{\beta}} \ldots i_{n-\alpha}}\formComma             
        \end{align*}
        with \emph{curl} $ \rot\matveloS = -\innerS[^2]{\nabla\matveloS}{\lct}\in\tangentS[^0] $,
        \emph{Hodge dual} $ *_{\beta}\arbt_{\sigma} = -\lct\underdot[\beta]\arbt_{\sigma}\in\tangentS[^{n-\alpha}] $
        \wrt\ $ \beta $-th dimension and
        \emph{Levi-Civita tensor} $ \lct\in\tangentS[^2] $, covariantly determined by 
        $ \lctC_{ij} = \sqrt{\det\g}\varepsilon_{ij} $ and \emph{Levi-Civita symbols} $ \varepsilon_{ij} $. 
    \end{thm}
    
\section{Scalar fields}\label{sec:scalar_fields}
    As already mentioned above, all introduced time derivatives \wrt\ material motions yield the same for 
    scalar fields $ f\in\sttensorbS[^0] $, \ie\ 
    \begin{align}\label{eq:scalar_derivatives}
        \matd f = \cond{}f = \dot{f} = \instcond{}f = \partial_t f + \nabla_{\relvelo} f = \partial_t f + \Lie_{\relvelo}f \in\sttensorbS[^0]
            \formComma
    \end{align}
    where $ \relvelo=\matveloS-\obveloS\in\tangentS $ is the relative,  $ \matveloS \in\tangentS$ the tangential material
    and  $ \obveloS \in\tangentS$ the tangential observer velocity.
    Note that
    we leave the shuffled flat operator $ \shflat[(|)]= \Id$ as well as the basis $ \sttenbase{(|)}= 1 $ \wrt\ empty shuffle $ (|)\in\shuffles{0}{0} $ blank.
    De facto up to small syntactical consideration, the spaces $ \sttensorbS[^0] $, $ \tangentstS[^0] $ and $ \tangentS[^0] $ are the same. 

\section{Vector fields}\label{sec:vector_fileds}
    Spacetime vector fields are closely related to so called 4-vectors \wrt\ our chosen embedding spacetime.
    We consider the spacetime vector field 
    $ \starbt=\starbtC^{I}\partial_{I}\stpara= \starbtC^{t}\partial_{t}\stpara+\starbtC^{i}\partial_{i}\stpara 
             = [\starbtC^{t},\{\starbtC^{i}\}]'_{\tangentstS}\in\tangentstS$,
    with observer direction $ \partial_{t}\stpara=[1,\obvelo]'_{\R^4} $
    and spatial basis direction $ \partial_{i}\stpara=[0,\partial_{i}\para]'_{\R^4} $.
    Thus Euclidean basis vectors $ \Eb_{t} $ (temporal) and $ \Eb_{a} $ (spatial) with $  a=x,y,z $ yield $ \starbt  $ as the Euclidean 4-vector
    $ \starbt=\tilde{\starbtC}^{t}\Eb_{t} +  \tilde{\starbtC}^{a}\Eb_{a}=[\tilde{\starbtC}^{t},\{\tilde{\starbtC}^{a}\}]_{\R^4}\in\tangentstR$,
    with $ \tilde{\starbtC}^{t}= \starbtC^{t}$ and 
    $ \tilde{\starbtC}^{a} = (\starbtC^{i}+\starbtC^{t}\obveloSC^{i})\partial_i\paraC^{a} + \starbtC^{t}\velonor\normalC^a$.
    Since the embedding frame is stationary, we see an observer dependence as long as $ \starbtC^I $ are considered as degree of freedoms
    even though we do not considering any dynamics at this point. 
    A different situation is appearing by using spacetime vector fields 
    $ \arbt=\bbrackets{\starbt}=\arbt_{\surf}\sttenbase{\surf} + \arbtC_{\transdir}\sttenbase{\transdir}
        =:\bbrackets{\arbtC_{\transdir} , \arbt_{\surf}}'\in\sttensorbS $, 
    \ie\ $ \starbt $ and $ \arbt $ are describing the same spacetime vector field isomorphically by 
    $ \starbt = [0,\arbt_{\surf}]'_{\tangentstS} + \arbtC_{\transdir}\transdir $ with transversal direction
    $ \transdir=[1,\velonor\normal]'_{\R^4}= [1,-\obveloS]'_{\tangentstS}\in\tangentstS$.  
    Hence $ \arbtC_{\transdir}=\starbtC^{t}\in\tangentS[^0] $ is valid as expected for a global time assumption.
    For the instantaneous part holds $ \arbt_{\surf}=\{\starbtC^{i}\}+\starbtC^{t}\obveloS\in\tangentS $
    and therefore the associated 4-vector uncovers $ \tilde{\starbtC}^{t} = \arbtC_{\transdir}$ temporally and
    $ \tilde{\starbtC}^{a} = \arbtC_{\surf}^i\partial_i\paraC^{a} + \arbtC_{\transdir}\velonor\normalC^a$ spatially.
    Considering
    the material derivative $ \nabla_{\matdir}\starbt $, 
    upper convected derivative $ \Lie_{\matdir}^{\sharp}\starbt= \nabla_{\matdir}\starbt - \nabla_{\starbt}\matdir$, 
    lower convected derivative  $ \Lie_{\matdir}^{\flat}\starbt= \nabla_{\matdir}\starbt + \starbt\nabla\matdir$
    and Jaumann derivative $ \frac{1}{2}(\Lie_{\matdir}^{\sharp}\starbt + \Lie_{\matdir}^{\flat}\starbt) $
    with material direction $ \matdir = [1,\matvelo]'_{\R^4}=[1,\relvelo]'_{\tangentstS}\in\tangentstS $,
    \autoref{thm:material_derivative} and \autoref{thm:convected_derivatives} are sufficient to bring these derivatives to $ \sttensorbS $ as
    stipulated in eq. \eqref{eq:tikzcd_invariant_operator} for $ n=1 $.
    \begin{concl}\label{concl:vector_fields}
        For spacetime vector fields $ \arbt=\bbrackets{\arbtC_{\transdir} , \arbt_{\surf}}'\in\sttensorbS $ the material $ \matd\ $, Jaumann $ \jaud\ $,
        upper convected $ \cond{\sharp}\ $ and lower convected derivative $\cond{\flat}: \sttensorbS \rightarrow \sttensorbS $ are
        \begin{align*}
            \matd\arbt 
                &= \btensor{\dot{\arbtC}_{\transdir} 
                                   + \zeta\velonor\dot{\velonor}\arbtC_{\transdir} 
                                              + \zeta\velonor\innerS{\bop_{\m}}{\arbt_{\surf}}\\
                            \dot{\arbt}_{\surf} - \velonor\arbtC_{\transdir}\bop_{\m}} \formComma
           &\jaud\arbt
                &= \btensor{\dot{\arbtC}_{\transdir}
                                    + \zeta\velonor\dot{\velonor}\arbtC_{\transdir}
                                              +\frac{\zeta}{2}\innerS{\instcond{\sharp}\matveloS}{\arbt_{\surf}} \\
                             \instjaud\arbt_{\surf} - \frac{\arbtC_{\transdir}}{2}\instcond{\sharp}\matveloS}\formComma\\
           \cond{\sharp}\arbt
                &=  \btensor{\dot{\arbtC}_{\transdir} \\
                             \instcond{\sharp}\arbt_{\surf} - \arbtC_{\transdir}\instcond{\sharp}\matveloS}\formComma
           &\cond{\flat}\arbt
                &= \btensor{\dot{\arbtC}_{\transdir}
                                    + 2\zeta\velonor\dot{\velonor}\arbtC_{\transdir}
                                              +\zeta\innerS{\instcond{\sharp}\matveloS}{\arbt_{\surf}} \\
                             \instcond{\flat}\arbt_{\surf}}
        \end{align*}
        with scalar time derivatives $\dot{\arbtC}_{\transdir}$ (resp. $\dot{\velonor})$ in $ \tangentS[^0] $, see eq. \eqref{eq:scalar_derivatives}, 
        and instantaneous 
        material $ \dot{\arbt}_{\surf} $, Jaumann $ \instjaud\arbt_{\surf} $,
        upper convected $ \instcond{\sharp}\arbt_{\surf} $ (resp. $\instcond{\sharp}\matveloS$)  and lower convected derivative $\instcond{\flat}\arbt_{\surf} $ given by
        \begin{align}\label{eq:instantaneous_vector_time_derivatives}
        \begin{aligned}
            \indexten{\dot{\arbt}_{\surf}}{^i}
                &= \partial_t\arbtC_{\surf}^i + \indexten{\nabla_{\relvelo}\arbt_{\surf} +\Bop\arbt_{\surf}}{^i}\formComma
            &\instjaud\arbt_{\surf}
                &= \dot{\arbt}_{\surf} - \frac{\rot\matveloS}{2}(*\arbt_{\surf})
                 = \dot{\arbt}_{\surf} - \frac{1}{2}\left( \nabla\matveloS - \left( \nabla\matveloS \right)^T \right)\arbt_{\surf}\formComma \\
            \indexten{\instcond{\sharp}\arbt_{\surf}}{^i}
                &= \partial_t\arbtC_{\surf}^i + \indexten{\nabla_{\relvelo}\arbt_{\surf} -\nabla_{\arbt_{\surf}}\relvelo}{^i}
                 = \indexten{\dot{\arbt}_{\surf} - \Bop_{\m}\arbt_{\surf}}{^i}\formComma
            &\indexten{\instcond{\flat}\arbt_{\surf}}{_i}
                &= \partial_t\indexten{\arbt_{\surf}}{_i} + \indexten{\nabla_{\relvelo}\arbt_{\surf} +\arbt_{\surf}\nabla\relvelo}{_i}
                 = \indexten{\dot{\arbt}_{\surf} + \Bop_{\m}^T\arbt_{\surf}}{_i}
        \end{aligned}
        \end{align}
        in $\tangentS$, where $ \bop_{\m} = \nabla\velonor + \shop\matveloS $, $ \Bop_{\m}=\nabla\matveloS - \velonor\shop $,
        $ \Bop=\nabla\obveloS - \velonor\shop $ and $ [*\arbt_{\surf}]^i = -\tensor{\lctC}{^i_k} \arbtC_{\surf}^k$.
    \end{concl}
    Note that eq. \eqref{eq:instantaneous_vector_time_derivatives} implements a set of time derivatives 
    $\tangentS \rightarrow \tangentS$ sufficiently for many situations in continuum mechanics, where instantaneous vector fields are part of the degrees of freedom,
    \ie\ the considered vector-valued quantities exhibit $\arbtC_{\transdir}=0$.

    \subsection{Material velocity and acceleration}\label{sec:material_direction}
        The \emph{material velocity} $ \matvelo=:\matveloS+\velonor\normal\in\tangentR $ of the surface can be obtained from a time-depending parametrization $ \para_{\m} $ of the moving surface by $ \matvelo:=\partial_t\para_{\m} $, similarly to \autoref{sec:spacetime_surface_and_tensor_bundle}.
        This special choice of the observer parametrization suffices the Lagrangian perspective of the material, though.
        The associated spacetime 4-vector is the (future-pointing) \emph{material direction} 
        $ \matdir:=[1,\matvelo]'_{\tangentstR}=\partial_t\stpara_{\m} $, where we define
        $ \stpara_{\m}:=[t,\para_{\m}]'_{\R^4} $.
        Hence, \wrt\ an arbitrary observer frame we obtain with $ \zeta(1+\velonor^2)= 1 $, that
        \begin{align*}
            \matdir = \tensorstS{\stgC^{tJ}\innerstR{\matdir}{\partial_J\para} \\
                                 \left\{ \stgC^{iJ}\innerstR{\matdir}{\partial_J\para} \right\}}
                    = \tensorstS{\zeta\left( 1+ \innerR{\matvelo}{\obvelo} - \innerS{\obveloS}{\matveloS} \right) \\
                                 \matveloS + \zeta\left( \innerS{\obveloS}{\matveloS} - \left( 1+ \innerR{\matvelo}{\obvelo} \right) \right)\obveloS}
                    = \tensorstS{1 \\ \relvelo}\formComma
        \end{align*}
        where $ \relvelo:=\matveloS - \obveloS=  \matvelo - \obvelo $ is the \emph{relative velocity}.
        Splitting $ \matdir=[0,\matveloS]'_{\tangentstS} + \transdir\in\tangentstS $ 
        into its instantaneous and transversal part yields the observer independent representation
        $ \bbrackets{\matdir}= \matveloS\sttenbase{\surf} + \sttenbase{\transdir}=\bbrackets{1,\matveloS}'\in\sttensorbS $.
        The \emph{material acceleration} can be obtained by the total derivative of the material velocity component-wise in $ \tangentR $,
        \ie\ $ \mataccelC^a = \totald\matveloC^a =:\indexten{\totald\matvelo}{^a} $, where 
        $ \mataccel=:\mataccelS + \mataccelnor\normal\in\tangentR $ with 
        \emph{material tangential acceleration} $ \mataccelS\in\tangentS $ 
        and \emph{material normal acceleration} $ \mataccelnor\in\tangentS[^0] $. 
        For this purpose we need the total derivative of the normal field $ \normal\in\tangentR $.
        With eq. \eqref{eq:dotfII} we have 
        $ \indexten{\totald\normal}{^a}:=\totald\normalC^a = \partial_t\normalC^a + \relveloC^k\partial_k\normalC^a $.
        Since $ \left\| \normal \right\|_{\tangentR}=1 $ it holds $ \totald\normal\in\tangentS $. 
        The product rule yields
        \begin{align*}
            \innerR{\totald\normal}{\partial_i\para}
                &= -\innerR{\normal}{\partial_i\obvelo} + \relveloC^k\innerR{\partial_k\normal}{\partial_i\para}
                 =-\left( \partial_i\velonor + \indexten{\shop\left( \obveloS + \relvelo \right)}{_{i}} \right)
                 =-\indexten{\nabla\velonor + \shop\matveloS}{_{i}}
                 =-\indexten{\bop_{\m}}{_i}\formPeriod
        \end{align*}
        With the event-wise 
        orthogonal projection $ \projsurf:\tangentR\rightarrow\tangentS $ 
        and the instantaneous derivative $ \dot{\obveloS}_{\m}\in\tangentS $ we obtain according to \autoref{prop:instantaneous_material_derivative}
        \begin{align*}
            \mataccelS
                &= \projsurf\left( \totald\matvelo \right)
                 = \projsurf\left( \totald\matveloS\right) + \velonor\totald\normal 
                 = \dot{\obveloS}_{\m} - \velonor\bop_{\m} 
                &&= \left\{\partial_t\matveloSC^i \right\} + \nabla_{\relvelo}\matveloS + \nabla_{\matveloS}\obveloS
                                             -\velonor\left( 2\shop\matveloS + \nabla\velonor \right)\formComma\\
            \mataccelnor
                &= \innerR{\totald\matvelo}{\normal}
                 = \totald\velonor - \innerR{\matvelo}{\totald\matvelo}
                 =\dot{\velonor} + \innerS{\matveloS}{\bop_{\m}}   
                &&= \partial_t\velonor + \nabla_{\relvelo+\matveloS}\velonor + \shop\left( \matveloS,\matveloS \right) \formComma
        \end{align*}
        which reveals the same result as \cite{Yavari2016}, most comparable in the representation of the last column. 
        The derivative $ \totald\matdir $ of the material direction does not lay tangential to the spacetime surface $ \stsurf $.
        Thus we now only evaluate the spacetime tangential part of $ \totald\matdir=[0,\mataccel]'_{\tangentstR} $ by means of
        orthogonal projection $ \stproj{\stsurf}:\tangentstR\rightarrow\tangentstS $. That is
        \begin{align*}
            \stproj{\stsurf}\left(\totald\matdir\right)
                &=\tensorstS{\stgC^{tt}\innerR{\mataccel}{\obvelo} + \stgC^{tj}\innerR{\mataccel}{\partial_j\para} \\
                             \left\{ \stgC^{it}\innerR{\mataccel}{\obvelo} + \stgC^{ij}\innerR{\mataccel}{\partial_j\para} \right\}}
                 = \tensorstS{\zeta\velonor\mataccelnor \\ 
                               \mataccelS - \zeta\velonor\mataccelnor\obveloS}
                 = \tensorstS{0\\\mataccelS} + \zeta\velonor\mataccelnor\transdir\formPeriod
        \end{align*}
        Therefore the spacetime \emph{tangential material acceleration direction} yields 
        $ \matd\bbrackets{\matdir}=\bbrackets{\stproj{\stsurf}(\totald\matdir)}\in\sttensorbS $, since by \autoref{concl:vector_fields} it holds
        \begin{align*}
            \matd\bbrackets{\matdir}
                &= \btensor{\zeta\velonor\left( \dot{\velonor} + \innerS{\bop_{\m}}{\obveloS} \right) \\
                            \dot{\obveloS}_{\m} - \velonor\bop_{\m}}
                 = \btensor{\zeta\velonor\mataccelnor \\ \mataccelS} \formPeriod
        \end{align*}
        It comes as no surprise that $ \cond{\sharp}\bbrackets{\matdir} = 0 \in\sttensorbS $, since the material direction is frozen within its own flow.
        For the sake of completeness it is 
        \begin{align*}
            \cond{\flat}\bbrackets{\matdir}
                &= 2\jaud\bbrackets{\matdir}
                 = \btensor{2\zeta\velonor\dot{\velonor} + \zeta\innerS{\cond{\sharp}\matveloS}{\matveloS}\\
                            \cond{\flat}\matveloS}
                 = \btensor{\zeta\left( \overdot{\left\| \matvelo \right\|_{\tangentR}^2} - \Bop_{\m}\left( \matveloS, \matveloS \right) \right)\\
                            \dot{\obveloS}_{\m} + \Bop_{\m}^{T}\matveloS} \in \sttensorbS\formPeriod
        \end{align*}
   
   \subsection{Force-free transport of instantaneous vector fields on a stretching spheroid}\label{sec:transportvectorstrechingspheroid}
\begin{figure}
            \centering
            \includegraphics[width=0.99\linewidth]{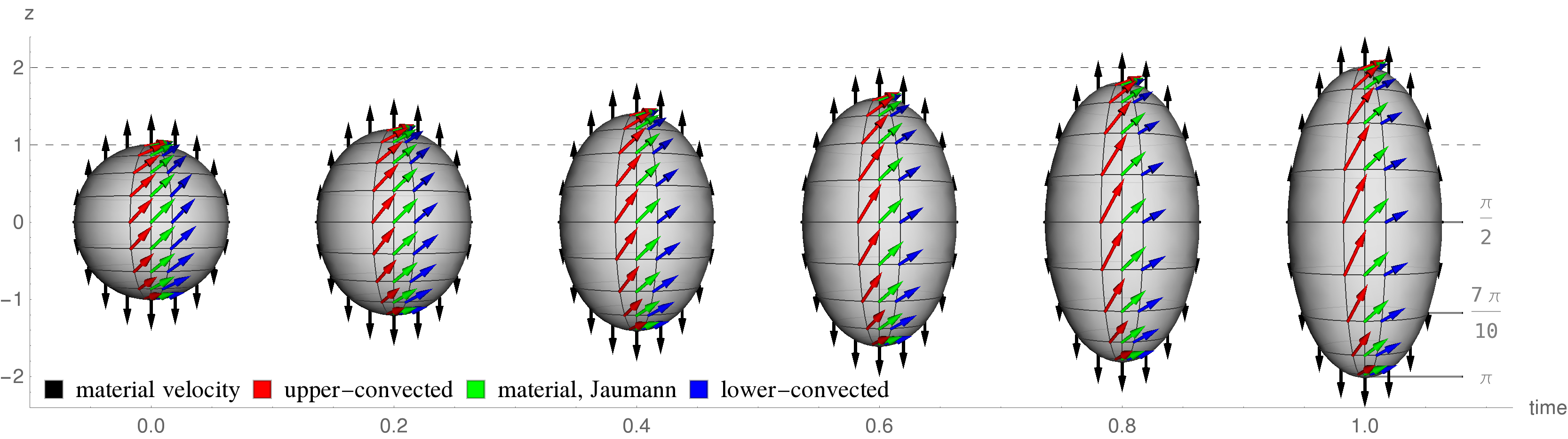}
            \includegraphics[width=0.35\linewidth]{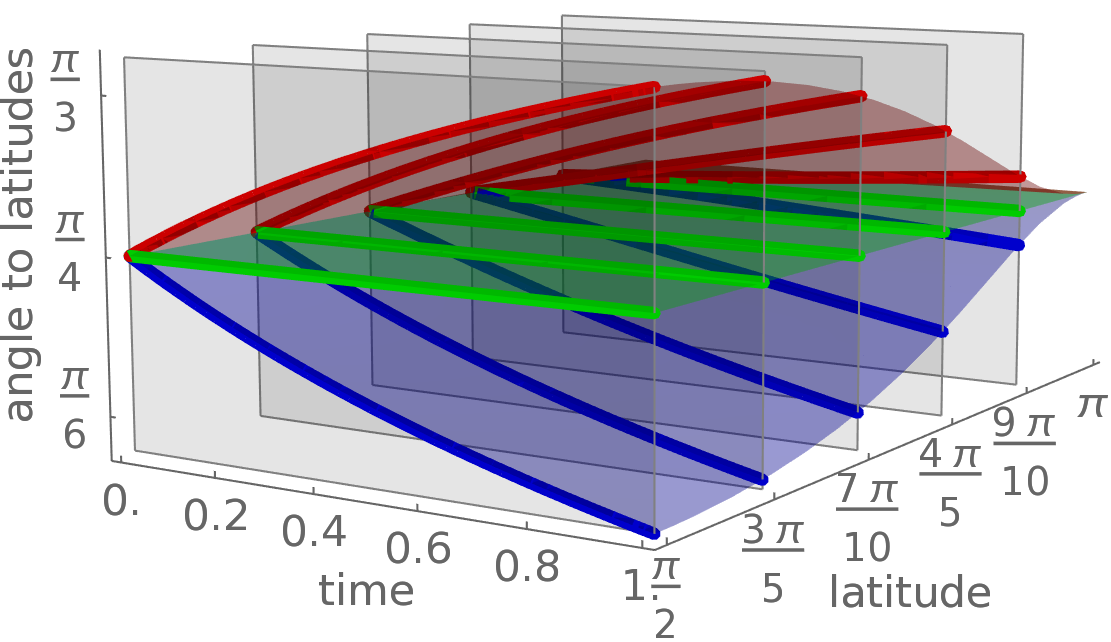}
            \hfill
            \includegraphics[width=0.35\linewidth]{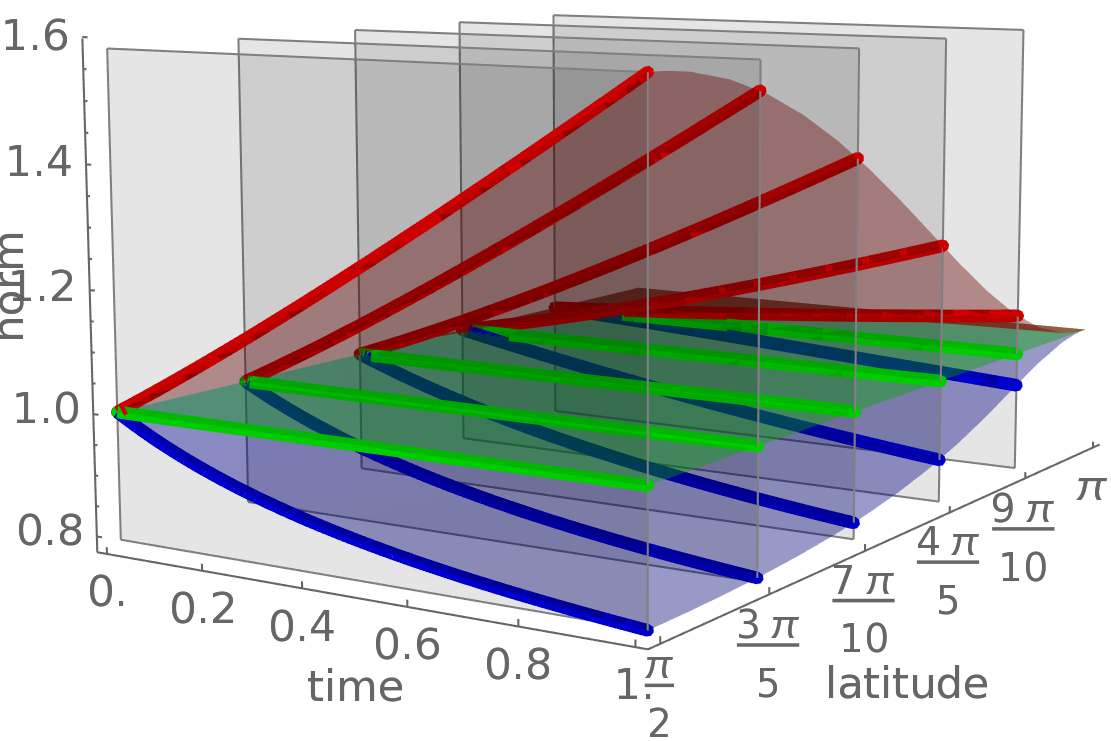}
            \hfill
            \raisebox{0.15\height}{\includegraphics[width=0.25\linewidth]{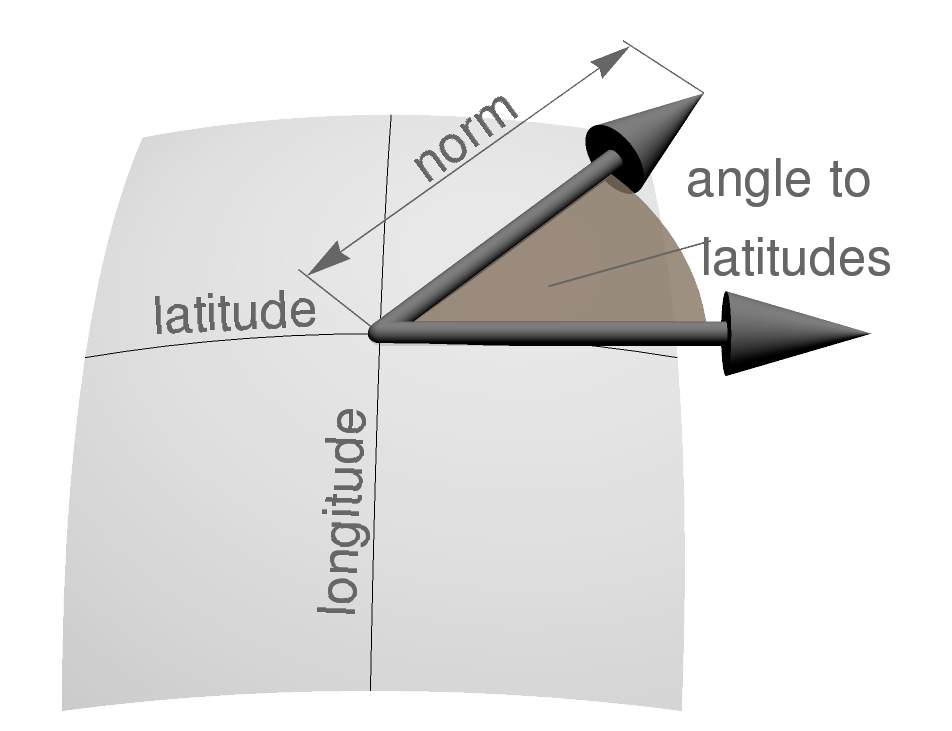}}
            \caption{(top) Evolution of the spheroid $ \surf $ at times $ t=0,0.2,0.4,0.6,0.8,1 $, 
                           its rotational invariant velocity field $ \matvelo $ and vector fields $ \arbt $ force-free transported 
                           \wrt\ upper-convected, material, Jaumann and lower-convected derivative.
                           The Jaumann and material transported vector fields are equal.
                           The motion of $ \surf $ yields a constant growing semi-axes \wrt\ Euclidean $ z $-direction.
                           The initial condition is $ \arbt_0 = \frac{1}{\sqrt{2}}[-1,\frac{1}{\sin y^1_{\m}}]'_{\tangentS|_{t=0}} $
                           \wrt\ Lagrangian observer Parametrization $ \para_{\m} $.
                      (bottom) Corresponding angle $ \phi^2(\arbt) $ to the latitudes and norm $ \|\arbt\| $ on the lower hemisphere,
                               where $ \phi^2(\arbt_0)=\frac{\pi}{4} $ and $ \|\arbt_{0}\|=1 $ holds.
                               Single latitudes used in the top image are emphasized here.}
            \label{fig:stretchingsphereeval}
        \end{figure}

        In this section we consider instantaneous vector fields $ \bbrackets{0,\arbt}'\in\sttensorbS $
        on a stretching spheroid
        $ \surf=\{[x,y,z]'\in\R^3 \mid x^2 + y^2 + (\frac{z}{1+t})^2=1\} $ for $ t\ge 0 $, see \autoref{fig:stretchingsphereeval} (top),
        and investigate the evolution of $ \arbt\in\tangentS $ within an instantaneous force-free transport equation
        \wrt\ time derivatives listed in \autoref{concl:vector_fields},
        \ie\  $ \matd[\surf]\bbrackets{0,\arbt}'= \dot{\arbt} = 0 $, 
        $ \jaud[\surf]\bbrackets{0,\arbt}'= \instjaud\arbt = 0 $,
        $ \cond[\surf]{\sharp}\bbrackets{0,\arbt}'= \instcond{\sharp}\arbt = 0  $ or
        $ \cond[\surf]{\flat}\bbrackets{0,\arbt}'= \instcond{\flat}\arbt = 0  $.
        Due to the observer-invariance we are free to choose arbitrary observer parametrizations which are sufficient for the stretching spheroid.
        A simple option is to use a Lagrangian observer realized by a single patch parametrization
        \begin{align*}
            \para_{\m}(t,y^1_{\m},y^2_{\m})
                &= \begin{bmatrix}
                        \sin y^1_{\m} \cos y^2_{\m}\\
                        \sin y^1_{\m} \sin y^2_{\m}\\
                        (1+t)\cos y^1_{\m}
                    \end{bmatrix}\in \R^3 \formComma
        \end{align*}
        where $ y^1_{\m}\in[0,\pi] $ is the latitude and $ y^2_{\m}\in[0,2\pi) $ the longitude coordinate.
        Regarding these coordinates we introduce the tangent angle fields $ \phi^i:\tangentS\rightarrow\tangentS[^0] $,
        \st\ for $ \arbt\in\tangentS $ holds $ (-1)^i\sqrt{g_{ii}}\|\arbt\|_{\tangentS}\cos\phi^i(\arbt) = \arbtC_i $,
        \ie\ point-wisely $ \phi^1(\arbt) $ is the angle of $ \arbt  $ to the longitudes and  $ \phi^2(\arbt) $ to the 
        latitudes in the tangent planes, see \autoref{fig:stretchingsphereeval} (bottom right).
        We observe that the tangential part of the material velocity $ \matvelo=\cos y^1_{\m} \Eb_{z}|_{\surf} $ is
        the potential field $ \matveloS = -\frac{1+t}{2}\nabla\sin^2 y_{\m}^1 $.
        This tangential vector field is curl-free and hence the considered material and Jaumann transport equations are equal.
        According to our choice of a Lagrangian observer the relative velocity $ \relvelo $ vanishes and the material and Jaumann
        derivative read 
        $ \tensor{[\dot{\arbt}]}{^i}= \tensor{[\instjaud\arbt]}{^i} = \partial_t\arbtC^i + \tensor{[\Bop_{\m}]}{^1_1}\delta^i_1 \arbtC^1 $
        and the upper-convected derivative $\tensor{[\instcond{\sharp}\arbt]}{^i}=\partial_t\arbtC^i  $
        contravariantly component-wise,
        where $ \tensor{[\Bop_{\m}]}{^1_1} = \frac{(1+t)\sin^2 y_{\m}^1}{1+t(2+t)\sin^2 y_{\m}^1} $ is the only non-vanishing
        component of $ \Bop_{\m}\in\tangentS[^1_1] $.
        The lower-convected derivatives can be written as $ \tensor{[\instcond{\flat}\arbt]}{_i}=\partial_t\arbtC_i $
        \wrt\ covariant components.
        With initial condition $ \arbt|_{t=0} = \arbt_0\in\tangentS|_{t=0} $ the solutions are
        \begin{align}
            \dot{\arbt}=\instjaud\arbt &= 0
                &&\Longrightarrow& \arbtC^1 &= \sqrt{g^{11}}\arbtC_0^1 = \frac{\arbtC_0^1}{\sqrt{1+t(2+t)\sin^2 y_{\m}^1}} \formComma
                                 & \arbtC^2 &= \arbtC_0^2\formComma \label{eq:matandjausolution_stretchingsphere}\\
            \instcond{\sharp}\arbt &= 0
                &&\Longrightarrow& \arbtC^1 &= \arbtC_0^1 \formComma
                                 & \arbtC^2 &= \arbtC_0^2\formComma \notag\\  
            \instcond{\flat}\arbt &= 0
                &&\Longrightarrow& \arbtC^1 &= g^{11}\arbtC_0^1 = \frac{\arbtC_0^1}{1+t(2+t)\sin^2 y_{\m}^1}\formComma
                                 & \arbtC^2 &= \arbtC_0^2\formPeriod  \notag                 
        \end{align}
        Note that only the material and Jaumann transported vector field preserve the length and the angles on the stretching spheroid $ \surf $,
        \ie\ $ \|\arbt\|= \|\arbt_0\|_{t=0}$ and $ \phi^i(\arbt) = \phi^i|_{t=0}(\arbt_0) $ is valid,
        see \autoref{fig:stretchingsphereeval} for an example.
        
    \subsection{Force-free transport of instantaneous vector fields on a rotating sphere}\label{sec:transportvectorrotatingsphere}
\begin{figure}
            \centering
            \includegraphics[width=0.99\linewidth]{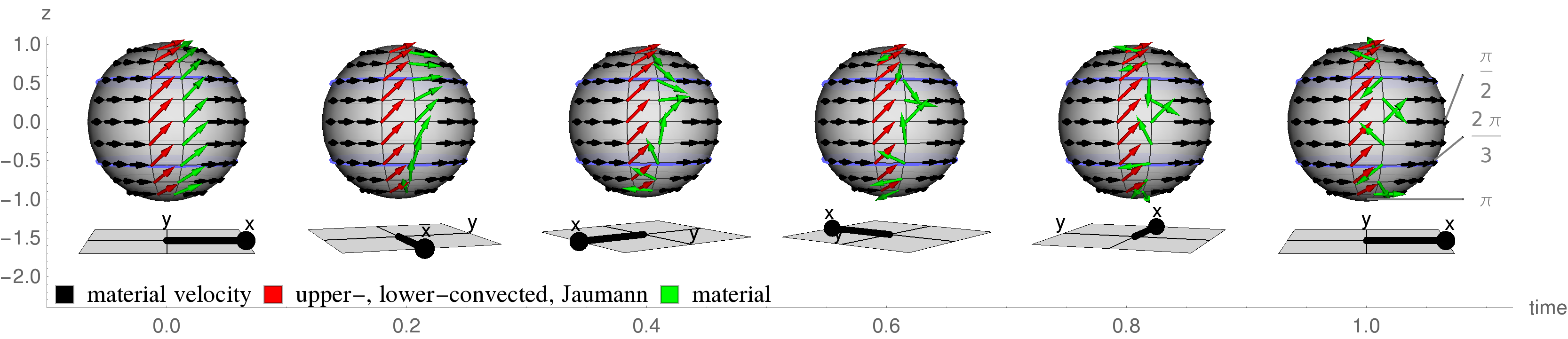}
            \includegraphics[width=0.7\linewidth]{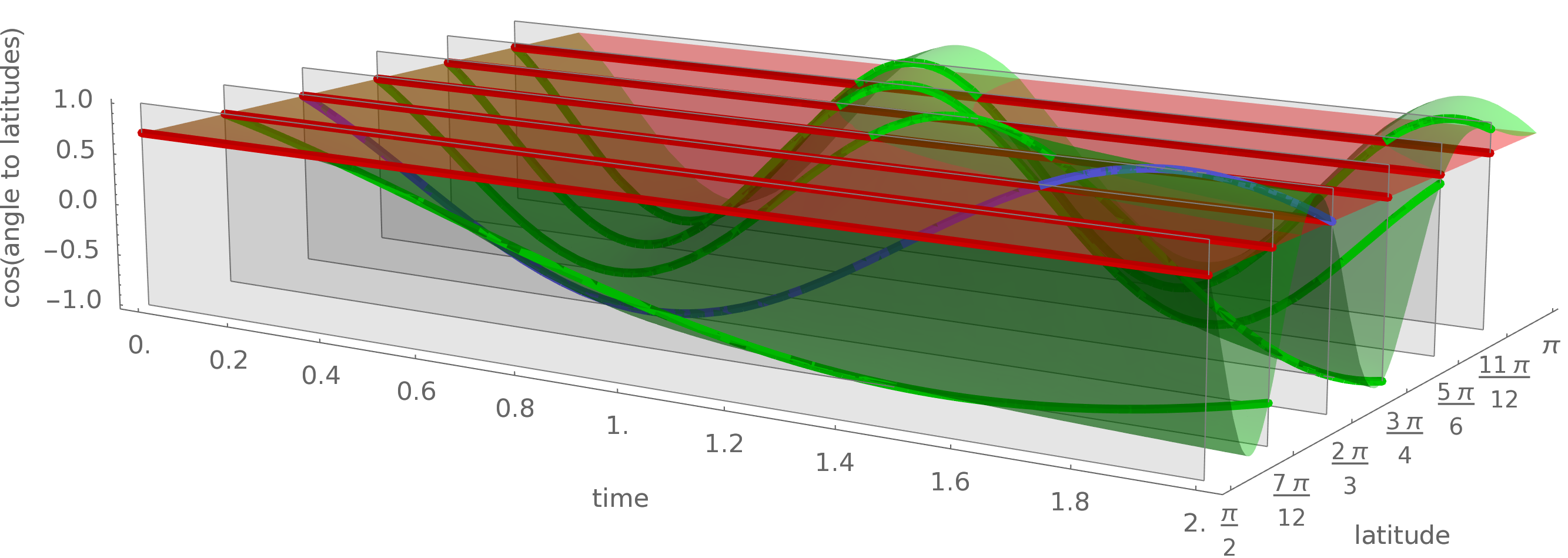}
            \hfill
            \raisebox{0.1\height}{\includegraphics[width=0.25\linewidth]{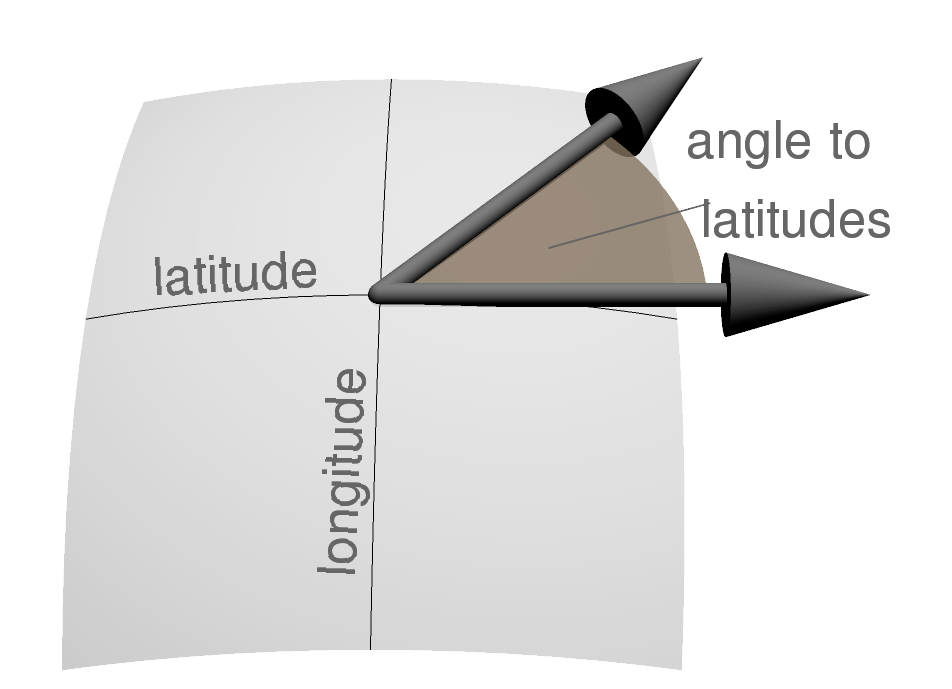}}
            \caption{(top) Rotating Sphere $ \surf $ at times $ t=0,0.2,0.4,0.6,0.8,1 $, 
                           its rotational invariant velocity field $ \matvelo $ and vector fields $ \arbt $ force-free transported 
                           \wrt\ upper-convected, lower-convected, Jaumann and material derivative.
                           The upper-convected, lower-convected and Jaumann transported vector fields are equal.
                           The perspective rotate consistently with the rotation of the sphere, \st\ the observed material points stay in front.
                           The initial condition is $ \arbt_0 = \frac{1}{\sqrt{2}}[-1,\frac{1}{\sin y^1_{\m}}]'_{\tangentS|_{t=0}} $
                           \wrt\ Lagrangian observer Parametrization $ \para_{\m} $.
                      (bottom) Corresponding cosine of angle $ \phi^2(\arbt) $ to the latitudes on one hemisphere,
                               where $ \phi^2(\arbt_0)=\frac{\pi}{4} $ holds.
                               Single latitudes used in the top image are emphasized here.
                               Especially blue lines represent 30 degree geographical latitude, where the material transported vector field 
                               needs two fully rotations of the sphere for one fully rotation in the tangent plane.}
            \label{fig:rotatingsphereeval}
        \end{figure}

        In this section we consider instantaneous vector fields $ \bbrackets{0,\arbt}'\in\sttensorbS $
        on a sphere
        $ \surf=\{[x,y,z]'\in\R^3 \mid x^2 + y^2 + z^2=1\} $, which is rotating constantly, see \autoref{fig:rotatingsphereeval} (top),
        and investigate the evolution of $ \arbt\in\tangentS $ within an instantaneous force-free transport equation similar to the example of the stretching spheroid above.
        We realize a Lagrangian observer for $ t \ge 0 $ by the single patch parametrization
        \begin{align*}
            \para_{\m}(t,y^1_{\m},y^2_{\m})
                &= \begin{bmatrix}
                        \sin y^1_{\m} \cos (y^2_{\m} + 2\pi t)\\
                        \sin y^1_{\m} \sin (y^2_{\m} + 2\pi t)\\
                        \cos y^1_{\m}
                    \end{bmatrix}\in \R^3 \formComma
        \end{align*}
        where $ y^1_{\m}\in[0,\pi] $ is the latitude and $ y^2_{\m}\in[0,2\pi) $ the longitude coordinate,
        \ie\ the orbital period is 1 unit of time.
        This is a rigid body motion, where the rate of deformation $ \partial_t g_{ij} $ vanish, 
        \ie\ the tensor field $ \Bop_{\m} = -\frac{\rot\matveloS}{2}\lct = -2\pi\cos y^1_{\m}\lct$ is antisymmetric and 
        hence both convected and the Jaumann transport equations are equal.
        The solution of $ \instcond{\sharp}\arbt=\instcond{\flat}\arbt=\instjaud\arbt=0 $ for 
        $ \arbt=\tangentS $ and $ \arbt|_{t=0} = \arbt_0\in\tangentS|_{t=0} $ is $ \arbtC^i = \arbtC_0^i $ \wrt\ the 
        considered Lagrangian observer frame.
        The solution for the material transport equation 
        $ \tensor{[\dot{\arbt}]}{^i} = \partial_t\arbtC^i - 2\pi\cos y^1_{\m}\tensor{\lctC}{^i_j}\arbtC^j = 0  $ is
        \begin{align*}
            \arbt 
                &= \cos\left( 2\pi t \cos y^1_{\m} \right) \arbt_0 - \sin\left( 2\pi t \cos y^1_{\m} \right)(*\arbt_0)
                 = \tensorS{\cos\left( 2\pi t \cos y^1_{\m} \right) \arbtC_0^1 
                            + \sin\left( 2\pi t \cos y^1_{\m} \right)\sin(y^1_{\m}) \arbtC_0^2\\
                            -\frac{\sin\left( 2\pi t \cos y^1_{\m} \right)}{\sin y^1_{\m}} \arbtC_0^1
                            + \cos\left( 2\pi t \cos y^1_{\m} \right) \arbtC_0^2} \formComma
        \end{align*}
        \ie\ the vector field circulates in the tangential plane, clockwisely at the upper and counterclockwisely at the lower hemisphere.
        This is the solution of a field of Foucault pendula, see \cite{Oprea1995}, 
        where $ \arbt $ represent the swing direction with a consistently chosen orientation.
        Note that all four transport equations preserve the length of the initial solution.
        For the convected solutions this is a consequence of the rigid body motion.
        The material transported solution does not preserve the angel to latitudes or longitudes, contrary to the other three solutions,
        see \autoref{fig:rotatingsphereeval}.
        Nevertheless, the material transported vector field experiences less directional changes \wrt\ the embedding space.
        
        We would like to use the opportunity to demonstrate observer-invariance of time derivatives at this simple example of a vector field on a rotating sphere.
        Let us occupy an Eulerian perspective instead of the Lagrangian above.
        For this purpose an obvious choice of a transversal observer parameterization is
        \begin{align*}
            \para_{\e}(t,y^1_{\e},y^2_{\e}) = \para_{\e}(y^1_{\e},y^2_{\e})
                &= \begin{bmatrix}
                        \sin y^1_{\e} \cos y^2_{\e} \\
                        \sin y^1_{\e} \sin y^2_{\e} \\
                        \cos y^1_{\e}
                    \end{bmatrix}\in \R^3 \formComma
        \end{align*}
        \ie\ the observer is stationary and hence it holds $\obvelo=\obveloS=0$, $\velonor=0$ and $\relvelo=\matveloS=2\pi\partial_2\para_{\e}$. 
        All convected derivatives are equal. 
        However, we have to solve the system of PDEs
        \begin{align*}
            0 = [\instcond{\sharp}\arbt]^i = [\instcond{\flat}\arbt]^i = [\instjaud\arbt]^i
              &= \partial_t \arbtC^i + 2\pi\left( \tensor{\arbtC}{^i_{|2}} + \cos y^1_{\e} \tensor{\lctC}{^i_k}\arbtC^k \right)
        \end{align*}
        instead of uncoupled ODEs for the material observer above. 
        The vector field $\arbt_{\C}= \arbtC_{\C}^i\partial_i \para_{\e}$ with components
        $\arbtC_{\C}^i(t,y^1_{\e},y^2_{\e}) = \arbtC_{0}^i(y^1_{\e}, y^2_{\e} - 2\pi t)$ solve these equations.
        Let us carry out a coordinate transformation into material coordinates at arbitrary shared points, 
        \st\ $\para_{\e}(y^1_{\e},y^2_{\e})= \para_{\m}(t,y^1_{\m},y^2_{\m})$ holds.
        Since $\partial_i \para_{\e}(y^1_{\e},y^2_{\e}) = \partial_i\para_{\m}(t,y^1_{\m},y^2_{\m})$
        and $\para_{\e}(y^1_{\m},y^2_{\m} + 2\pi t) = \para_{\m}(t,y^1_{\m},y^2_{\m})$ is valid,
        the solution vector field becomes $\arbt_{\C}= \tilde{\arbtC}_{\C}^i\partial_i \para_{\m}$ with components
        $\tilde{\arbtC}_{\C}^i(t,y^1_{\m},y^2_{\m}) = \arbtC_{\C}^i(t,y^1_{\m},y^2_{\m} + 2\pi t) = \arbtC_{0}^i(y^1_{\m}, y^2_{\m} )$,
        which is the already known solution above.
        Therefore, $\arbt_{\C}$ solve these force-free transport equation for a material as well as a transversal observer
        and it does not make a difference whether we occupy a Lagrangian or an Eulerean perspective, the outcome is equal.
        The same is true for the transport equation \wrt\ material derivative.
        With the transversal observer these equations result in PDEs
        $0=[\dot{\arbt}]^i = \partial_t\arbtC^i + 2\pi\tensor{\arbtC}{^i_{|2}} $ with solution
        $\arbt = \cos\left( 2\pi t \cos y^1_{\e} \right) \arbt_{\C} - \sin\left( 2\pi t \cos y^1_{\e} \right)(*\arbt_{\C})$,
        which is the same solution as for the material observer \wrt\ coordinates transformation.

    \subsection{Force-free transport of instantaneous vector fields on a helically stretching spheroid}\label{sec:transportvectorstretchrotsphere}
\begin{figure}
            \centering
            \includegraphics[width=0.99\linewidth]{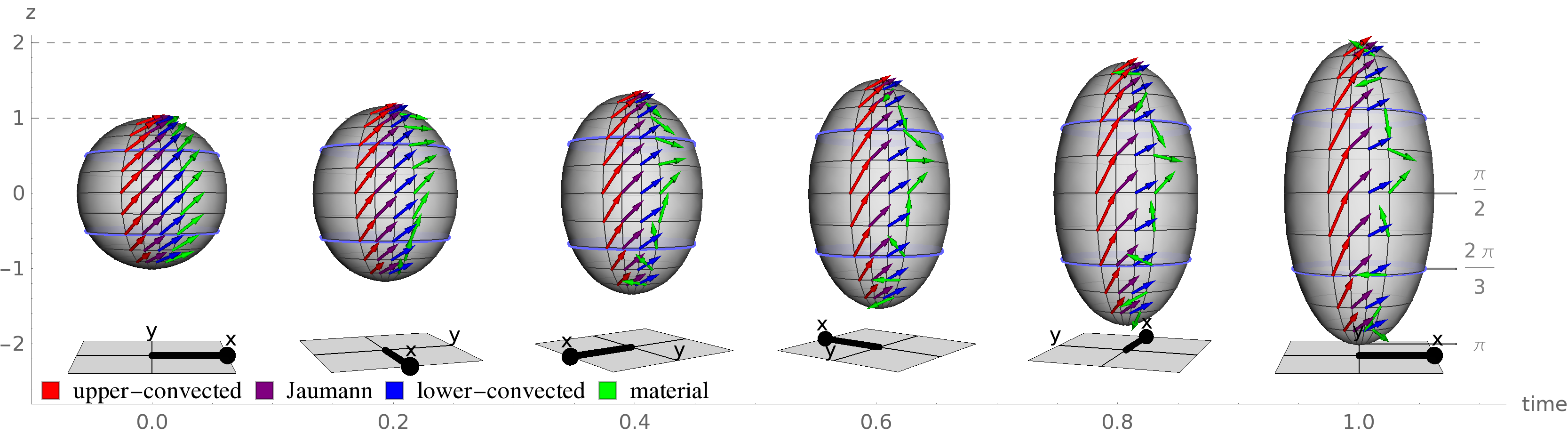}
            \includegraphics[width=0.5\linewidth]{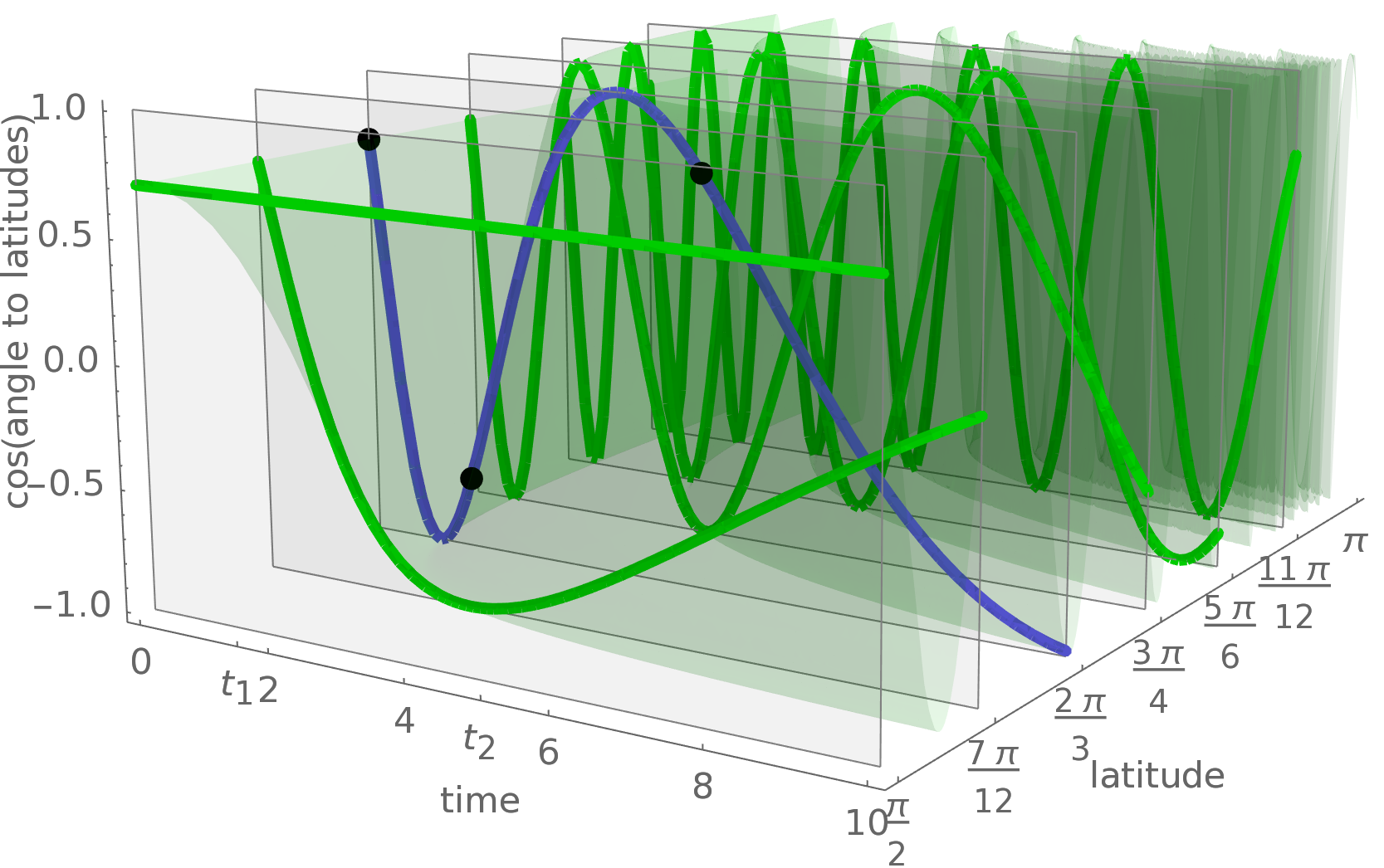}
            \hfill
            \raisebox{0.07\height}{\includegraphics[width=0.2\linewidth]{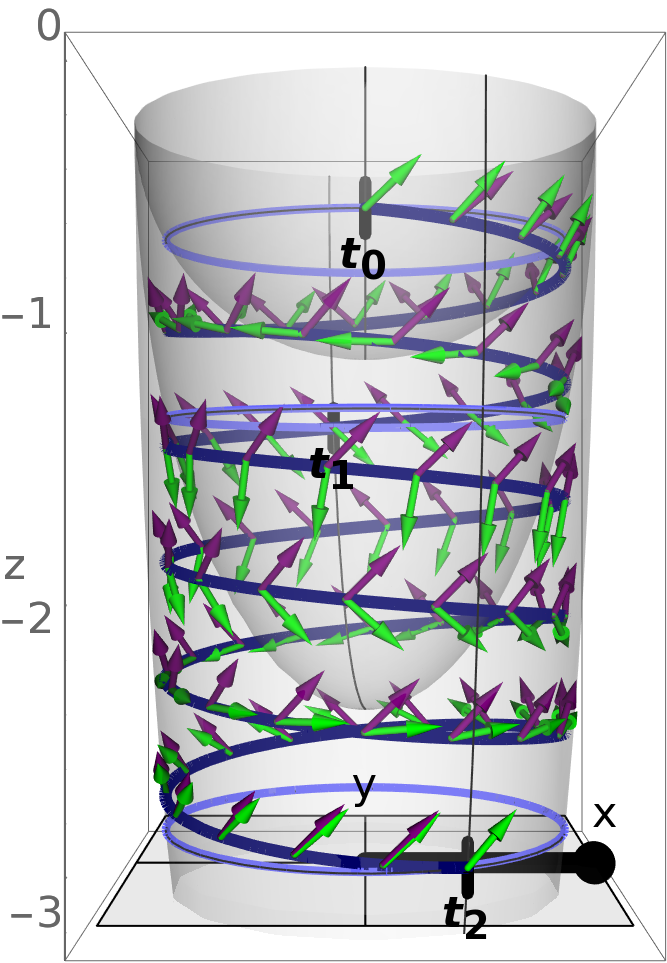}}
            \hfill
            \raisebox{0.07\height}{\includegraphics[width=0.2\linewidth]{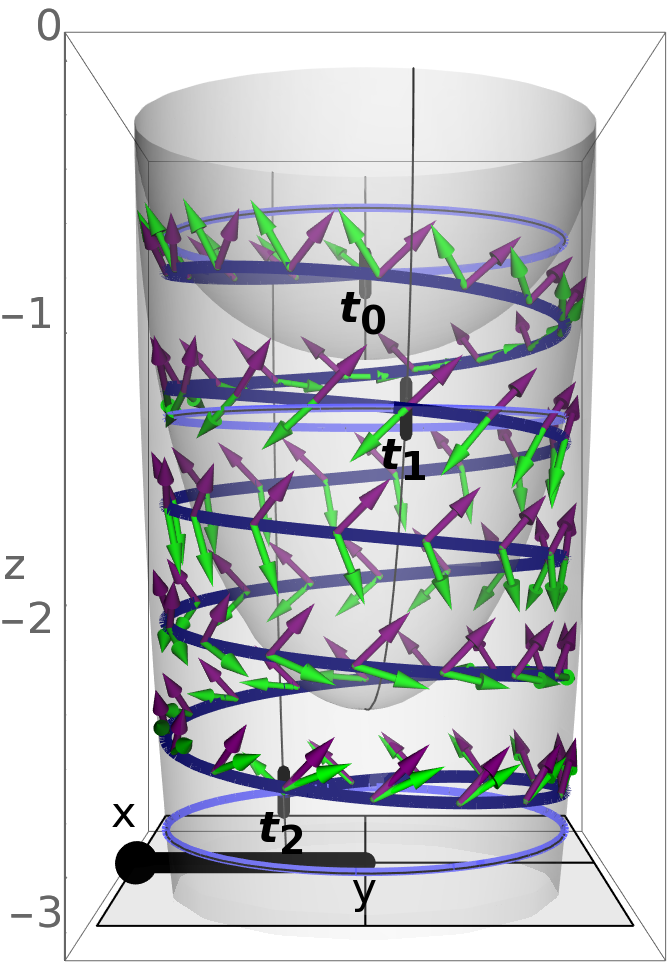}}
            \caption{(top) Helically stretching spheroid $ \surf $ at times $ t=0,0.2,0.4,0.6,0.8,1 $ 
                           and vector fields $ \arbt $ force-free transported 
                           \wrt\ upper-convected, lower-convected, Jaumann and material derivative.
                           The perspective rotate consistently with the rotation of the sphere, \st\ the observed material points stay in front.
                           The initial condition is $ \arbt_0 = \frac{1}{\sqrt{2}}[-1,\frac{1}{\sin y^1_{\m}}]'_{\tangentS|_{t=0}} $
                           \wrt\ Lagrangian observer Parametrization $ \para_{\m} $.
                      (bottom left) Corresponding cosine of angle $ \phi^2(\arbt) $ 
                               to the latitudes only for the material transported vector field on lower hemisphere for $ 0\le t \le 10 $,
                               where $ \phi^2(\arbt_0)=\frac{\pi}{4} $ holds.
                               The other three solutions are corresponding to the stretching spheroid in \autoref{fig:stretchingsphereeval}.
                               Single latitudes used in the top image are emphasized here.
                               Especially blue lines represent 30 degree geographical latitude and 
                               we mark the times $ t_0=0 $ at the beginning, $ t_1\approx 1.53 $ of tangential half- and $ t_2 \approx 5.08$ of full-circulation.
                      (bottom right) Opposite views of lower hemisphere at times $ t_0$, $ t_1 $, $ t_2 $, 
                                     the trajectory of one material particle (dark blue) at 30 degree geographical latitude (light blue)
                                     for $ t_0 \le t \le t_2 $ and the evolution of the material and Jaumann transported vector there.
                                     At times $ t_{2\alpha} $ of tangential full-circulations both vectors are equal, whereas 
                                     at times $ t_{2\alpha+1} $ of tangential half-circulations they have different signs for $ \alpha\in\mathbb{N} $.}
            \label{fig:strechrotsphereeval}
        \end{figure}

        We consider a spheroid $ \surf  $ as in \autoref{sec:transportvectorstrechingspheroid}, 
        but with a helically stretch, \st\ every material particle carries out a uniform helical motion and corresponding 
        single patch Lagrangian observer parametrization
        \begin{align}\label{eq:parastretchrotsphere}
            \para_{\m}(t,y^1_{\m},y^2_{\m})
                &= \begin{bmatrix}
                        \sin y^1_{\m} \cos (y^2_{\m} + 2\pi t)\\
                        \sin y^1_{\m} \sin (y^2_{\m} + 2\pi t)\\
                        ( 1 + t )\cos y^1_{\m}
                    \end{bmatrix}\in \R^3 \formComma
        \end{align}
        where $ y^1_{\m}\in[0,\pi] $ is the latitude and $ y^2_{\m}\in[0,2\pi) $ the longitude coordinate,
        see \autoref{fig:strechrotsphereeval}.
        Basically, this is an orthogonal superposition of the motions in \autoref{sec:transportvectorstrechingspheroid}
        and \autoref{sec:transportvectorrotatingsphere}.
        Since the rotational part is a rigid body motion, the instantaneous force-free transported vector field $ \arbt\in\tangentS $ is equal
        to the solutions on a pure stretching spheroid, as in  \autoref{sec:transportvectorstrechingspheroid} \wrt\ convected time derivatives including the Jaumann derivative.
        The behavior of the material transported vector field is not equally easy to obtain.
        The transport equation reads
        \begin{align}\label{eq:stretchrot_transport_eq}
            \partial_t \arbtC^i + \frac{(1+t)\sin^2 y_{\m}^1}{1+t(2+t)\sin^2 y_{\m}^1}\delta^i_1 \arbtC^1
                                - \frac{2\pi\cos y^1_{\m}}{\sqrt{1+t(2+t)\sin^2 y_{\m}^1}}\tensor{\lctC}{^i_j}\arbtC^j
                &= 0\formComma
            &\arbt|_{t=0} &= \arbt_0\in\tangentS|_{t=0}\formPeriod
        \end{align}
        Our ansatz is a tangential clockwise circulation of the Jaumann solution 
        $ \arbt_{\!\text{J}} := [\frac{\arbtC_0^1}{\sqrt{1+t(2+t)\sin^2 y_{\m}^1}},\arbtC_0^2 ]'_{\tangentS}$ 
        \eqref{eq:matandjausolution_stretchingsphere},
        \ie\ 
        \begin{align}\label{eq:stretchrot_ansatz}
                \arbt
                    &= \cos\left( 2\pi f_{y^1_{\m}}(t) \right) \arbt_{\!\text{J}} - \sin\left( 2\pi f_{y^1_{\m}}(t) \right)(*\arbt_{\!\text{J}})\formComma
                & f_{y^1_{\m}}(0) = 0\formPeriod
        \end{align}
        Applying eq. \eqref{eq:stretchrot_ansatz} on eq. \eqref{eq:stretchrot_transport_eq} gives an ODE for $ f_{y^1_{\m}} $. It reads together with its solution
        \begin{align}\label{eq:fy1}
            f_{y^1_{\m}}'(t) &= \frac{\cos y^1_{\m}}{\sqrt{1+t(2+t)\sin^2 y_{\m}^1}}\formComma
            &f_{y^1_{\m}}(t) &= \cot y^1_{\m} \ln\frac{(1+t)\sin y_{\m}^1 + \sqrt{1+t(2+t)\sin^2 y_{\m}^1}}{1 + \sin y_{\m}^1} \formPeriod
        \end{align}
        We observe that the time depending behavior of $ f_{y^1_{\m}} $ is qualitatively the inverse hyperbolic sine almost everywhere 
        instead of a linear function as in \autoref{sec:transportvectorrotatingsphere}.
        For instance, at 30 degree geographical latitude, \ie\  $ y^1_{\m}= \frac{\pi}{2}\pm\frac{\pi}{6} $,
        the time for the $ \alpha $th tangential half-circulation, is
        $ t_{\alpha}=\frac{1}{\sqrt{3}}\sinh(\frac{\sqrt{3}}{2}\alpha + \ln(2+\sqrt{3})) - 1 $, instead of $t_{\alpha}=\alpha$ on the pure rotating sphere.
        However, similarly to the former example the material transported vector field minimizes the directional change \wrt\ the embedding space.
        See \autoref{fig:strechrotsphereeval} for an example.

\section{2-tensor fields}\label{sec:two_tensor_fields}
    The reasoning for spacetime 2-tensor fields are similar to those for vector fields in \autoref{sec:vector_fileds}.
    We merely summarize the conversions between the spacetime 2-tensor spaces $ \sttensorbS[^2] $, $ \tangentstS[^2] $ and 
    $ \tangentstR[^2] $, that is
    \begin{align*}
        \arbt &= \bbrackets{\starbt}
                = \btensor{\arbtC_{\transdir\transdir} & \arbt_{\transdir\surf} \\
                            \arbt_{\surf\transdir}     & \arbt_{\surf\surf}}
               = \btensor{ \starbtC^{tt}
                                    & \left\{\starbtC^{tj} + \starbtC^{tt}\obveloSC^{j}\right\}\\
                           \left\{\starbtC^{it} + \starbtC^{tt}\obveloSC^{i}\right\} 
                                    &\left\{\starbtC^{ij}  + \starbtC^{jt}\obveloSC^{i}
                                            + \starbtC^{it}\obveloSC^{j} + \starbtC^{tt}\obveloSC^{i}\obveloSC^{j}\right\}} \in \sttensorbS[^2] 
    \end{align*}
    \begin{align*}
        \starbt &= \bbrackets{\arbt}^{-1}
               =  \tensorstS[^2]{\starbtC^{tt}                & \left\{\starbtC^{tj}\right\}\\
                                 \left\{\starbtC^{it}\right\} & \left\{\starbtC^{ij}\right\}}
               = \tensorstS[^2]{\arbtC_{\transdir\transdir}
                                        &\arbt_{\transdir\surf} - \arbtC_{\transdir\transdir}\obveloS \\
                                \arbt_{\surf\transdir} -\arbtC_{\transdir\transdir}\obveloS
                                        & \arbt_{\surf\surf} - \arbt_{\surf\transdir}\otimes\obveloS
                                          -\obveloS\otimes\arbt_{\transdir\surf} + \arbtC_{\transdir\transdir}\obveloS\otimes\obveloS} \\
              &= \tensorstR[^2]{\starbtC^{tt}
                                    &\starbtC^{tj}\partial_j\para + \starbtC^{tt}\obvelo \\
                                \starbtC^{it}\partial_i\para + \starbtC^{tt}\obvelo
                                    &\starbtC^{ij}\partial_i\para\otimes\partial_j\para + \starbtC^{it}\partial_i\para\otimes\obvelo
                                     +\starbtC^{tj}\obvelo\otimes\partial_j\para + \starbtC^{tt}\obvelo\otimes\obvelo}\\
               &= \tensorstR[^2]{\arbtC_{\transdir\transdir}
                                    &\arbt_{\transdir\surf} + \velonor\arbtC_{\transdir\transdir}\normal \\
                                 \arbt_{\surf\transdir} +\velonor\arbtC_{\transdir\transdir}\normal
                                    &\arbt_{\surf\surf} + \velonor\arbt_{\surf\transdir}\otimes\normal
                                     +\velonor\normal\otimes\arbt_{\transdir\surf} + \velonor^2\arbtC_{\transdir\transdir}\normal\otimes\normal}  
    \end{align*} 
    with surface tensor fields $ \arbtC_{\transdir\transdir}\in\tangentS[^0] $, $ \arbt_{\transdir\surf}, \arbt_{\surf\transdir} \in\tangentS$
    and $ \arbt_{\surf\surf}\in\tangentS[^2] $.
    The considered time derivatives are the material derivative $ \nabla_{\matdir}\starbt $, 
    upper-upper convected derivative 
    $ \Lie_{\matdir}^{\sharp\sharp}\starbt= \nabla_{\matdir}\starbt - (\nabla\matdir)\starbt - \starbt(\nabla\matdir)^T$, 
    lower-lower convected derivative  
    $ \Lie_{\matdir}^{\flat\flat}\starbt= \nabla_{\matdir}\starbt + (\nabla\matdir)^T\starbt + \starbt(\nabla\matdir)$,
    upper-lower convected derivative
    $ \Lie_{\matdir}^{\sharp\flat}\starbt= \nabla_{\matdir}\starbt - (\nabla\matdir)\starbt + \starbt(\nabla\matdir) $,
    lower-upper convected derivative
    $ \Lie_{\matdir}^{\flat\sharp}\starbt= \nabla_{\matdir}\starbt + (\nabla\matdir)^T\starbt - \starbt(\nabla\matdir)^T$
    and Jaumann derivative $ \frac{1}{2}(\Lie_{\matdir}^{\sharp\sharp}\starbt + \Lie_{\matdir}^{\flat\flat}\starbt) $ in $ \tangentstS[^2] $.
    The associated derivatives in $ \sttensorbS[^2] $ are determined by \autoref{thm:material_derivative} and \autoref{thm:convected_derivatives}.
    \begin{concl}
         For spacetime tensor fields 
         $ \arbt=\bsmalltensor{\arbtC_{\transdir\transdir} & \arbt_{\transdir\surf} \\
                                \arbt_{\surf\transdir}     & \arbt_{\surf\surf}}\in\sttensorbS[^2] $ the material $ \matd\ $, Jaumann $ \jaud\ $,
         upper-upper convected $ \cond{\sharp\sharp}\ $, lower-lower convected $ \cond{\flat\flat} $, upper-lower convected $ \cond{\sharp\flat} $ 
         and lower-upper convected derivative $\cond{\flat\sharp}: \sttensorbS[^2] \rightarrow \sttensorbS[^2] $ are
         \begin{align*}
            \matd\arbt
                &= \btensor{\dot{\arbtC}_{\transdir\transdir} + 2\zeta\velonor\dot{\velonor} \arbtC_{\transdir\transdir}
                                                              + \zeta\velonor\innerS[^2]{\bop_{\m}}{\arbt_{\surf\transdir}+ \arbt_{\transdir\surf}}
                                & \dot{\arbt}_{\transdir\surf} + \zeta\velonor\dot{\velonor}\arbt_{\transdir\surf} 
                                                                + \zeta\velonor \arbt_{\surf\surf}^T\bop_{\m}
                                                                -\velonor\arbtC_{\transdir\transdir}\bop_{\m}\\
                            \dot{\arbt}_{\surf\transdir} + \zeta\velonor\dot{\velonor}\arbt_{\surf\transdir} 
                                                          + \zeta\velonor \arbt_{\surf\surf}\bop_{\m}
                                                         -\velonor\arbtC_{\transdir\transdir}\bop_{\m}
                                & \dot{\arbt}_{\surf\surf} - \velonor \arbt_{\surf\transdir}\otimes\bop_{\m}
                                                            - \velonor \bop_{\m}\otimes\arbt_{\transdir\surf}
                            } \formComma\\
           \jaud\arbt
               &= \btensor{\dot{\arbtC}_{\transdir\transdir} + 2\zeta\velonor\dot{\velonor} \arbtC_{\transdir\transdir}
                                                             + \frac{\zeta}{2}\innerS[^2]{\instcond{\sharp}\matveloS}{\arbt_{\surf\transdir}
                                                                                                                    + \arbt_{\transdir\surf}}
                               & \instjaud\arbt_{\transdir\surf} + \zeta\velonor\dot{\velonor}\arbt_{\transdir\surf} 
                                                               + \frac{\zeta}{2} \arbt_{\surf\surf}^T\instcond{\sharp}\matveloS
                                                               -\frac{\arbtC_{\transdir\transdir}}{2}\instcond{\sharp}\matveloS\\
                           \instjaud\arbt_{\surf\transdir} + \zeta\velonor\dot{\velonor}\arbt_{\surf\transdir} 
                                                         + \frac{\zeta}{2} \arbt_{\surf\surf}\instcond{\sharp}\matveloS
                                                        -\frac{\arbtC_{\transdir\transdir}}{2}\instcond{\sharp}\matveloS
                               & \instjaud\arbt_{\surf\surf} - \frac{1}{2} \arbt_{\surf\transdir}\otimes\instcond{\sharp}\matveloS
                                                          - \frac{1}{2} \instcond{\sharp}\matveloS\otimes\arbt_{\transdir\surf}
                           } \formComma\\
           \cond{\sharp\sharp}\arbt
              &= \btensor{\dot{\arbtC}_{\transdir\transdir}
                                & \instcond{\sharp}\arbt_{\transdir\surf} - \arbtC_{\transdir\transdir}\instcond{\sharp}\matveloS\\
                          \instcond{\sharp}\arbt_{\surf\transdir}- \arbtC_{\transdir\transdir}\instcond{\sharp}\matveloS
                                & \instcond{\sharp\sharp}\arbt_{\surf\surf} - \arbt_{\surf\transdir}\otimes\instcond{\sharp}\matveloS
                                                                            - \instcond{\sharp}\matveloS\otimes\arbt_{\transdir\surf}
                         }\formComma\\
           \cond{\flat\flat}\arbt
              &=  \btensor{ \dot{\arbtC}_{\transdir\transdir} + 4\zeta\velonor\dot{\velonor} \arbtC_{\transdir\transdir}
                                                              +\zeta\innerS[^2]{\instcond{\sharp}\matveloS}{\arbt_{\surf\transdir}
                                                                                                          + \arbt_{\transdir\surf}}
                                & \instcond{\flat}\arbt_{\transdir\surf} + 2\zeta\velonor\dot{\velonor}\arbt_{\transdir\surf}
                                                                          +\zeta\arbt_{\surf\surf}^T\instcond{\sharp}\matveloS\\
                            \instcond{\flat}\arbt_{\surf\transdir} + 2\zeta\velonor\dot{\velonor}\arbt_{\surf\transdir}
                                                                   +\zeta\arbt_{\surf\surf}\instcond{\sharp}\matveloS 
                                & \instcond{\flat\flat}\arbt_{\surf\surf} 
                           }\formComma\\
           \cond{\sharp\flat}\arbt
              &= \btensor{ \dot{\arbtC}_{\transdir\transdir} + 2\zeta\velonor\dot{\velonor} \arbtC_{\transdir\transdir}
                                                            +\zeta\innerS[^2]{\instcond{\sharp}\matveloS}{\arbt_{\transdir\surf}}
                              & \instcond{\flat}\arbt_{\transdir\surf} \\
                           \instcond{\sharp}\arbt_{\surf\transdir} + 2\zeta\velonor\dot{\velonor}\arbt_{\surf\transdir}
                                                                   +\zeta\arbt_{\surf\surf}\instcond{\sharp}\matveloS
                                                                    - \arbtC_{\transdir\transdir}\instcond{\sharp}\matveloS 
                              &\instcond{\sharp\flat}\arbt_{\surf\surf}- \instcond{\sharp}\matveloS\otimes\arbt_{\transdir\surf}
                          }\formComma\\
           \cond{\flat\sharp}\arbt
              &= \btensor{ \dot{\arbtC}_{\transdir\transdir} + 2\zeta\velonor\dot{\velonor} \arbtC_{\transdir\transdir}
                                                             +\zeta\innerS[^2]{\instcond{\sharp}\matveloS}{\arbt_{\surf\transdir}}
                             &\instcond{\sharp}\arbt_{\transdir\surf} + 2\zeta\velonor\dot{\velonor}\arbt_{\transdir\surf}
                                                                      +\zeta\arbt_{\surf\surf}^T\instcond{\sharp}\matveloS
                                                                      - \arbtC_{\transdir\transdir}\instcond{\sharp}\matveloS \\
                           \instcond{\flat}\arbt_{\surf\transdir}
                                &\instcond{\flat\sharp}\arbt_{\surf\surf}- \arbt_{\surf\transdir}\otimes\instcond{\sharp}\matveloS
                          }
         \end{align*}
         with scalar time derivatives, see eq. \eqref{eq:scalar_derivatives}, 
         instantaneous vector time derivatives listed in \autoref{concl:vector_fields} and instantaneous 
         material $ \dot{\arbt}_{\surf\surf} $, Jaumann $ \instjaud\arbt_{\surf\surf} $,
         upper-upper convected $ \instcond{\sharp\sharp}\arbt_{\surf\surf} $, lower-lower convected $ \instcond{\flat\flat}\arbt_{\surf\surf}  $, 
         upper-lower convected $ \instcond{\sharp\flat}\arbt_{\surf\surf} $ 
         and lower-upper convected derivative $ \instcond{\flat\sharp}\arbt_{\surf\surf} $
         \begin{align}\label{eq:instantaneous_time_derivaties_twotensors}
         \begin{aligned}
            \indexten{\dot{\arbt}_{\surf\surf}}{^{ij}}
                  &= \partial_t\arbtC_{\surf\surf}^{ij} + \indexten{\nabla_{\relvelo}\arbt_{\surf\surf} 
                                                                     +\Bop\arbt_{\surf\surf} + \arbt_{\surf\surf}\Bop^{T} }{^{ij}}\formComma\\
           \instjaud\arbt_{\surf\surf}
               &= \dot{\arbt}_{\surf\surf} - \frac{\rot\matveloS}{2}(*_{1}\arbt_{\surf\surf} + *_{2}\arbt_{\surf\surf})
                = \dot{\arbt}_{\surf\surf} - \frac{1}{2}\left( \nabla\matveloS - \left( \nabla\matveloS \right)^T \right)
                                              \left( \arbt_{\surf\surf} + \arbt_{\surf\surf}^T 
                                                        - \left(\trace\arbt_{\surf\surf}\right)\IdS \right)\formComma \\
           \indexten{\instcond{\sharp\sharp}\arbt_{\surf\surf}}{^{ij}}
                &= \partial_t\arbtC_{\surf\surf}^{ij} 
                   + \indexten{\nabla_{\relvelo}\arbt_{\surf\surf} - \left( \nabla\relvelo \right)\arbt_{\surf\surf}
                                                                   - \arbt_{\surf\surf}\left( \nabla\relvelo \right)^T}{^{ij}}
                 = \indexten{\dot{\arbt}_{\surf\surf} - \Bop_{\m}\arbt_{\surf\surf} - \arbt_{\surf\surf}\Bop_{\m}^T}{^{ij}} \formComma\\
           \indexten{\instcond{\flat\flat}\arbt_{\surf\surf}}{_{ij}}
                &= \partial_t\indexten{\arbt_{\surf\surf}}{_{ij}}
                   + \indexten{\nabla_{\relvelo}\arbt_{\surf\surf} + \left( \nabla\relvelo \right)^T\arbt_{\surf\surf}
                                                                   + \arbt_{\surf\surf}\left( \nabla\relvelo \right)}{_{ij}}
                 = \indexten{\dot{\arbt}_{\surf\surf} + \Bop_{\m}^T\arbt_{\surf\surf} + \arbt_{\surf\surf}\Bop_{\m}}{_{ij}}\formComma\\
           \indexten{\instcond{\sharp\flat}\arbt_{\surf\surf}}{^i_j}
                &= \partial_t\indexten{\arbt_{\surf\surf}}{^i_j}
                   +\indexten{\nabla_{\relvelo}\arbt_{\surf\surf} - \left( \nabla\relvelo \right)\arbt_{\surf\surf}
                                                                  +  \arbt_{\surf\surf}\left( \nabla\relvelo \right)}{^i_j} 
                 =\indexten{\dot{\arbt}_{\surf\surf}-\Bop_{\m}\arbt_{\surf\surf}+ \arbt_{\surf\surf}\Bop_{\m}}{^i_j} \formComma\\   
           \indexten{\instcond{\flat\sharp}\arbt_{\surf\surf}}{_i^j}
                &=  \partial_t\indexten{\arbt_{\surf\surf}}{_i^j}
                    + \indexten{\nabla_{\relvelo}\arbt_{\surf\surf} + \left( \nabla\relvelo \right)^T\arbt_{\surf\surf}
                                                                    - \arbt_{\surf\surf}\left( \nabla\relvelo \right)^T}{_i^j}
                 = \indexten{\dot{\arbt}_{\surf\surf} + \Bop_{\m}^T\arbt_{\surf\surf} - \arbt_{\surf\surf}\Bop_{\m}^T}{_i^j} 
        \end{aligned}
         \end{align}
         in $\tangentS[^2]$, where $ \bop_{\m} = \nabla\velonor + \shop\matveloS $, $ \Bop_{\m}=\nabla\matveloS - \velonor\shop $
         $ \Bop=\nabla\obveloS - \velonor\shop $, $ [*_{1}\arbt_{\surf\surf}]^{ij} =  -\tensor{\lctC}{^i_k} \arbtC_{\surf\surf}^{kj} $
         and $ [*_{2}\arbt_{\surf\surf}]^{ij} =  -\tensor{\lctC}{^j_k} \arbtC_{\surf\surf}^{ik} $.
    \end{concl}
    The instantaneous time derivatives \eqref{eq:instantaneous_time_derivaties_twotensors}  
    give an observer-independent instantaneous rate for $ \arbt_{\surf\surf}\in\tangentS[^2] $ \wrt\ instantaneous 2-tensor fields
    $ \arbt=\bsmalltensor{0 & 0 \\ 0 & \arbt_{\surf\surf}} $.
    If we consider $ \arbt_{\surf\surf} $ as \eg\ Cauchy (instantaneous) stress tensor, \ie\ 
    $ \arbt_{\surf\surf}\in\symS[^2]:=\tangentS[^2]\big/ T $, the space of symmetric tangential 2-tensors fields, 
    then $ \instjaud\arbt_{\surf\surf}\in\symS[^2] $ is the Jaumann rate, 
    $ \instcond{\sharp\sharp}\arbt_{\surf\surf}\in\symS[^2] $ the Oldroyd rate
    and $ \instcond{\flat\flat}\arbt_{\surf\surf}\in\symS[^2] $ the Cotter-Rivlin rate, see \eg\ \cite{Szabo1989}.
    
  \subsection{Force-free transport of instantaneous Q-tensor fields on a helically stretching spheroid}
\begin{figure}
        \centering
        \includegraphics[width=0.99\linewidth]{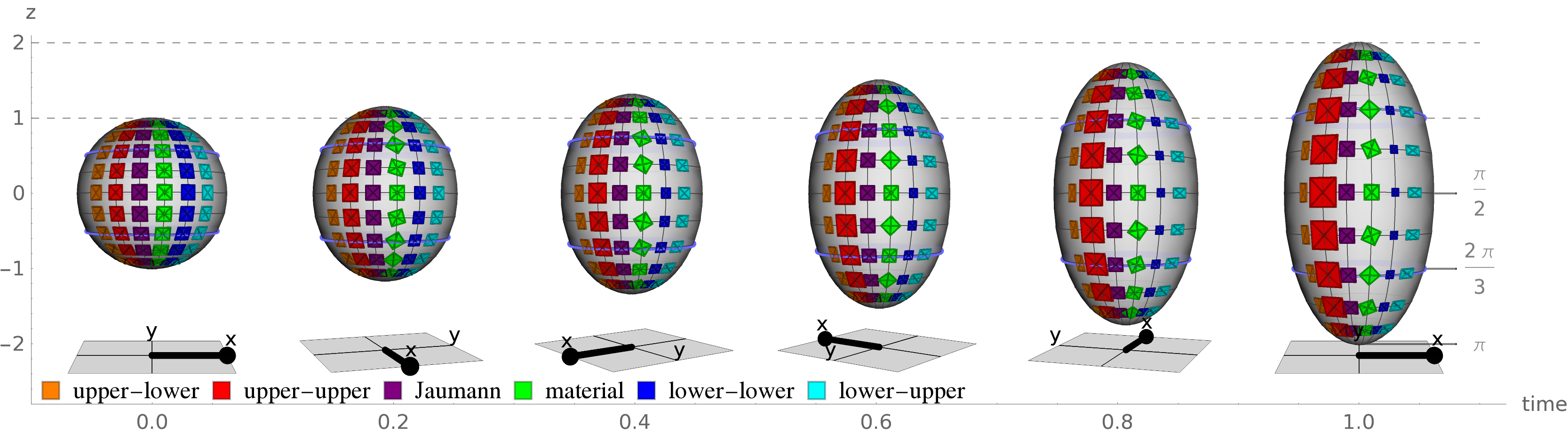}
        \caption{Helically stretching spheroid $ \surf $ at times $ t=0,0.2,0.4,0.6,0.8,1 $ 
                       and tensor fields $ \vararbt $ force-free transported 
                       \wrt\ different derivatives.
                       The perspective rotate consistently with the rotation of the sphere, \st\ the observed material points stay in front.
                       The initial condition is the Q-tensor field 
                       $ \vararbt_0=\qmap|_{t=0}(\arbt_0)$ 
                       with eigenvector  
                       $ \arbt_0 = \frac{1}{\sqrt{2}}[-1,\frac{1}{\sin y^1_{\m}}]'_{\tangentS|_{t=0}}$
                       \wrt\ Lagrangian observer Parametrization $ \para_{\m} $,
                       \ie\ $ \vararbtC^{11}_0= \vararbtC^{22}_0= 0$ and $ \vararbtC^{12}_0= \vararbtC^{21}_0= -\frac{1}{\sin y^1_{\m}}$.
                       Tensors are depicted as rectangular tensor glyphs,
                       where the diagonals are along the eigenvectors and scaled by absolute value of corresponding eigenvalues.
                       As a consequence Q-tensors appear as squares.
                       Since the special choice of initial condition, the transported tensor fields \wrt\ $ \instcond{\sharp\sharp} $ and
                       $ \instcond{\flat\flat}$ fulfill condition \eqref{eq:tracecondforqtensor} and hence stay Q-tensor fields metastably, 
                       \cf\ \cite{NitschkeMendeley2020},
                       beside as it is for the Jaumann and material transported Q-tensor fields generally.
                       The angles between eigenvectors and the latitudes corresponding to the considered vector fields in former examples,
                       see \autoref{fig:stretchingsphereeval} 
                       ($ \instjaud\arbt=0 $ for $ \instjaud\vararbt=0 $, $ \instcond{\sharp\sharp}\vararbt=0 $, $ \instcond{\flat\flat}\vararbt=0$;
                        $\instcond{^\sharp}\arbt=0$ for $ \instcond{\sharp\flat}\vararbt=0 $;
                        $\instcond{^\flat}\arbt=0$ for $ \instcond{\flat\sharp}\vararbt=0 $)
                       and \autoref{fig:strechrotsphereeval} ($ \dot{\arbt}=0$ for  $\dot{\vararbt}=0 $).
                       Though the eigenvalues \wrt\ the four convected derivatives are incompatible with $ \|\arbt\| $ for corresponding transported vector fields. 
                  }
        \label{fig:strechrotsphereevaltensor}
    \end{figure}

    We consider the same moving spheroid as in \autoref{sec:transportvectorstretchrotsphere} with Lagrangian observer parametrization as in eq.
    \eqref{eq:parastretchrotsphere}. 
    However, instead of instantaneous vector fields we are interested in transport of so-called instantaneous \emph{Q-tensors} 
    $ \QS \subset \tangentS[^2]$.
    They are trace-free and symmetric 2-tensors on the surface, 
    \ie\ $ \QS := \{ \vararbt\in\symS \mid \trace\vararbt = 0 \} $.
    Besides its importance in applications, such as liquid crystals, Q-tensor fields are also comparable to vector fields $ \tangentS $.
    Both consider in a pointwise sence two-dimensional vector spaces, though they are not isomorphic.
    Observing the surjection $ \qmap: \tangentS \rightarrow \QS $ with
    $ \qmap(\arbt)=\frac{2}{\|\arbt\|}(\arbt\otimes\arbt - \frac{\|\arbt\|^2}{2}\IdS) $, we deduce
    that $ \vararbt=\qmap(\arbt) $ holds if $ \arbt $ is the eigenvector to eigenvalue $ \|\arbt\| $ of $ \vararbt $
    and hence $ \qmap(-\arbt)=\qmap(\arbt) $ is valid.
    Therefore if vector fields in $ \tangentS $ are representing \emph{polar vector fields} with tangential $ 2\pi $-periodicity, 
    than we would call Q-tensor fields a representation of \emph{apolar vector fields} with tangential $ \pi $-periodicity.
    Based on this semantic interconnection we are able to investigate consistencies between force-free transport of Q-tensor fields and 
    vector fields in \autoref{sec:transportvectorstretchrotsphere}.
    Unfortunately, all kernels of convected instantaneous time derivatives 
    $ \instcond{\sharp\sharp} $, $ \instcond{\flat\flat} $, $ \instcond{\sharp\flat} $ and $ \instcond{\flat\sharp} $ are not
    laying in $ \QS $ only.
    Considering the initial solution $ \vararbt_{0}\in\QS|_{t=0} $,
    the first two only guarantee symmetric solutions and the last two trace-free solutions.
    For instance,
    the solution of $ \instcond{\sharp\sharp}\vararbt = 0 $ has to 
    fulfill $ \frac{d}{dt}\!\trace\vararbt = 0 $ in order that $ \vararbt\in\QS $ is valid.
    Hence by metric compatibility, this condition reads
    \begin{align}\label{eq:tracecondforqtensor}
        0 &=\frac{d}{dt}\!\trace\vararbt
           = \trace\dot{\vararbt}=\trace\left(\instcond{\sharp\sharp}\vararbt + \Bop_{\m}\vararbt + \vararbt\Bop_{\m}^T\right)
           = \innerS[^2]{\Bop_{\m}+\Bop_{\m}^T}{\vararbt}\formComma
    \end{align}
    which can only been implemented generally if the rate of deformation tensor is a multiple of the identity $ \IdS $, 
    \eg\ for an uniformly expanding surface.
    However, for the considered spheroid in Lagrangian coordinates, we deduce from eq. \eqref{eq:tracecondforqtensor} the condition
    $ 0 = \vararbtC^{11} =  [\vararbt_0]^{11} $ and  $ \left([\arbt_0]^1\right)^2 = \left([\arbt_0]^2\right)^2\sin^2 y_{\m}^1 $ 
    if $ \vararbt_0 = \qmap|_{t=0}(\arbt_0) $,
    see \eg\ \autoref{fig:strechrotsphereevaltensor}.
    For $ \instcond{\flat\flat}\vararbt = 0 $ we get the same condition.
    Similarly to the trace, anti-symmetry is a scalar valued measurement on surface 2-tensors and can be defined by 
    $ \innerS[^2]{\cdot}{\lct}=\trace\circ*_{1}:\tangentS[^2]\rightarrow\R $.
    Since the instantaneous material time derivative is also compatible with the Levi-Civita tensor, \ie\ $ \dot{\lct}=0 $, 
    we obtain the condition $ 0 =  \frac{d}{dt}\!\innerS[^2]{\vararbt}{\lct} =  \innerS[^2]{\dot{\vararbt}}{\lct}$ 
    to ensure a symmetric solution for transport equations and symmetric initial tensor fields.
    The solutions of the upper-upper and lower-lower convected transport equations operate this obviously.
    However, this is not true for $ \instcond{\sharp\flat}\vararbt = 0 $ generally, since
    \begin{align}\label{eq:symcondforqtensor}
        0 &=  \frac{d}{dt}\!\innerS[^2]{\vararbt}{\lct} =  \innerS[^2]{\dot{\vararbt}}{\lct}
           = \innerS[^2]{\instcond{\sharp\flat}\vararbt + \Bop_{\m}\vararbt - \vararbt\Bop_{\m}}{\lct}
           = \innerS[^2]{*_1\Bop_{\m} + *_2\Bop_{\m}}{\vararbt^T}
           = -2\innerS[^2]{\proj{\QS}\Bop_{\m}}{*_1\vararbt}\formComma
    \end{align}
    where $ \proj{\QS}:\tangentS[^2]\rightarrow\QS $ is the uniquely defined orthogonal Q-tensor projection.
    Note that we used in the last identity that $ *_1\circ *_1 = -\IdS $ and $ (*_1\circ *_2)(\cdot) =  \trace(\cdot)\IdS - (\cdot)^T$ holds.
    Especially at the considered moving spheroid we deduce from eq. \eqref{eq:symcondforqtensor} that
    $ 0 = \tensor{\vararbtC}{^1_2}=\tensor{[\vararbt_0]}{^1_2} $ and $ 0=[\arbt_0]^1[\arbt_0]^2 $ respectively, 
    if $ \vararbt|_{t=0} = \vararbt_{0} = \qmap(\arbt_0)\in\QS|_{t=0} $ and $ \arbt_0\in\tangentS|_{t=0} $. 
    For $ \instcond{\flat\sharp}\vararbt = 0 $ we get the same condition.
    Eventually, by lack of generality, these four convected derivatives are not recommended to apply untreated 
    in a theory using apolar vector fields on a moving surface.
    For $ \dot{\vararbt}=0 $ the conditions in eqs. \eqref{eq:tracecondforqtensor} and  \eqref{eq:symcondforqtensor} are fulfilled
    obviously.
    For the Jaumann transport $ \instjaud\vararbt=0 $, we have to show that 
    $ 0=\innerS[^2]{*_1\vararbt + *_2\vararbt}{\IdS}$ and $0 = \innerS[^2]{*_1\vararbt + *_2\vararbt}{\lct} $ holds, ultimately if 
    $ \rot\matveloS $ is not vanishing everywhere at all times.
    But this is generally true, since $ *_1\vararbt + *_2\vararbt\in\QS $ is valid for all $ \vararbt\in\tangentS[^2] $, and 
    $ \IdS $, as well as $ \lct $, lays orthogonal to $ \QS $ in $ \tangentS[^2] $.
    Moreover, the solution of both apolar transport equations are consistent to their polar counterpart, \ie 
    \begin{align}
        \forall\arbt|_{t=0} = \arbt_{0}\in\tangentS|_{t=0}, \vararbt|_{t=0} = \vararbt_{0}=\qmap|_{t=0}(\arbt_{0})\in\QS|_{t=0}:
        \quad \dot{\vararbt}=0 \Leftrightarrow \dot{\arbt}=0
        \text{ and } \instjaud\vararbt = 0 \Leftrightarrow \instjaud\arbt = 0
    \end{align}
    with solutions $ \arbt\in\tangentS $ and $ \vararbt=\qmap(\arbt)\in\QS $.
    The reverse directions can easily be seen by calculating $ \overdot{\qmap(\arbt)} $ or $ \instjaud\qmap(\arbt)$, respectively, since
    all summands are containing $ \dot{\arbt} $ or $ \instjaud\arbt $, respectively.
    For the forward direction, we use the identities $ \vararbt\arbt=\|\arbt\|\arbt $ and $ \vararbt(*\arbt)=-\|\arbt\|(*\arbt) $.
    This yields $ \|\arbt\|\dot{\arbt} = \vararbt\dot{\arbt} - \frac{\innerstS{\arbt}{\dot{\arbt}}}{\|\arbt\|^2}\arbt $
    or $ \|\arbt\|\instjaud\arbt = \vararbt\instjaud\arbt - \frac{\innerstS{\arbt}{\instjaud\arbt}}{\|\arbt\|^2}\arbt $, respectively.
    By testing these equations with the orthogonal system $\{\arbt,*\arbt\} $ 
    we obtain that $ \dot{\arbt}=0 $ or $ \instjaud\arbt=0 $, respectively.
    As a consequence in the present example of moving spheroid, there is an one-to-one correspondence \wrt\ the results of \autoref{sec:transportvectorstretchrotsphere} up to sign, 
    see \eg\ \autoref{fig:strechrotsphereeval} in connection with \autoref{fig:strechrotsphereevaltensor}.
    
    We still give a full summary of the tensor-valued results on the spheroid.
    For initial condition
    \begin{align*}
        \vararbt|_{t=0} 
            &= \begin{bmatrix} \alpha_0\sin^2 y_{\m}^1 & \beta_0 \\
                                    \beta_0 & -\alpha_0\end{bmatrix}_{\tangentS[^2]|_{t=0}} \in\QS|_{t=0}
       &&\text{with}
       &\alpha_0 &= \frac{\left([\arbt_0]^1\right)^2 - \left([\arbt_0]^2\right)^2\sin^2 y_{\m}^1}{\|\arbt_0\|\sin^2 y_{\m}^1}\formComma
       &\beta_0 &= \frac{2[\arbt_0]^1[\arbt_0]^2}{\|\arbt_0\|}\formComma
    \end{align*}
    where $ \arbt_{0}\in\tangentS|_{t=0} $, and the component $ \gC^{11}=\frac{1}{1+t(2+t)\sin^2y_{\m}^1} $ of the inverse metric tensor, we conclude that
    \begin{align*}
        \instcond{\sharp\sharp}\vararbt &= 0
            &&\Rightarrow
            &\vararbt&=\tensorS[^2]{ \alpha_0\sin^2 y_{\m}^1 & \beta_0 \\
                                            \beta_0 & -\alpha_0}\!\!\!\!\!\!\!\!\in\symS[^2]\formComma
       &\instcond{\flat\flat}\vararbt &= 0
            &&\Rightarrow
            &\vararbt&=\tensorS[^2]{ (\gC^{11})^2\alpha_0\sin^2 y_{\m}^1 & \gC^{11}\beta_0 \\
                                            \gC^{11}\beta_0 & -\alpha_0}\!\!\!\!\!\!\!\!\in\symS[^2]\formComma\\
      \instcond{\sharp\flat}\vararbt &= 0
          &&\Rightarrow
          &\vararbt&=\tensorS[^2]{ \gC^{11}\alpha_0\sin^2 y_{\m}^1 & \beta_0 \\
                                          \gC^{11}\beta_0 & -\alpha_0}\!\!\!\!\!\!\!\!\in\TFS\formComma
      &\instcond{\flat\sharp}\vararbt &= 0
          &&\Rightarrow
          &\vararbt&=\tensorS[^2]{ \gC^{11}\alpha_0\sin^2 y_{\m}^1 & \gC^{11}\beta_0 \\
                                          \beta_0 & -\alpha_0}\!\!\!\!\!\!\!\!\in\TFS\formComma \\
      \instjaud\vararbt &= 0
          &&\Rightarrow
          &\vararbt&= \tensorS[^2]{ \gC^{11}\alpha_0\sin^2 y_{\m}^1 & \sqrt{\gC^{11}}\beta_0 \\
                                    \sqrt{\gC^{11}}\beta_0 & -\alpha_0}\!\!\!\!\!\!\!\!\in\QS\formComma
      &\dot{\vararbt} &= 0
            &&\Rightarrow
            &\vararbt&= \boldsymbol{\Omega} \vararbt_{\text{J}}\boldsymbol{\Omega}^T \in \QS \formComma\\
                 &&&&&=:\vararbt_{\text{J}}\formComma
          &&&&& \boldsymbol{\Omega} &= \cos(2\pi f_{y_{\m}^1})\IdS + \sin(2\pi f_{y_{\m}^1})\lct\in\SOS\formComma
    \end{align*}
    where the time depending function $ f_{y_{\m}^1} $ is given in eq. \eqref{eq:fy1},
    $ \TFS := \{ \vararbt\in\tangentS[^2] \mid \trace\vararbt = 0 \} $ is the space of trace-free tensor fields and  
    $ \SOS :=  \{ \vararbt\in\tangentS[^2] \mid \vararbt\vararbt^T=\IdS\text{ and } \det\vararbt =  1  \}$
    is the space of tangential rotation tensor fields.
    An example of these solutions can be seen in \autoref{fig:strechrotsphereevaltensor}.

\section{Discussion}
    We developed various time derivatives for $ n $-tensor fields on a moving surface.
    Fundamentally, they are based on the assumption of a curved classical Newtonian spacetime providing arbitrary observers.
    Using a (2+1)-dimensional Ricci calculus, the time derivatives are observer-invariant by construction.
    We translated these descriptions of time derivatives into a 2-dimensional instantaneous calculus without loss of information,
    \st\ we are able to use common time-depending surface calculus independently of the choice of an observer.
    This brings us in a comfortable situation to develop time-discrete schemes, which can be handled with established numerical tools.
    Note that identifying the observer velocity as \emph{mesh velocity} \wrt\ an instantaneous discretization, we inevitably end up in an
    \emph{Arbitrary Lagrangian-Eulerian (ALE)} method on surfaces, 
    see \eg\ \cite{ElliottStyles_MJoM_2012,Torres-SanchezMillanArroyo_JoFM_2019,Sahu2020} for finite element discretizations of some applications on fluid and biological interfaces.
    Moreover \autoref{prop:instantaneous_material_derivative} gives us the opportunity for an embedded $ \R^3 $ Euclidean calculus 
    to express time derivatives and to discretize them, see \cite{Reuther2020,Nitschke2019a}.
    
    We want to point out that though we calculate the spacetime observer metric tensor $ \stg=\stg(\g,\velonor^2,\obveloS) $ \eqref{eq:spacetime_metric}
    from time-depending surface quantities, 
    also the inverse is true, \ie\ equivalently, if we have a given spacetime observer metric we can calculate $ \g $, $ \velonor^2 $
    and $ \obveloS $ from it as well.
    This makes $ \velonor^2 $ an intrinsic scalar quantity from the perspective of spacetime, where the square reflects the invariance of the orientation of the normals, which are extrinsic. 
    More surprisingly, the tensor quantity $ \velonor\shopC_{ij}=\frac{1}{2}(\obveloSC_{i|j}+\obveloSC_{j|i}- \partial_t\gC_{ij}) $ is also an intrinsic
    quantity, \ie\ it does not depend on a spacetime embedding.
    Hence an inhabitant of a moving surface is able to sense the shape of its world up to scalar scaling as long as $ \velonor^2 \neq 0 $.  
    Note that, throughout the spacetime calculus, the shape-operator has always the prefactor of normal velocity. 
    In conclusion, though we use quantities known for their extrinsic origin in spatial differential geometry, the presented spacetime calculus is entirely intrinsic. 
    
    In \autoref{sec:vector_fileds} the focus was on vector fields.
    We calculated the material time derivative applying on the material direction, as an example of a non-instantaneous vector field, 
    and referred this to the tangential and normal material acceleration, which can be used as inertia term in observer-invariant Navier-Stokes equations on free surfaces
    concerning the change of kinetic energy, see \cite{Yavari2016}. 
    We discussed the transport of instantaneous vector fields, \eg\ polar fields, on moving spheroids in absence of any forces.
    The results are fairly intuitive and give anticipations of the choice of time derivatives in a modeling process for instantaneous vector quantities.
    If we reduce the problem formulation to a mechanical system and assume that pointwisely every vector quantity can be seen as an arrow on 
    a frictionless bearing located at the foot, than we can advocate to use the material derivative as long as the arrows belong to a mass, 
    \ie\ there exists a kind of inertia along the arrow, \st\ directional changes tend to be minimized. 
    In contrast, if the ''arrow quantity'' does not correlate with a mass, \eg\ possible statistical directional quantities,
    then the Jaumann derivative is recommended. 
    The upper convected derivative could be useful for vector field quantities, which are adjacent to the material or describing the material itself,
    \st\ the vector field tend to be ``frozen'' in the material flow and hence obey every material motion including stretching and compression in absence of 
    any opposite forces.
    Basically, the lower convected derivative yields a similar behavior, but for covector fields, \eg\ if the considered quantity is used for linear mappings
    $ (\tangentS\rightarrow\R)=(\tangentS)^*\cong\tangentS[_1] $ pointwisely.
    Another decision guidance is the algebraic closeness of the inverse derivative concerning the solution of a PDE containing an observer-invariant time derivative 
    \wrt\ subspaces of vector fields.
    For instance, the kernels of upper and lower convected derivatives on directional fields, which are normalized vector fields, obviously do not lay in the space of directional fields generally. Therefore convected derivatives should not be used untreated in this situation.
    Similar conclusions arise for 2-tensor fields, which we approached in \autoref{sec:two_tensor_fields}.
    Here we investigated Q-tensor fields as a linear subbundle in general and on a moving spheroid.
    This restriction also exposed a violation of algebraic closeness \wrt\ to the kernel of convected time derivatives 
    and hence should only used carefully.
    Also here applies that the choice of time-derivative depends on specific modeling aspects.
    Q-tensor fields are chosen because they allow a direct comparison with vector fields discussed above. 
    Here we anticipate a similar behavior of polar (common vector fields) and apolar (Q-tensor fields) vector fields, which is only be fulfilled by the material and Jaumann time derivative. 
    The situation is much more complex for general instantaneous 2-tensor fields,
    were we can also guarantee that the  eigensystem and transport equation commute for material and Jaumann derivative only, 
    \ie\ for $\vararbt\in\tangentS[^2]$, $ \vararbt\arbt_\alpha = \lambda_\alpha\arbt_\alpha \in\tangentS  $ and $ \alpha = 1,2 $ holds
    $ \dot{\vararbt} = 0 $, or $\instjaud\vararbt= 0$, respectively, if $ \dot{\lambda}_\alpha = 0 $ and
    $ \dot{\arbt}_{\alpha} = 0 $, or $\instjaud\arbt_\alpha = 0$, respectively.
    However, \wrt\ convected derivatives, the solutions of instantaneous force-free transport equation cannot be predicted by the tensor field eigensystem so easily.
    For this purpose we provide an application in \cite{NitschkeMendeley2020}, where the reader can experiment with and is encourage to test several initial conditions and eigensystem behaviors on a helical stretching spheroid. 
    The initial setting is the one in \autoref{fig:strechrotsphereevaltensor}.
    The reader need the free Wolfram CFD Player \cite{Wolfram2019}, version 12.0.0 is recommended.
    
    For establishing force-free transport equations on instantaneous tensors fields we only exerted the pure instantaneous part of the spacetime equation.
    We did that for practical reasons mainly, since the restriction of the solution to instantaneous tensor field would overdetermine the system generally,
    \ie\ there are more equations than degrees of freedom.
    Maybe we can interpret the additional equations as kind of pseudo-forces depending of the instantaneous quantity of interest and given by restriction.
    For instance, on vector fields $ \arbt\in\tangentS $ yields
    $ \dot{\arbt} =0 \Leftrightarrow \matd \bbrackets{0, \arbt}' = \bbrackets{\zeta\velonor\innerS{\bop_{\m}}{\arbt}, 0}'\in\sttensorbS $
    for equal initial solutions,
    where $ \innerS{\bop_{\m}}{\arbt}= \nabla_{\arbt}\velonor + \shop(\matveloS,\arbt) = \innerR{\nabla_{\arbt}\matvelo}{\normal} $
    could be read as a mechanism forcing the spacetime vector field to be maintained instantaneous.
    Note that only the pure upper-convected derivative $ \cond{\sharp^n} $ yields vanishing non-instantaneous components on instantaneous tensor fields.
    Beside velocity vector fields, we discussed only instantaneous tensor fields in the example sections, 
    thought we think that also less restrictive spacetime tensor fields could be applicable,
    \eg\ Newtonian spacetime surface equivalents of the four-momentum or electromagnetic field tensor.
    
    We developed the convected time derivatives by the incompatibility of musical isomorphisms $ \flat $ and $ \sharp $ for the spacetime Lie-derivative.
    There is at least another set of incompatible isomorphisms, which are compatible for the material derivative though.
    These are the Hodge isomorphisms, a generalization of the well-known Hodge isomorphism on differential forms, 
    see \cite[Ch.~6]{Abraham2012}, \ie\ antisymmetric tensor fields in alternative terms.
    For instance, the Hodge isomorphism $ \circledast:\tangentstS[^2]\rightarrow \operatorname{ASym}^3_{1,2}\!\stsurf $ on spacetime 2-tensor fields,
    where $ \operatorname{ASym}^3_{1,2}\!\stsurf=\{\starbt\in\tangentstS[^3]\mid \starbtC^{IJK}=-\starbtC^{JIK} \}$ holds,
    yields $ [\circledast\varstarbt]^{IJK} = -\tensor{\lctC}{^{IJ}_L}\varstarbtC^{LK} $ and 
    $ [\circledast^{-1}\starbt]^{IJ} = -\frac{1}{2}\tensor{\lctC}{^I_{KL}}\starbt^{KLJ} $ with spacetime Levi-Civita tensor field $ \lct\in\tangentstS[^3] $.
    Hence $ \cond{\circ}:= \bbrackets{\circledast^{-1}\condst{\flat\flat\sharp}(\circledast\bbrackets{\cdot}^{-1})}:\sttensorbS[^2]\rightarrow\sttensorbS[^2] $ gives the \emph{Truesdell derivative} on moving surfaces with
    $ \cond{\circ}\vararbt= \cond{\sharp\sharp}\vararbt + (\operatorname{div}\matveloS - \velonor\trace\shop + \zeta\velonor\dot{\velonor})\vararbt $.
    
    The general proceeding in this paper is not restricted to embedding a 3-dimensional Riemannian spacetime manifold into a 4-dimensional Euclidean space.
    It could also be worthwhile to embed a $ m $-dimensional pseudo-Riemannian spacetime manifold into a $ M $-dimensional pseudo-Riemannian manifold with $ M > m $, \eg\ the vacuum solution of Einsteins equation, a 4-dimensional Lorentzian manifold, 
    embedded into canonical space, a 5-dimensional pseudo-Riemannian manifold with vanishing Ricci curvature 
    and the index of the metric tensor is $ 1\pm1 $ depending on the cosmological constant, see \eg\ \cite{Wesson2011}.
    Thought we are faced with possible new issues, which need attention,
    \eg\ dealing with singularities, higher dimensional co-normal space, \ie\ $ \velonor $ is no longer a scalar field,
    no absolute time, \ie\ observer-invariant instantaneous spaces cannot be considered as Newtonian slices in spacetime.

\appendix

\section{Shuffles}\label{sec:shuffles}
	Shuffles turn out to be a valuable tool for clear distinctions between transversal and instantaneous properties.
    Shuffles are permutations in $ \operatorname{S}_n $ defined by 
	\begin{align*}
		\shuffles{n}{\alpha}
			&:= \left\{\sigma\in \operatorname{S}_n \mid 
								\sigma(1)<\ldots<\sigma(\alpha) \text{ and }
								\sigma(\alpha+1)<\ldots\sigma(n) \right\}
	\end{align*}
	for all $ 0 \le \alpha \le n $, see \cite[Ch.~6.1]{Abraham2012}. 
    Syntactically, we write either 
    \begin{align*}
        &&\sigma &= \big(\underset{\text{transversal}}{\underbrace{\sigma(1)\ldots\sigma(\alpha)}}
                        \mid \underset{\text{instantaneous}}{\underbrace{\sigma(\alpha+1)\ldots\sigma(n)}}\big) \\
        \text{or} &&\sigma 
               &= \Lambda^n_\alpha(\sigma(1))\ldots\Lambda^n_\alpha(\sigma(n))
        \quad\text{ with word}
            &\Lambda^n_\alpha &:= \underset{\alpha\text{-times}}{\underbrace{\transdir\ldots\transdir}}
                                    \underset{(n-\alpha)\text{-times}}{\underbrace{\surf\ldots\surf}}\formPeriod
    \end{align*}
    In the first notation, the front entries concern transversal parts and the rear entries instantaneous parts,
    whereas $ \transdir $ stands for a transversal part and $ \surf $ for an instantaneous part in the latter notation.
    For $ \sigma\in\shuffles{5}{2} $ we write $ \sigma=(3\,5\mid 1\,2\,4) = \surf\surf\transdir\surf\transdir $ for instance.
    We describe the one pure instantaneous shuffle as $ \surf^n := (\mid 1\,\ldots\,n)\in\shuffles{n}{0} $. 
    The identity shuffle $ \shId{n}{\alpha}:=(1\ldots\alpha\mid\alpha+1\ldots n) $ is justified by 
    $ \shId{n}{\alpha}=\sigma^{-1}\circ\sigma $ formally, despite the fact that the permutation $ \sigma^{-1} $ is not a shuffle for all
    shuffles $ \sigma\in\shuffles{n}{\alpha} $ generally.
    Combinatorial reasoning gives that there exists $ |\shuffles{n}{\alpha}|=\binom{n}{\alpha} $ shuffles for fixed $ \alpha $.
    This yields the total amount of $ \sum_{\alpha=0}^{n}|\shuffles{n}{\alpha}| = 2^n$.
    We can derive a shuffle from another by converting the $ \beta $th instantaneous part into transversal part 
    for $ \alpha>0 $ and vice versa for $ \alpha < n $, namely
    \begin{align}\label{eq:shuffle_relations}
        \begin{aligned}
            \sigma^{\beta} &:= \left( \sigma(1)\ldots\widehat{\sigma(\beta)}\ldots\sigma(\alpha) 
                               \mid \sigma(\alpha+1) \ldots \sigma(\alpha+\rbeta-1)\sigma(\beta)\sigma(\alpha+\rbeta) \ldots \sigma(n) \right)\\
            \sigma_{\beta} &:= \left( \sigma(1)\ldots\sigma(\rbeta-1)\sigma(\alpha+\beta)\sigma(\rbeta)\ldots\sigma(\alpha) 
                                   \mid \sigma(\alpha+1) \ldots \widehat{\sigma(\alpha+\beta)} \ldots \sigma(n) \right)
        \end{aligned} 
    \end{align}
    for $ \sigma\in\shuffles{n}{\alpha} $, \ie\ $ \sigma^{\beta}\in\shuffles{n}{\alpha-1} $, $ \sigma_{\beta}\in\shuffles{n}{\alpha+1} $,
    $ (\sigma^{\beta})_{\rbeta}= \sigma $ and $ (\sigma_{\beta})^{\rbeta}= \sigma $, respectively, 
    since the partial shift results in $ \sigma^{\beta}(\alpha+\rbeta)=\sigma(\beta) $ and $ \sigma_{\beta}(\rbeta)=\sigma(\alpha+\beta) $, respectively.
    The example above results in $ \sigma^2 = (3\mid 1\,2\,4\,5)= \surf\surf\transdir\surf\surf $, 
    $ \sigma_2 = (2\,3\,5\mid 1\,4) = \surf\transdir\transdir\surf\transdir  $
    and $ (\sigma^2)_4 = (\sigma_2)^1 = \sigma $.
    Note that the associated Hasse diagram to present all $ 2^n $ shuffles based on one of the two relations above yields 
    a $ n $-dimensional hypercube graph, similarly to power sets ordered by inclusion.
    Occasionally, we filter the transversal components of $ \sigma\in\shuffles{n}{\alpha} $ within another shuffle  $ \tsigma\in\shuffles{n}{\talpha} $
    and renumber the remaining elements with $ \{1,\ldots,n-\alpha\} $ by maintaining the order, \ie\ $ \bsigma:=\tsigma\setminus\sigma|_{\{1,\ldots,\alpha\}}\in\shuffles{n-\alpha}{\balpha} $
    .
    For instance, if $ \tsigma=(2\,5\mid 1\,3\,4) $ then our leading example above yields
    $ \bsigma= \tsigma\setminus\{3\,5\}=(2\mid 1\,3)\in\shuffles{3}{1}$, where the remaining 4 is renamed to 3. 
    As a consequence it holds 
    \begin{align*}
        \left\{ 1\le\beta\le n-\alpha \mid (\tsigma^{-1}\circ\sigma)(\alpha+\beta)\le\talpha \right\}
            &= \left\{ 1\le\bbeta\le n-\alpha \mid \bsigma^{-1}(\bbeta)\le\balpha \right\} \formPeriod
    \end{align*}
    These are all instantaneous indices \wrt\ $ \sigma $, whose permuted elements are transversal in $ \tsigma $.   
    Especially for the example above, this set becomes $ \{2\} $.
    
    Once introduced we like to use the shuffles also for the shuffled flat operator $ \shflat[\sigma] $ established in \autoref{sec:convected_derivatives}.
    Thus for $ \sigma\in\shuffles{n}{\alpha} $ the $ \alpha $ front elements indicate indices which stay up and the $ n-\alpha $ rear elements advertise indices used for lowering. Related to above we deploy the word 
    $ \Lambda^n_\alpha:= \sharp\ldots\sharp\flat\ldots\flat  $ for syntactical assignment, 
    but for the operator directly, \ie\ $ \shflat[\sigma]= \Lambda^n_\alpha(\sigma(1))\ldots\Lambda^n_\alpha(\sigma(n))$.
    Hence the leading example in this section becomes $ \shflat[(3\,5\mid 1\,2\,4)]=\flat\flat\sharp\flat\sharp $.
    Analogously to above we define $ \flat^{n}:=\shflat[\shId{n}{0}]$ to purpose lowering all indices
    and $ \sharp^n:= \shflat[\shId{n}{n}]$ in addition. 
 
\section{Linear mapping \wrt\ single dimensions of spacetime tensor fields}\label{sec:linear_maps}
    We consider in this section linear maps $ \varstarbt\underdot[l]:\tangentstS[^n]\rightarrow\tangentstS[^n] $, 
    which afflicts only the $ l $-th dimension of spacetime $ n $-tensors
    $ \starbt\in\tangentstS[^n] $.
    Such an vector space endomorphism is fully determined by a 2-tensor field $ \varstarbt\in\tangentstS[^2] $ and the rule of calculation
    $ [\varstarbt\underdot[l]\starbt]^{I_1\ldots I_n} = \tensor{\varstarbtC}{^{I_l}_{K}}\starbtC^{I_1\ldots I_{l-1}KI_{l+1}\ldots I_n} $.
    We use the orthogonal decompositions $ \starbt=\sum_{\talpha=0}^n\sum_{\tsigma\in\shuffles{n}{\alpha}}\starbt_{\tsigma} $ for 
    $ \starbt_{\tsigma}=\stproj{\tsigma}\starbt\in\tanproj{\tsigma}{\stsurf} $ and
    $ \varstarbt=\varstarbt_{\surf\surf} + \transdir\otimes\varstarbt_{\transdir\surf} 
                    + (\varstarbt_{\surf\transdir} + \varstarbtC_{\transdir\transdir}\transdir)\otimes\transdir $,
    as well as their pendants $ \arbt=\bbrackets{\starbt}\in\sttensorbS[^{n}] $ and $ \vararbt=\bbrackets{\varstarbt}\in\sttensorbS[^{2}] $,
    with proxy tensors $ \arbt_{\tsigma} = \bbrackets[\tsigma]{\starbt_{\tsigma}}\in\tangentS[^{n-\talpha}] $
    and $ \vararbt_{\tsigma} = \bbrackets[\tsigma]{\varstarbt_{\tsigma}}\in\tangentS[^{2-\talpha}] $ for all appropriated $ \tsigma $.
    Additionally, we write $ \hat{\starbt}_{\tsigma}=\phi_{\tsigma}\starbt_{\tsigma}\in\taninststS[^{n-\talpha}] $ and 
    $ \hat{\varstarbt}_{\tsigma}=\phi_{\tsigma}\varstarbt_{\tsigma}\in\taninststS[^{2-\talpha}] $, \cf\ \eqref{eq:tikzcd_shuffle_spaces}. 
    We observe that the image of $ \varstarbt\underdot[l]\starbt_{\tsigma} $ have a very narrow image compared to $ \tangentstS[^n] $
    and the half of summands vanish according to the $ l $-th dimension of $ \starbt_{\tsigma} $ is either transversal or instantaneous.
    
    The transversal case, where $ \instproj\underdot[l]\starbt_{\tsigma}= 0 $ holds, yields
    \begin{align*}
        \indexten{\varstarbt\underdot[l]\starbt_{\tsigma}}{^{I_1\ldots I_n}}
         &= \frac{1}{\zeta}\left( \hat{\varstarbtC}_{\surf\transdir}^{I_{\tsigma(\beta)}} + 		 \hat{\varstarbtC}_{\transdir\transdir}\transdirC^{I_{\tsigma(\beta)}}\right)
            \transdirC^{I_{\tsigma(1)}} \cdots \widehat{\transdirC^{I_{\tsigma(\beta)}}} \cdots \transdirC^{I_{\tsigma(\talpha)}}
            \hat{\starbtC}_{\tsigma}^{I_{\tsigma(\talpha+1)}\ldots I_{\tsigma(n)}}
    \end{align*}
    with a positive $ \tbeta\le\talpha $ \st\ $ l=\tsigma(\tbeta) $.
    Hence $ \varstarbt\underdot[l]\starbt_{\tsigma} $ is only in 
    $ \tanproj[\tbeta]{\tsigma}{\stsurf}\oplus\tanproj{\tsigma}{\stsurf} \subset \tangentstS[^n] $ and thus
    we have to consider two cases of $ \tsigma $ which give non-vanishing $ \bbrackets[\sigma]{\varstarbt\underdot[l]\starbt_{\tsigma}}\in\tangentS[^{n-\alpha}] $ 
    for a fixed $ \sigma\in\shuffles{n}{\alpha} $.
    This is on the one hand $ \tsigma=\sigma $, which results in
    $ \bbrackets[\sigma]{\varstarbt\underdot[l]\starbt_{\sigma}} 
            = \frac{\vararbtC_{\transdir\transdir}}{\zeta} \arbtC_{\sigma}^{i_1\ldots i_{n-\alpha}}$
    for $ \beta=\tbeta $, \ie\ $ l= \sigma(\beta) $.
    And on the other hand $ \tsigma=\sigma_{\beta} $ with $ \rbeta=\tbeta $, \ie\ it holds
    $ \bbrackets[\sigma]{\varstarbt\underdot[l]\starbt_{\sigma_{\beta}}}
            =\frac{1}{\zeta}\vararbtC_{\surf\transdir}^{i_{\beta}} \arbtC_{\sigma_{\beta}}^{i_1\ldots \hat{i_{\beta}} \ldots i_{n-\alpha}}$ and
    $(\sigma_{\beta})^{\rbeta}=\sigma$ for $ l=¸\sigma_{\beta}(\rbeta)=\sigma(\alpha+\beta) $.
    
    The instantaneous case, where $ \transproj\underdot[l]\starbt_{\tsigma}= 0 $ holds, yields
    \begin{align*}
        \indexten{\varstarbt\underdot[l]\starbt_{\tsigma}}{^{I_1\ldots I_n}}
            &= \left( \indexten{\hat{\varstarbt}_{\surf\surf}}{^{I_{\tsigma(\talpha+\tbeta)}}_{K}} 
                      + \transdirC^{I_{\tsigma(\talpha+\tbeta)}}\indexten{\hat{\varstarbt}_{\transdir\surf}}{_{K}} \right)
                 \transdirC^{I_{\tsigma(1)}} \cdots \transdirC^{I_{\tsigma(\talpha)}}
                 \hat{\starbtC}_{\tsigma}
                    ^{I_{\tsigma(\talpha+1)}\ldots  I_{\tsigma(\talpha+\tbeta -1)} K  I_{\tsigma(\talpha+\tbeta +1)} \ldots I_{\tsigma(n)}}
    \end{align*}
    with a positive $ \tbeta\le n-\talpha $ \st\ $ l=\tsigma(\talpha+\tbeta) $.
    Hence $ \varstarbt\underdot[l]\starbt_{\tsigma} $ is only in 
    $ \tanproj{\tsigma}{\stsurf}\oplus\tanproj{\tsigma_{\tbeta}}{\stsurf} \subset \tangentstS[^n] $ and thus
    we have to consider two cases of $ \tsigma $ which give non-vanishing $ \bbrackets[\sigma]{\varstarbt\underdot[l]\starbt_{\tsigma}}\in\tangentS[^{n-\alpha}] $ 
    for a fixed $ \sigma\in\shuffles{n}{\alpha} $.
    Once again one case is $  \tsigma=\sigma $, which gives
    $ \bbrackets[\sigma]{\varstarbt\underdot[l]\starbt_{\sigma}}  
        = \indexten{\vararbt_{\surf\surf}}{^{i_{\beta}}_{k}} 
            \arbtC_{\sigma}^{i_1 \ldots i_{\beta-1} k i_{\beta+1} \ldots i_{n-\alpha}}$
    with $ \beta=\tbeta $, \ie\ $ l=\sigma(\alpha+\beta) $.
    The other non-trivial case is $ \tsigma=\tsigma^{\beta} $ with $ \rbeta=\tbeta $,
    \ie\ $ \bbrackets[\sigma]{\varstarbt\underdot[l]\starbt_{\sigma^{\beta}}}
            = \indexten{\vararbt_{\transdir\surf}}{_{k}} \arbtC_{\sigma^{\beta}}^{i_1\ldots i_{\rbeta-1} k i_{\rbeta} \ldots i_{n-\alpha}}$,
    $ (\sigma^\beta)_{\rbeta}=\sigma $ and $ l= \sigma^{\beta}(\alpha + \rbeta) = \sigma(\beta) $.
    
    Adding up these two times two cases yields
    $\bbrackets{\varstarbt\underdot[l]\starbt}
        = \sum_{\talpha=0}^n\sum_{\tsigma\in\shuffles{n}{\alpha}} \bbrackets[\sigma]{\varstarbt\underdot[l]\starbt}
                \sttenbase{\sigma}$ with
    \begin{align}\label{eq:QdotlR}
        \bbrackets[\sigma]{\varstarbt\underdot[l]\starbt}
            &=
               \begin{cases}
                    \indexten{\vararbt_{\transdir\surf}}{_{k}} \arbtC_{\sigma^{\beta}}^{i_1\ldots i_{\rbeta-1} k i_{\rbeta} \ldots i_{n-\alpha}}
                        + \frac{\vararbtC_{\transdir\transdir}}{\zeta} \arbtC_{\sigma}^{i_1\ldots i_{n-\alpha}}
                            &\text{if } \exists! \beta \le \alpha:\ l = \sigma(\beta) \formComma\\
                    \indexten{\vararbt_{\surf\surf}}{^{i_{\beta}}_{k}}\arbtC_{\sigma}^{i_1 \ldots i_{\beta-1} k i_{\beta+1} \ldots i_{n-\alpha}}
                        + \frac{1}{\zeta}\vararbtC_{\surf\transdir}^{i_{\beta}} \arbtC_{\sigma_{\beta}}^{i_1\ldots \hat{i_{\beta}} \ldots i_{n-\alpha}}
                            &\text{if } \exists! \beta \le n-\alpha:\ l = \sigma(\alpha+\beta) \formPeriod
               \end{cases}
    \end{align}
    In \autoref{sec:convected_derivatives} we consider the sum
    $ \sum_{l=1}^{\talpha} \varstarbt\underdot[\tsigma(l)]\starbt -  \sum_{l=\talpha +1 }^{n} \varstarbt^{T}\underdot[\tsigma(l)]\starbt$
    for a given shuffle $ \tsigma\in\shuffles{n}{\talpha} $.
    Obviously, all $ \beta=1,\ldots,n $ fulfill the conditions in \eqref{eq:QdotlR}, hence it is also suitable to sum over
    $ \beta $ instead $ l $.
    By linearity of $ \bbrackets{\cdot} $ and validity of 
    $ \bbrackets{\varstarbt^T} = \vararbt_{\surf\surf}^{T}\sttenbase{\surf\surf} 
                                 + \vararbt_{\surf\transdir}\sttenbase{\transdir\surf} 
                                 + \vararbt_{\transdir\surf}\sttenbase{\surf\transdir}
                                 + \vararbtC_{\transdir\transdir}\sttenbase{\transdir\transdir} \in \sttensorbS[^2]$ we justify the following lemma.
    \begin{lem}\label{lem:sumofQdotsR}
        For $ \tsigma\in\shuffles{n}{\talpha} $, $ \varstarbt\in\tangentstS[^2] $, $ \starbt\in\tangentstS[^n] $,
        $ \vararbt=\bbrackets{\varstarbt}= \vararbt_{\surf\surf}\sttenbase{\surf\surf} + \vararbt_{\transdir\surf}\sttenbase{\transdir\surf} 
                                            + \vararbt_{\surf\transdir}\sttenbase{\surf\transdir}
                                            + \vararbtC_{\transdir\transdir}\sttenbase{\transdir\transdir} \in \sttensorbS[^2]$ 
        and $ \arbt=\bbrackets{\starbt}=\sum_{\alpha=0}^{n}\sum_{\sigma\in\shuffles{n}{\alpha}}\arbt_{\sigma}\sttenbase{\sigma}
                    \in\sttensorbS[^n] $ holds
        \begin{align*}
            \bbrackets{\sum_{l=1}^{\talpha} \varstarbt\underdot[\tsigma(l)]\starbt -  \sum_{l=\talpha +1 }^{n} \varstarbt^{T}\underdot[\tsigma(l)]\starbt}
                &= \sum_{\alpha=0}^{n}\sum_{\sigma\in\shuffles{n}{\alpha}}\bbrackets[\sigma]{\sum_{l=1}^{\talpha} \varstarbt\underdot[\tsigma(l)]\starbt 
                                   -  \sum_{l=\talpha +1 }^{n} \varstarbt^{T}\underdot[\tsigma(l)]\starbt}\sttenbase{\sigma} 
                 =: \sum_{\alpha=0}^{n}\sum_{\sigma\in\shuffles{n}{\alpha}}\boldsymbol{s}_{\sigma}\sttenbase{\sigma} \in\sttensorbS[^n]\formComma
         \end{align*}
         where $ \boldsymbol{s}_{\sigma}\in\tangentS[^{n-\alpha}] $ and
         \begin{align*}
             \boldsymbol{s}_{\sigma}
                  &= \sum_{\beta=1}^{\alpha} \left(\begin{Bmatrix} + \\ - \end{Bmatrix}
                           \Big[\vararbt_{\left\{\begin{smallmatrix} \transdir\surf \\ \surf\transdir \end{smallmatrix}\right\}}\Big]_{k}
                                 \arbtC_{\sigma^{\beta}}^{i_1\ldots i_{\rbeta-1} k i_{\rbeta} \ldots i_{n-\alpha}}
                               \begin{Bmatrix} + \\ - \end{Bmatrix} \frac{\vararbtC_{\transdir\transdir}}{\zeta} \arbtC_{\sigma}^{i_1\ldots i_{n-\alpha}}
                              \right)
                     &&\left\{\begin{matrix}
                            \text{if }	(\tsigma^{-1}\circ\sigma)(\beta)\le\talpha\formComma \\
                            \text{otherwise,}
                         \end{matrix}\right. \\
                  &\quad +\sum_{\beta=1}^{n-\alpha} \left( 
                           \begin{Bmatrix} \indexten{\vararbt_{\surf\surf}}{^{i_{\beta}}_k} \\ 
                                         - \indexten{\vararbt_{\surf\surf}}{_k^{i_{\beta}}}\end{Bmatrix} 
                            \arbtC_{\sigma}^{i_1 \ldots i_{\beta-1} k i_{\beta+1} \ldots i_{n-\alpha}}
                            \begin{Bmatrix} + \\ - \end{Bmatrix} \frac{1}{\zeta}
                              \vararbtC_{\left\{\begin{smallmatrix} \surf\transdir \\ \transdir\surf \end{smallmatrix}\right\}}^{i_{\beta}}
                              \arbtC_{\sigma_{\beta}}^{i_1\ldots \hat{i_{\beta}} \ldots i_{n-\alpha}}\right)
                   &&\left\{\begin{matrix}
                               \text{if }	(\tsigma^{-1}\circ\sigma)(\alpha+\beta)\le\talpha\formComma \\
                               \text{otherwise.}
                            \end{matrix}\right. 
        \end{align*}
    \end{lem}

\section{Spacetime and surface quantities}   

\subsection{Tangential derivative of velocity}\label{sec:Bop_N_bop}
    In this section we investigate the tangential derivative $ \nablatan $, see \eg\ \cite{Jankuhn2018}, of a velocity field 
    $ \Wb\in\tangentR $ and associate it with frequently used surface quantities $ \bop,\bop_{\m}\in\tangentS $ and
    $ \Bop,\Bop_{\m}\in\tangentS[^2] $ throughout this paper for either $ \Wb=\obvelo $ or $ \Wb=\matvelo $. 
    The tangential derivative 
    $ \nablatan\Wb:= (\projsurf\cdot\nabla_{\!\R^3}|_{\surf})\Wb = (\nabla_{\!\R^3}\Wb)|_{\surf}\cdot\projsurf \in \tangentR$
    is defined \wrt\ tangential projection $ \projsurf:= \Id[\R^3]\!|_{\surf} - \normal\otimes\normal:\tangentR \rightarrow \tangentS $,
    \ie\ $ \projsurf|_{\tangentS}=\IdS $.
    This means that $ \nablatan\Wb $ is only right-tangential. 
    The left-tangential and -normal part can be calculated with aid of thin film coordinates in a vicinity of $ \surf $, 
    see \cite{Nestler2019, Nitschke2019a}, directly by evaluating $ \innerR{\partial_j \Wb}{\partial_i\para} $ for the tangential part and $  \innerR{\partial_i \Wb}{\normal} $ for the normal part
    or coordinate-free with only applying the product rule. 
    Ultimately, for $ \Wb=\wb+\velonor\normal $ and $ \wb\in\tangentS $, all calculations lead to
    \begin{align*}
        \begin{aligned}
            \nablatan\Wb &= \nabla\wb - \velonor\shop + \normal\otimes\left( \nabla\velonor + \shop\wb \right)\\
                            &=:\begin{cases}
                                    \Bop + \normal\otimes\bop    & \text{if }\Wb=\obvelo\phantom{\matvelo} \text{ (observer velocity)}\\
                                    \Bop_{\m} + \normal\otimes\bop_{\m}    & \text{if }\Wb=\matvelo\phantom{\obvelo}   \text{ (material velocity)}           
                        \end{cases}\formPeriod
        \end{aligned}
    \end{align*}
    Hence, it holds $ \BopC_{ij} = \innerR{\partial_j \obvelo}{\partial_i\para}= \obveloSC_{i|j} - \velonor \shopC_{ij} $
    and $ \bopC_{i}= \innerR{\partial_i \obvelo}{\normal}= \velonor_{|i} + \shopC_{ij}\obveloSC^j $ \wrt\ observer velocity.
    The same applies \wrt\ material velocity.
    
\subsection{Rate of surface metric tensor}\label{sec:rate_of_metric_tensor}
    Actually, the rate of metric tensor $ \partial_t g_{ij} $ is twice the rate of observer deformation tensor, see \eg\ \cite{Arroyo2009}.
    A given embedding of $ \surf $ under observer parametrization $ \para $ and \ref{sec:Bop_N_bop} leads to
    \begin{align*}
        \partial_t g_{ij} &= \innerR{\partial_i \obvelo}{\partial_j\para} + \innerR{\partial_i\para}{\partial_j \obvelo}
                           = \BopC_{ji} + \BopC_{ij} \formPeriod
    \end{align*}
    From this it follows for the inverse metric tensor $ g^{ij}= g^{ik}g^{il}g_{kl} $ that
    \begin{align*}
        \partial_t g^{ij} &= 2\partial_t g^{ij} +  g^{ik}g^{il}\partial_t g_{kl}
                           = - g^{ik}g^{il}\partial_t g_{kl}
                           = -\left( \BopC^{ji} + \BopC^{ij} \right)\formPeriod
    \end{align*}
    
\subsection{Acceleration}\label{sec:accel}
    The \emph{observer acceleration} is $ \obaccel:= \obaccelS + \obaccelnor\normal := \partial_t\obvelo  $ with
    \emph{tangential observer acceleration} $ \obaccelS\in\tangentS $ 
    and scalar-valued \emph{normal observer acceleration} $ \obaccelnor\in\tangentS[^0]$ \wrt\ observer parametrization $ \para $.
    By means of \ref{sec:Bop_N_bop}, \ref{sec:rate_of_metric_tensor} and orthogonality $ \partial_t\normal\bot\normal $,
    we calculate
    \begin{align*}
        \obaccelSC_{i} 
            &= \innerR{\obaccel}{\partial_i \para}
              = \partial_t\obveloSC_i - \obveloSC^k\innerR{\partial_k\para}{\partial_i\obvelo} - \velonor\innerR{\normal}{\partial_i\obvelo}
              =  \partial_t\obveloSC_i - \indexten{\Bop^{T}\obveloS + \velonor\bop}{_i}\\
        \obaccelSC^{i} 
            &= \partial_t\obveloSC^i + \indexten{\Bop\obveloS - \velonor\bop}{^i}\\
        \obaccelnor &= \innerR{\obaccel}{\normal}
                     = \obveloSC^k\innerR{\partial_k\para}{\normal} + \partial_t\velonor
                     = \partial_t\velonor + \innerS{\bop}{\obveloS}\formPeriod
    \end{align*}
    We notice that the normal acceleration is observer dependent in contrast to the normal velocity.

\subsection{Spacetime Christoffel symbols}\label{sec:christoffel_symbols}
    Usually, Christoffel symbols are calculated by partial derivatives of the metric tensor $ \stg $.
    For a given parametrization $ \stpara $ of the spacetime manifold $ \stsurf $ it is easier to develop the Christoffel symbols of first kind
    by $ \gamma_{IJK} = \innerstR{\partial_I \partial_J \stpara}{\partial_K \stpara} $ though.  
    Using \ref{sec:Bop_N_bop} and \ref{sec:accel} leads to
    \begin{align*}
        \gamma_{ijk} &= \innerR{\partial_i \partial_j \para}{\partial_k \para} = \Gamma_{ijk}
        &\gamma_{ijt} &= \innerR{\partial_i \partial_j \para}{\obvelo} = \Gamma_{ijk}\obveloSC^k + \velonor\shopC_{ij} \\
        \gamma_{itk} &= \innerR{\partial_i \obvelo}{\partial_k \para} = \BopC_{ki}
        &\gamma_{itt} &= \innerR{\partial_i \obvelo}{\obvelo} = \BopC_{ki}\obveloSC^{k} + \velonor\bopC_{i} \\
        \gamma_{ttk} &= \innerR{\obaccel}{\partial_k \para} = \obaccelSC_{k}
        &\gamma_{ttt} &= \innerR{\obaccel}{\obvelo} = \innerS{\obaccelS}{\obveloS} + \obaccelnor\velonor
    \end{align*}
    up to symmetry in the two first indices.
    Rising rear indices, \ie\ $ \gamma^{t}_{IJ} = \stgC^{tK}\gamma_{IJK} = \zeta(\gamma_{IJt} - \obveloSC^k\gamma_{IJk}) $
    and $ \gamma^{l}_{IJ} = \stgC^{lK}\gamma_{IJK} = g^{lk}\gamma_{IJk} - \obveloSC^l\gamma_{IJt} $,
    accomplish the Christoffel symbols of second kind
    \begin{align*}
         \gamma_{ij}^{t} &= \zeta\velonor \shopC_{ij}
			&\gamma_{ij}^{k} &= \Gamma_{ij}^{k} - \gamma_{ij}^{t}\obveloSC^{k} \\
			\gamma_{tj}^{t} &= \zeta\velonor\bopC_j 
                                = \zeta\velonor\indexten{\nabla\velonor + \shop\obveloS}{_j} 
			&\gamma_{tj}^{k} &= \tensor{\BopC}{^{k}_{j}} - \gamma_{tj}^{t}\obveloSC^{k} 
                              = \indexten{\nabla\obveloS - \velonor\shop}{^k_j} - \gamma_{tj}^{t}\obveloSC^{k}  \\
			\gamma_{tt}^{t} &= \zeta\velonor\obaccelnor 
                              = \zeta\velonor\left(\partial_t\velonor + \innerS{\bop}{\obveloS}\right)
			&\gamma_{tt}^{k} &= \obaccelSC^{k} - \gamma_{tt}^{t}\obveloSC^{k}     
                              = \partial_t\obveloSC^k + \indexten{\Bop\obveloS - \velonor\bop}{^k} - \gamma_{tt}^{t}\obveloSC^{k}
    \end{align*} 
    up to symmetry in the lower indices.
    
 \subsection{Gradient of material direction}\label{sec:gradmatdir}
    In this section, we determine the gradient $ \nabla\matdir\in\tangentstS[^1_1] $ of the material 
    direction $ \matdir=[1,\relvelo]'_{\tangentstS[^1]}\in\tangentstS $ and its orthogonal spacetime representation $ \bbrackets{\nabla\matdir}\in\sttensorbS[^2] $, where $ \relvelo=\matveloS-\obveloS $ is the relative velocity.
    The calculations of the components $ \indexten{\nabla\matdir}{^{I}_{K}} = \partial_k\transdirC_{m}^{I} + \gamma_{KJ}^{I}\transdirC_{m}^{J} $
    are straightforward using the Christoffel symbols in \ref{sec:christoffel_symbols} and lead to
    \begin{align*}
        \indexten{\nabla\matdir}{^{t}_{k}}
            &= \zeta\velonor\indexten{\bop_{\m}}{_{k}} \formComma
         &\indexten{\nabla\matdir}{^{i}_{k}}
            &= \indexten{\Bop_{\m} - \zeta\velonor\obveloS\otimes\bop_{\m}}{^{i}_{k}} \\
         \indexten{\nabla\matdir}{^{t}_{t}}
            &= \zeta\velonor\left( \dot{\velonor} + \innerS{\bop_{\m}}{\obveloS} \right) \formComma
         &\indexten{\nabla\matdir}{^{i}_{t}}
            &= \partial_t\matveloSC^{i} + \indexten{\Lie^{\sharp}_{\relvelo}\matveloS + \Bop_{\m}\obveloS - \velonor\bop_{\m}
                                                     -\zeta\velonor\left( \dot{\velonor} + \innerS{\bop_{\m}}{\obveloS} \right)\obveloS}{^{i}}
                                                            \formComma
    \end{align*} 
    where we used that $ \bop + \shop\relvelo = \bop_{\m} $, $ \Bop + \nabla\relvelo=\Bop_{\m} $,
    $ \partial_t\velonor + \innerS{\bop}{\matveloS} = \dot{\velonor} + \innerS{\bop_{\m}}{\obveloS} $,
    $ \Bop\matveloS-\velonor\bop = \Lie^{\sharp}_{\relvelo}\matveloS + \Bop_{\m}\obveloS - \velonor\bop_{\m} $
    and $  \Lie^{\sharp}_{\relvelo}\matveloS = \nabla_{\relvelo}\matveloS - \nabla_{\matveloS}\relvelo $.
    Rising the right-hand index gives 
    \begin{align*}
        \nabla\matdir
                &= \begin{bmatrix}
                        0 & 0 \\ 0 &\Bop_{\m}
                    \end{bmatrix}_{\tangentstS[^2]}
                    +\zeta\begin{bmatrix}
                        0\\ \instcond{\sharp}\matveloS - \velonor\bop_{\m}
                    \end{bmatrix}_{\tangentstS[^1]} \otimes \transdir
                    +\zeta\transdir\otimes\begin{bmatrix}
                                                0\\\velonor\bop_{\m}
                                            \end{bmatrix}_{\tangentstS[^1]}
                    +\zeta^2\velonor\dot{\velonor}\transdir\otimes\transdir
    \end{align*}
    with $ \indexten{\instcond{\sharp}\matveloS}{^i} = \partial_t\matveloSC^{i} + \indexten{\Lie^{\sharp}_{\relvelo}\matveloS}{^i} $.
    Ultimately, we deduce from this that 
    \begin{align}\label{eq:gradmatdir}
    \bbrackets{\nabla\matdir} &= \Bop_{\m}\sttenbase{\surf\surf} + \zeta( \instcond{\sharp}\matveloS - \velonor\bop_{\m} )\sttenbase{\surf\transdir}
                                      + \zeta\velonor\bop_{\m}\sttenbase{\transdir\surf} + \zeta^2\velonor\dot{\velonor}\sttenbase{\transdir\transdir}
                        \formPeriod
    \end{align}

\vspace*{1cm}
\noindent
{\bf Acknowledgements}: AV was supported by DFG through FOR3013. 

\bibliographystyle{elsarticle-num} 
\bibliography{refs.bib}

\end{document}